\documentclass[11pt]{article}
\usepackage{lmodern}
\usepackage{jheppub}
\usepackage{caption} % Ensures standard caption defaults
\usepackage{hyperref}
\usepackage{amsmath,amssymb,amsthm,amscd}
\usepackage{mathtools}
\usepackage[all,cmtip]{xy}
\usepackage{xcolor}
\usepackage{blkarray}
\usepackage{float} 
\usepackage{youngtab}
\usepackage{ytableau}
\usepackage[shortlabels]{enumitem}
\Ylinethick{0.3pt}
\Yboxdim{5pt}
\usepackage{tikz,textcomp}
\usetikzlibrary{decorations.pathreplacing,calc,fadings,fit,shapes,arrows,positioning,chains,matrix}
\usepackage{textgreek}
\DeclareFontFamily{U}{wncy}{}
\DeclareFontShape{U}{wncy}{m}{n}{<->wncyr10}{}
\DeclareSymbolFont{mcy}{U}{wncy}{m}{n}
\DeclareMathSymbol{\Sh}{\mathord}{mcy}{"58} 
\usepackage{silence}%This removes the warnings 
\numberwithin{equation}{section}

\usepackage{multirow}
\usepackage{booktabs}
\def\bea{\begin{eqnarray}}
\def\eea{\end{eqnarray}}
\def\be{\begin{equation}}
\def\ee{\end{equation}}
\def\ba{\begin{align}}
\def\ea{\end{align}}
\def\bse{\begin{subequations}}
\def\ese{\end{subequations}}
\def\nn{\nonumber}
\newcommand{\cM}{\mathcal{M}}
\newcommand{\IP}{\mathbb{P}}
\newcommand{\IF}{\mathbb{F}}
\newcommand{\eps}{\epsilon}
\newcommand{\ch}{\text{ch}}
\newcommand{\sgn}{{\rm sgn}}

\newcommand{\yadj}{y_{\rm adj}}
\newcommand{\madj}{m_{\rm adj}}
\newcommand{\sfq}{\mathsf{q}}
\newcommand{\sfp}{\mathsf{p}}
\newcommand{\cN}{\mathcal{N}}

\newcommand{\cO}{\mathcal{O}}
\newcommand{\cE}{\mathcal{E}}
\newcommand{\IC}{\mathbb{C}}
\newcommand{\IZ}{\mathbb{Z}}
\newcommand{\I}{{\rm i}}
\newcommand{\de}{{\rm d}}
\newcommand{\Gr}{{\rm Gr}}
\newcommand{\Tr}{{\rm Tr}}

\usepackage{longtable}

\usepackage{graphicx}
\usepackage{subcaption}

\title{Refined Vafa--Witten invariants for toric surfaces from supersymmetric localization in 5D gauge theory}
\abstract{
We study the partition function of five-dimensional $\mathcal{N}=1$ $U(N)$ supersymmetric Yang--Mills (SYM) theory with an adjoint hypermultiplet of mass $m_{\rm adj}$ on a toric K\"ahler surface $S$ times a circle of radius $\boldsymbol{\beta}$. Extending earlier work in $\mathcal{N}=2^*$ SYM theory on $S$, and in pure $\cN=1$ SYM on $S\times \mathbb{S}^1_{\boldsymbol{\beta}}$, we find that the path integral localizes to an integral along the Cartan torus of the product of Nekrasov 5D partition functions for each affine patch. Restricting to the gauge group $U(2)$ for simplicity, the integrand has an infinite set of poles of degree at most $\chi(S)-2$.  With a natural prescription for integrating around such poles, we find that the contributing poles are in one-to-one correspondence with the torus-fixed points in the moduli space of semi-stable torsion-free sheaves on $S$.
Moreover, for non-even first Chern class, their contributions are independent of the equivariant parameters $\epsilon_1,\epsilon_2$ and add up to the $\chi_{y^2}$--genus of that moduli space, where $y^2=e^{-\boldsymbol{\beta} m_{\rm adj}}$, and hence coincide with the refined Vafa--Witten invariants. For even Chern class, the partition function depends on the equivariant parameters $\epsilon_1,\epsilon_2$ as well as $y$, and its relation to rational, refined Vafa--Witten invariants remains unclear.
}

\author[a]{Osama Khlaif,}
\affiliation[a]{Philippe Meyer Institute, Physics Department, École Normale Supérieure (ENS), Université PSL, 24 rue Lhomond, F-75231 Paris, France}
\emailAdd{osama.khlaif@phys.ens.fr}

\author[b]{Boris Pioline}
\affiliation[b]{Laboratoire de Physique Th\'eorique et Hautes
Energies (LPTHE), UMR 7589 CNRS-Sorbonne Universit\'e,
Campus Pierre et Marie Curie,
4 place Jussieu, F-75005 Paris, France}
\emailAdd{pioline@lpthe.jussieu.fr}

\author[c,d,e]{Alessandro Tanzini}
\affiliation[c]{SISSA, Via Bonomea 265, 34136 Trieste, Italy}
\affiliation[d]{INFN, Sezione di Trieste, Trieste, Italy}
\affiliation[e]{Institute for Geometry and Physics, IGAP, via Beirut 2, 34136 Trieste, Italy}
\emailAdd{tanzini@sissa.it}

\begin{document}

\maketitle
\flushbottom

%%%%%%%%%%%%%%%%%%%%%%%%%%%%%%%%%%%%%%%%%
%%%%%%%%%%%%%%%%%%%%%%%%%%%%%%%%%%%%%%%%%
%%%%%%%%%%%%%%%%%%%%%%%%%%%%%%%%%%%%%%%%%
\section{Introduction and summary}

Supersymmetric quantum field theories often possess protected observables, which can be computed exactly by exploiting their analytic structure or by using localization techniques.  These observables are often sensitive to the topology of the manifold on which the Euclidean theory is defined and related to topological invariants of mathematical interest. In the case of four-dimensional,
$\cN=2$ supersymmetric Yang--Mills (SYM) theories with gauge group $U(2)$, this has had a dramatic impact on the theory of Donaldson invariants of four-manifolds, including its reformulation in terms of Seiberg--Witten (SW) invariants~\cite{Witten:1988ze,Witten:1994cg} and computation scheme using the $u$-plane integral~\cite{Moore:1997pc}. For four-dimensional
SYM theories with $\cN=4$ supersymmetry and gauge groups $U(N)$ of arbitrary rank, this has led to the definition of Vafa--Witten (VW) invariants, and to the conjecture that their generating series should be modular invariant, as a consequence of the S-duality symmetry of 
$\cN=4$ super-Yang--Mills theory~\cite{Vafa:1994tf}. When the four-manifold $S$ is a K\"ahler surface, these invariants can be defined mathematically in terms of intersection theory on the moduli space of stable coherent sheaves on $S$, or more generally the moduli stack of semi-stable Higgs 
pairs~\cite{Tanaka:2017jom,Tanaka:2017bcw}. When the four-manifold $S$ has $b_2^+(S)=1$, as is the case for Fano surfaces, these invariants depend on the K\"ahler class of the metric, and exhibit wall-crossing behavior as the polarization is varied~\cite{Gottsche:1996aoa}. Consequently, the 
generating series of VW invariants is expected to be mock modular, with a precise 
modular anomaly~\cite{Vafa:1994tf,Alexandrov:2019rth,Dabholkar:2020fde,
Alexandrov:2020bwg,Alexandrov:2020dyy}.

Five-dimensional Yang--Mills theories are notoriously hard to define, being strongly coupled in the UV. However, their supersymmetric counterparts can be constructed by reducing the six-dimensional $(0,2)$ SCFT on a circle, or by compactifying M-theory on a toric Calabi--Yau singularity. In particular, one can 
consider $\cN=1$ five-dimensional super Yang--Mills theory with gauge group $U(2)$ on a five-manifold which is the product of a four-manifold $S$ times a circle $\mathbb{S}^1_{\boldsymbol{\beta}}$ of radius $\boldsymbol{\beta}$. As anticipated in~\cite{Nekrasov:1996cz}, this leads to a K-theoretic version of Donaldson
%Thomas (DT) 
invariants, introduced from the mathematical viewpoint in~\cite{Nakajima:2005fg,Gottsche:2006bm} and
studied most recently from a physics perspective in~\cite{Kim:2025fpz}. These K-theoretic invariants are now given by the  $L^2$-index of a suitable Dirac operator on the moduli space of $U(2)$ instantons, coupled to a suitable determinant line bundle. 

In this paper, we instead consider five-dimensional $\cN=1^*$ SYM theory on the product $S\times \mathbb{S}^1_{\boldsymbol{\beta}}$, where the star means that the gauge theory has an additional hypermultiplet of mass $\madj$ transforming in the adjoint representation of the gauge group. In the limit $\madj\to 0$, the resulting theory has $\cN=2$ supersymmetry, and its reduction on a circle gives $\cN=4$ SYM theory. The adjoint matter multiplet naturally arises in the reduction from the six-dimensional $(0,2)$ SCFT. It is therefore natural to expect that it computes a K-theoretic version of the VW invariants. In the case of Fano surfaces, which we focus on this paper, there is no contribution from the monopole branch and these K-theoretic invariants are identified with the $\chi_{y^2}$-genus of the moduli space of semi-stable sheaves on $S$ (as anticipated in~\cite{Hollowood:2003vt,Manschot:2021qqe})\footnote{Note that for a toric surface, the cohomology of the moduli space of semi-stable sheaves is supported on Dolbeault degree $(p,p)$, hence the $\chi_{y^2}$-genus coincides with the Poincar\'e polynomial. Similarly, in the limit $y\to 1$, the holomorphic Euler characteristic coincides with the Euler number.}. For general projective surfaces, they should correspond to the K-theoretic VW invariants introduced in~\cite{Thomas:2018lvm}. The case of four-dimensional $\cN=2^*$ SYM, arising in the limit $\boldsymbol{\beta}\to 0$ of our set-up, was considered in~\cite{Manschot:2021qqe}, where the $u$-plane integral with Donaldson--Witten (DW) integrals was developed.

Our analysis will instead use supersymmetric localization, assuming that $S$ is a toric Fano surface and generalizing the approach developed in~\cite{Bershtein:2015xfa,Bershtein:2016mxz,Bonelli:2020xps} in four dimensions and extended in \cite{Kim:2025fpz} to five dimensions. In this approach, first suggested by Nekrasov in \cite{nekrasov2006localizing}, the equivariant partition function of 5D $\cN=1^*$ SYM theory on $S \times \mathbb{S}^1_{\boldsymbol{\beta}}$ is given by the product of the 5D Nekrasov partition functions on each affine $\IC^2$ patch, integrated over the Cartan torus of the gauge group $G$, which we assume to be $U(2)$ for simplicity. The relevant version of Nekrasov partition function with a massive adjoint, including both the classical, one-loop, and instanton factors, was constructed in~\cite{Iqbal:2008ra,Poghossian:2008ge} and revisited in~\cite{Hwang:2014uwa,Brennan:2018yuj}. 
A crucial tool to study the analytic properties of the integrand is AGT correspondence \cite{Alday_2010} and its extension to 5D supersymmetric gauge theories, relating the 5D-instanton partition function to $\sfq$-Virasoro conformal blocks \cite{AwataYamada2010}. In this paper, we generalize Zamolodchikov's recursion relations to~$\sfq$-Virasoro conformal blocks on the torus with one puncture, corresponding via $\sfq$-AGT to 5D $\mathcal{N}=1^*$ $U(2)$ gauge theory.
For a toric Fano surface of Euler number $\chi$\footnote{In practice, we consider $S=\IP^2$ with $\chi=3$, $S=\IP^1\times \IP^1$ 
with $\chi=4$ and the Hirzebruch surface $S=\IF_1$, obtained by blowing up $\IP^2$ at one point, also having $\chi=4$.},  after multiplying out the factors from each of the $\chi$ affine patches, the integrand exhibits an infinite set of poles of order $\chi-2$, labeled by the toric fluxes $(\sfp_\ell)_{\ell=1,\dots, \chi}$ in each patch. Due to these poles, the integral is a non-holomorphic function of the complexified gauge coupling $\tau=\frac{\theta}{2\pi}+\frac{4\pi\I}{g_{\rm YM}^2}$. In the holomorphic limit $g_{\rm YM}\to 0$ keeping $\tau$ fixed, each pole contributes via its residue, weighted with a sign depending on the polarization $J$ (i.e., the K\"ahler class on $S$). Generalizing~\cite{Bershtein:2015xfa,Bonelli:2020xps}, we observe a one-to-one correspondence between the contributing fluxes $(\sfp_\ell)$ (or rather, their Weyl orbit under $\sfp_\ell\mapsto - \sfp_\ell$) and Klyachko's classification of fixed points under the toric action on the space of slope-stable torsion-free sheaves~\cite{klyachko1990equivariant,Kool2015}. 

For contributions with first Chern class $c_1\neq 0$ modulo 2, upon identifying $\yadj \equiv e^{-\boldsymbol{\beta}\madj}=y^2$,
the sum over poles reproduces, as expected, the $\chi_{y^2}$-genus of the moduli space of stable coherent sheaves on $S$ computed in
\cite{Yoshioka1994,Yoshioka1995} and tabulated in \cite[Appendix A] {Beaujard:2020sgs}. In particular, the dependence on the equivariant parameters 
$(\epsilon_1,\epsilon_2)$ entirely disappears, as a consequence of the projectivity and smoothness of the corresponding moduli spaces. The 
supersymmetric localization method therefore perfectly reproduces the expected generating series of refined Vafa--Witten invariants, at least up to the order where 
we have computed them. 

For even $c_1$, due to the existence of strictly semi-stable sheaves (or, in physics parlance, marginal bound states),  
the relation between the results from supersymmetric localization and refined (or motivic) Vafa--Witten invariants is unfortunately less transparent. 
In general, we find that after summing over all poles,
and weighting strictly semi-stable contributions with a coefficient $1/2$ as implied by supersymmetric localization results, the dependence on the equivariant parameters $y_1=e^{-\boldsymbol{\beta}\epsilon_1}, y_2=e^{-\boldsymbol{\beta}\epsilon_2}$ does not drop out, except if one first takes the unrefined limit $\yadj\to 1$\footnote{This follows from the enhancement of supersymmetry in this limit, which implies that the fluctuation determinants around fixed points cancel, leaving a sum over fixed points with unit weight.}. Moreover, in the unrefined limit we find agreement with the rational Vafa--Witten invariants only after
adding by hand a contribution $-1/(4\overline{\eta}^{2\chi})$, where $\overline{\eta}$ is the normalized Dedekind eta function defined in \eqref{eta bar}, 
which we attribute to contributions with degenerate flux vectors such that $c_2(\vec\sfp)=0$. For $S=\IP^2$,
this is related to the constant term $-1/12$ in the generating series of Hurwitz class numbers $H_0(\tau)$, which is crucial for ensuring (mock) modular properties, but whose
physical or mathematical origin is largely mysterious. We speculate that a proper equivariant, refined extension of this degenerate contribution would allow us to remove
the dependence of our refined Vafa--Witten invariants on the equivariant parameters, at least for odd $c_2$ where the moduli space of semi-stable sheaves is expected to be smooth and projective, but we have not found a prescription that does the job.\footnote{In particular, it is natural to expect that $-1/(4\overline{\eta}^{2\chi})$ should be replaced
by $C\, (h_1^S)^2$, where $h_1^S$ is the $U(1)$ $\mathcal{N}=1^*$ instanton partition function in \eqref{ZU(1) = h1S} and $C$ is a $\sfq$-independent coefficient, which could depend on $\yadj, y_1,y_2$.
However, choosing $C$ such that the refined VW invariant at order $\sfq$ (i.e. $c_2=1$) vanishes, the dependence on equivariant parameters does {\it not}
drop out at higher orders in $\sfq$, nor does it lead
to the correct refined VW invariants in the non-equivariant limit. See the discussion below \eqref{P2q1y10} for more details. }

Despite this puzzle, our results pave the way to an extension of 5D localization techniques in several directions. Firstly, it is interesting to extend our analysis to higher rank gauge groups; in that case, one should use the Jeffrey--Kirwan residue prescription to deal with multi-dimensional contour integrals~\cite{BonelliToAppear}. Secondly, our results can be extended to compute topological correlators in the 5D $\mathcal{N}=1^*$ theory for any compact smooth toric surface, providing a K-theoretic version of the $\cN=2^*$ invariants introduced in~\cite{Manschot:2021qqe}, which would now interpolate between standard K-theoretic Donaldson--Witten invariants and Vafa--Witten invariants, while maintaining modularity.
At a more basic level,  our techniques can be also be used to compute K-theoretic DW invariants for any compact toric surfaces beyond the cases already considered in~\cite{Gottsche:2006bm,Kim:2025fpz}. It would be also interesting to include charged hypermultiplets in the fundamental representation.\footnote{For more recent developments on the 4D $\mathcal{N}=2$ SYM theories with fundamental matter from the $u$-plane integral perspective, see~\cite{Aspman:2021vhs,Aspman:2021kfp,Aspman:2021evt,Aspman:2022sfj,Aspman:2023ate,Furrer:2026byd} and references therein.}  To this end one would need to generalize Zamolodchikov's recursion relation to $\sfq$-Virasoro conformal block on the sphere with four punctures. 
Fintushel--Stern blow-up formulae for Donaldson invariants can be recovered from the non-equivariant limit of blow-up formulae of $\mathcal{N}=2$ SYM theory \cite{Nakajima:2003uh}.
Much in the same vein, one could derive the blow-up formulae for refined Vafa--Witten invariants~\cite{Gottsche_1999,Li_1999,kuhn2025blowupformulainstantonvafawitten} from the non-equivariant limit of blow-up formulae for 5D $\mathcal{N}=1^*$ supersymmetric theory, yet to be determined. Blow-up formulae for $\mathcal{N}=2^*$ supersymmetric theory in 4D are known~\cite{Bershtein:2021uts, Bonelli:2025bmt}, and it would be interesting to uplift them to 5D.  
We hope to report some progress in these directions in the future.

The outline of this article is as follows. In Section \ref{sec_toricloc}, we review the toric localization
method for computing Euler numbers of the  moduli space of slope-stable rank $N$ coherent sheaves on a toric surface $S$,
with special emphasis
on the rank $N = 2$ case. In Section~\ref{sec_susyloc}, we review the supersymmetric localization method
in 5D $\cN=1$ SYM theories on $S\times \mathbb{S}^1_{\boldsymbol{\beta}}$ with a partial twist along $S$,
and explain how the inclusion of a massive adjoint hypermultiplet leads to the equivariant
$\chi_{y^2}$-genus of the moduli space of Gieseker-stable coherent sheaves on $S$.
In Section \ref{sec_5Dpart}, we construct the partition function of 5D $\cN=1^*$ SYM theory  first on $\IC^2$,
and then on an arbitrary toric surface by gluing together partition functions on each affine patch.
In Section \ref{sec_U1part}, we work out the partition function in the Abelian case $G=U(1)$, where no integral over the Coulomb
branch is necessary, and reproduce G\"ottsche's formula for the generating series of Poincar\'e polynomials of the
Hilbert scheme of $k$ points on $S$. In the remaining Sections \ref{sec:VW2P2} and \ref{sec:VW2Fn}, we compute the partition function 
of 5D $\cN=1^*$ SYM with gauge group $U(2)$ on $S=\IP^2$ and $\IF_n$, respectively, and compare the 
result from supersymmetric localization to earlier computations of refined VW invariants of the corresponding surfaces.
Finally, we collect in the Appendices technical details on the derivation of some formulae quoted in the main text, and  explicit results for the Hirzebruch surfaces $\IF_0$ and $\IF_1$ at low instanton number.

\section{Localization on moduli spaces of torsion-free sheaves on toric surfaces}

\label{sec_toricloc}

In this section, we briefly review mathematical results on equivariant vector bundles and torsion-free sheaves on toric surfaces, and on 
fixed points of the toric action on the moduli space of slope-stable rank $N$ coherent sheaves on a toric surface $S$. Our aim is to compute the Euler characteristics of the moduli space of torsion-free sheaves on $S$ following the toric localization method of \cite{Kool2015}, 
which we shall then compare to the results of supersymmetric localization in later sections. We start
with a telegraphic review of toric surfaces, referring to~\cite{cox2024toric} for a more systematic introduction.

\subsection{Lightning review of toric surfaces}

A $d$-dimensional toric variety $S$ is a complex manifold with an action of the algebraic torus $(\IC^\times)^d$ having a dense open orbit. It can be decomposed into a set of complex tori $(\IC^\times)^{d-k}$ with $0\leq k\leq d$ associated to $k$-dimensional cones in a fan in $\IZ^d$. For $d=2$, the fan includes $\chi:=\chi(S)$ two-dimensional cones, describing the affine $\IC^2$ patches, and the same number of one-dimensional cones, which describe the toric divisors $D_\ell, \ell=1\dots\chi$ along which these affine patches are glued. Let $(\rho_1,\dots,\rho_\chi)$ be the primitive vectors in $\IZ^2$ generating the one-dimensional cones, in clockwise order. We identify the rays modulo $\chi$, so that $\rho_\ell=\rho_{\ell+\chi}$. The intersection product $D_\ell \cdot D_{\ell'}$ vanishes unless $\ell-\ell'\in \{-1,0,1\}$ modulo $\chi$, and satisfies:
\be\label{divisor-intersection}
D_\ell \,\cdot\, D_{\ell+1}\,=\,1~, \qquad D_\ell \,\cdot\, D_\ell \,=\, -\, h_\ell~,
\ee
where the integers $h_\ell$ are determined by the relations:
\be
\rho_{\ell-1}\,+\,\rho_{\ell+1} \,=\, h_\ell\, \rho_\ell~, \qquad \forall \,\ell\,=\,1,\,\cdots,\,\chi~.
\ee
For any vector $\rho\in \IZ^2$, the following sum vanishes:
\be
\sum_{\ell=1}^{\chi}\, (\rho,\rho_\ell)\, D_\ell\,=\,0~,
\ee 
consistent with the fact that the number of linearly independent divisors is 
$\chi-2=b_2(S)$. Here, $(\cdot,\cdot)$ is the standard inner product on $\IZ^2$. The first Chern class of the toric surface is given by the sum of all toric divisors:
\be
c_1(S) \,=\, \sum_{\ell=1}^{\chi}\, D_\ell~, \qquad D_\ell\, \cdot\, c_1(S) \,=\, 2\, -\, h_\ell~.
\ee
Let $K_S$ and $\mathcal{O}_S$ be the canonical line bundle and the structure sheaf over $S$, respectively. Using the relations $K_S^2=2\chi+3\sigma$, $\chi(\cO_S)=\frac14(\chi+\sigma)=1$ we find that the intersection numbers $h_\ell$ satisfy: 
\begin{equation}\label{sum inter numbs}
    \sum_{\ell=1}^{\chi}\,h_{\ell}\,=\,-\, 3\,\sigma(S)\,=\, -\,3\,(2\,-\,b_2(S))\,=\,3\,(\chi\,-\,4)~,
\end{equation}

\medskip
\noindent
\textbf{$\Omega$-background parameters.}
We now introduce equivariant parameters $(\epsilon_1,\epsilon_2)$ corresponding to the $\IC^\times \times \IC^\times$ action in the first affine patch $\ell=1$. 
Since the affine coordinates in consecutive patches are related by: 
\be
z_1^{(\ell+1)}\,= \,(z_1^{(\ell) })^{h_\ell} \,z_2^{(\ell) }~ , \qquad
z_2^{(\ell+1)} \,=\, 1/ z_1^{(\ell)}~,
\ee
the $\IC^\times \times \IC^\times$ action in the other patches is given by $(\eps_{1,\ell},\eps_{2,\ell})$ satisfying the recursion:
\begin{equation}\label{defn toric weights}
\begin{split}
    &(\eps_{1,1}\,,\, \eps_{2,1}) \,=\, (\eps_1\,,\, \eps_2)~,\\
    &(\eps_{1,\ell+1}\,,\,\eps_{2,\ell+1}) \,=\, (h_{\ell+1} \,\eps_{1,\ell}\, +\, \eps_{2,\ell}\,,\,- \,\eps_{1,\ell})~.
\end{split}
\end{equation}
It will be important to note that the toric weights $(\eps_{1,\ell},\eps_{2,\ell})$ satisfy the following identities:
\begin{align}\label{toric-weights-ident}
\begin{split}
    &\sum_{\ell=1}^{\chi} \frac{1}{\eps_{1,\ell}\,\eps_{2,\ell}} \,=\, \frac{1}{\eps_{1,\ell}}\, +\, \frac{1}{\eps_{2,\ell+1}} \,=\, 0~,\\
    &\frac{\eps_{1,\ell}}{\eps_{2,\ell}} \,+\, \frac{\eps_{2,\ell-1}}{\eps_{1,\ell-1}} \,=\,-\, h_{\ell}~,
\end{split}
    \end{align}
as a direct consequence of~\eqref{defn toric weights}.\footnote{Let us remark here that the first identity can also be realized via equivariant localization -- see for instance~\protect\cite[Theorem 13.3.7]{cox2024toric}:
\protect\begin{equation}
    \int_{S}\,1 \,=\,(-1)^{\chi} \,\sum_{\ell=1}^\chi\,\frac{1}{\eps_{1,\ell}\,\eps_{2,\ell}}\,=\,0~,  
\end{equation}
where the first equality follows from the Atiyah--Bott localization formula, and the
vanishing in the second equality follows from the degree of the integrand.}

In the context of 5D gauge theories compactified on $\mathbb{S}^1_{\boldsymbol{\beta}}$, it will be convenient to use the 
K-theoretic version of the equivariant parameters given, in each affine patch, by: 
\be
\label{defy12l}
    y_{1,\ell}\,\equiv\,e^{-\boldsymbol{\beta}\,\eps_{1,\ell}}~, \qquad  y_{2,\ell}\,\equiv\,e^{-\boldsymbol{\beta}\,\eps_{2,\ell}}~, \qquad \forall\ell\,=\,1,\cdots,\chi~.
\ee
In terms of these toric characters, the recursion~\eqref{defn toric weights} becomes:
\begin{equation}\label{toric characters}
    y_{1,\ell+1}\,=\,y_{1,\ell}^{h_{\ell+1}}\,y_{2,\ell}~, \qquad y_{2,\ell+1} \,=\,y_{1,\ell}^{-1}~.
\end{equation}
Analogous to~\eqref{toric-weights-ident}, we have the following two identities:
\begin{equation}\label{Identity I,II}
\begin{split}
            &\sum_{\ell=1}^\chi\left(\frac{1}{1\,-\,y_{1,\ell}}\,+\,\frac{1}{1\,-\,y_{2,\ell}}\right)\,=\,\chi~,\\
            &\sum_{\ell=1}^{\chi} \frac{1}{(1\,-\,y_{1,\ell})\,(1\,-\,y_{2,\ell})}\,=\,1~.
\end{split}
\end{equation}
The first identity follows from using the second relation in~\eqref{toric characters} as follows:
\begin{equation}
        \begin{split}
            \sum_{\ell=1}^\chi\left(\frac{1}{1\,-\,y_{1,\ell}}\,+\,\frac{1}{1\,-\,y_{2,\ell}}\right)\,&=\,\sum_{\ell=1}^\chi\left(\frac{1}{1\,-\,y_{1,\ell}}\,+\,\frac{1}{1\,-\,y_{2,\ell+1}}\right)\\
            &=\,\sum_{\ell=1}^\chi\left(\frac{1}{1\,-\,y_{1,\ell}}\,+\,\frac{1}{1\,-\,y^{-1}_{1,\ell}}\right)~\\
            &=\sum_{\ell=1}^\chi\left(\frac{1}{1\,-\,y_{1,\ell}}\,+\,\frac{-y_{1,\ell}}{1\,-\,y_{1,\ell}}\right)\\
            &=\sum_{\ell=1}^\chi 1~.
        \end{split}
    \end{equation}
Meanwhile, the second identity in~\eqref{Identity I,II} can be viewed as the K-theoretic uplift of the first relation in~\eqref{toric-weights-ident}.\footnote{In particular, one can derive this by computing the holomorphic Euler characteristic of $S$ via equivariant localization. Recall from the line before~\eqref{sum inter numbs}: $\chi(S, \mathcal{O}_S)=1$. Meanwhile, the l.h.s. of the identity above is the result we get from equivariant localization of the Riemann--Roch--Hirzebruch formula~\protect\cite[Theorem 13.3.4]{cox2024toric}.}

Additionally, using the two identities in~\eqref{Identity I,II}, one can directly derive the following identity:
\begin{equation}\label{identity III}
\begin{split}
    \sum_{\ell=1}^{\chi} \frac{y_{1,\ell}\,y_{2\,\ell}}{(1\,-\,y_{1,\ell})\,(1\,-\,y_{2,\ell})}\, =\, 1~.
    \end{split}  
\end{equation}
This will be of importance to us later on.

\subsection{Toric localization on moduli space of stable sheaves}\label{subsec:toric-localization}

The classification of equivariant vector bundles and torsion-free sheaves on toric surfaces has a long history, going back to~\cite{klyachko1990equivariant,klyachko1991vector,klyachko1996stable}, later extended in~\cite{Knutson:1997yt,perling2004moduli,payne2008moduli,kool2011fixed,Kool2015}. Here we restrict to the simple case of a smooth toric surface $S$. The key idea is that an
equivariant bundle (or equivalently, locally-free sheaf) 
$\cE$ splits into a sum of line bundles in each affine patch, and is 
determined globally by its behavior near the toric divisors $D_\ell, \ell=1,\dots,\chi$ bounding each patch. This is encoded in a set of non-decreasing filtrations  $(E_i^{(\ell)})_{i\in \IZ}$ of the generic rank $N$ fiber: 
\be
0 \,=\,  E_{-\infty}^{(\ell)} \,\subset\, \dots\, \subset\,
E_i^{(\ell)} \,\subset\, E_{i+1}^{(\ell)} \,\subset\, \dots\, E_{\infty}^{(\ell)}\,=\,\IC^N~,
\ee 
subject to certain compatibility conditions between $(E_i^{(\ell)})$ and $(E_i^{(\ell+1)})$ associated to pairs of divisors bounding the same affine patch~\cite[Theorem 1.2.2]{klyachko1991vector}. Following~\cite{Kool2015},
we introduce  integers $u_\ell\in\IZ,v_{a,\ell}\in \IZ_+$ where  $\ell=1,\dots,\chi$, $a=1,\dots N-1$ and nested subspaces $p_{a,\ell}\in \Gr(a,N)$ such that: 
\be
E_i^{(\ell)} \,=\, \begin{cases} 
0~, & \mbox{if}\  i\,<\,u_\ell~,\\
p_{1,\ell}~, &  \mbox{if}\  u_\ell\,\leq \,i \, < \,u_\ell \,+\, v_{1,\ell}~, \\
p_{2,\ell}~, &  \mbox{if}\  u_\ell\,+\, v_{1,\ell} \,\leq\, i\,<\, u_\ell \,+\, v_{1,\ell}\,+\, v_{2,\ell}~,\\
~~\vdots &\\
\IC^{N}~, & \mbox{if}\  u_\ell\,+\, v_{1,\ell}\,+\, \dots \,+\, v_{N-1,\ell}\, \leq \,i~.
\end{cases}
\ee
The subspaces $p_{1,\ell}\subset \dots \subset p_{N-1,\ell} \subset \IC^N$ define a point in the partial flag variety ${\rm Flag}(v_{1,\ell},\dots, v_{N-1,\ell})$, sometimes called \textit{configuration data}. Note that $v_{a,\ell}$ is allowed to vanish, in which case $p_{a,\ell}$ is undefined.

The  Chern character of $\cE$ is then given by~\cite[Proposition 3.3]{Kool2015}: 
\be
\label{c1Klyachko}
c_1(\cE) \,=\,  \sum_{i\in\IZ}\,
\sum_{\ell=1}^{\chi} \,i\, \dim\left( \frac{ E_i^{(\ell)}}{E_{i-1}^{(\ell)}}\right) \, D_\ell
\,=\, -\,\sum_{\ell=1}^\chi\, \left( N\, u_\ell \,+\, \sum_{a=1}^{N-1} \,(N\,-\,a)\, v_{a,\ell} \right)\, D_\ell~,
\ee
and, 
\begin{multline}
\label{ch2Klyachko}
{\rm ch}_2(\cE)\, =\, \frac12 \,\left(\, \sum_{\ell=1}^\chi \,u_\ell \,D_\ell  \,\right)^2\, 
+\,\frac12 \, \sum_{a=1}^{N-1}\, \left[ \, \sum_{\ell=1}^\chi \, ( u_\ell \,+\, \sum_{b=1}^a \,v_{b,\ell})\, D_\ell \,\right]^2\\ 
-\,\sum_{\ell=1}^\chi\, \sum_{a,b=1}^{N-1} \,v_{a,\ell} \,v_{b,\ell+1} \,\left( \min({a},{b}) \,-\, \delta_{a,b,\ell,\ell+1} \right)~,
\end{multline}
where $\delta_{a,b,\ell,\ell'}:=\dim(p_{a,\ell} \cap p_{b,\ell'})$, sometimes called \textit{incidence data}.
Using linear equivalence, we can eliminate $D_1$ and $D_2$ in favor of $D_3,\dots, D_\chi$. Therefore, we can parameterize the first Chern class by:
\begin{equation}\label{c1E}
    c_1(\cE)\,=\,\sum_{\ell=3}^{\chi}\, f_\ell\, D_\ell~,
\end{equation}
 where $f_\ell$ are integers. Moreover, by fixing the equivariant structure -- i.e., by tensoring by a suitable line bundle $\cO(\sum n_\ell D_\ell)$ -- we can set  $u_1=u_2=0$, and determine the remaining integers $u_\ell$ in terms of the $f_\ell$'s via~\eqref{c1Klyachko}. The second Chern class is then expressed entirely in terms of the integers $v_{a,\ell}$ and $\delta_{a,b,\ell,\ell'}$
for $\ell'=\ell+1$, 
given 
by:\footnote{This is $Q({\bf v})+R({\bf v},{\bf \delta})$ in  \protect\cite{Kool2015}. Note that the second sum starts from $a=0$, in which case the sum $\sum_{b=1}^a$ is understood to vanish.}
{\small\be
\label{c2Klyachko}
\begin{split}
c_2(\cE) &\,=\,
 \frac12 \,\left( \sum_{\ell=3}^\chi\, f_\ell\, D_\ell \right)^2
\,-\,\frac{1}{2\,N^2}\, \sum_{a=0}^{N-1}\,
\left[\sum_{\ell=3}^\chi\,
\left( -\,f_\ell \,+\, \sum_{b=1}^a \,N\, v_{b,\ell} 
\,-\,\sum_{b=1}^{N-1}\,(N\,-\,b)\, v_{b,\ell}\,  \right.\right.
\\ 
&\left.\left.
+ \left\{ \sum_{b=1}^a \,N\, v_{b,1}\, 
-\,\sum_{b=1}^{N-1}\,(N\,-\,b)\, v_{b,1} \right\} \,\xi_\ell
\,+\, \left\{ \sum_{b=1}^a\, N\, v_{b,2}\, 
-\,\sum_{b=1}^{N-1}\,(N\,-\,b)\, v_{b,2}\right\} \,\eta_\ell
\right) \,D_\ell 
\right]^2\\
&\,+\,\sum_{\ell=1}^\chi \,\sum_{a,b=1}^{N-1}\, v_{a,\ell}\, v_{b,\ell+1}\, \left( \min({a},{b}) \,-\, \delta_{a,b,\ell,\ell+1} \right)~,
\end{split}\ee}
where $(\xi_\ell,\eta_\ell)$ is the $2\times \chi$ matrix of integer coordinates of the primitive vectors $(\rho_1,\dots,\rho_\chi)$ in the toric fan.

\medskip

Let us now denote by 
$\mathcal{N}_S^J(N,c_1,c_2)$ the moduli space  of slope-stable locally-free sheaves of rank $N$ and fixed Chern classes $(c_1,c_2)$, where $J$ is a choice of polarization (or K\"ahler class) on $S$. Recall that a coherent sheaf $\cE$ is slope stable
if $\mu(\mathcal{F})<\mu(\cE)$ for any proper subsheaf $0\neq \mathcal{F}\subsetneq\cE$, where $\mu(\mathcal{F}):=\frac{J\cdot c_1(\mathcal{F})}{{\rm rk}(\mathcal{F})}$ is the slope with 
respect to the polarization $J$.
As explained in~\cite{Kool2015}, the Euler number of 
$\mathcal{N}_S^J(N,c_1,c_2)$
can be computed by toric localization, summing over the components of the fixed locus $\mathcal{N}_S^J(N,c_1,c_2)^T$ under the toric action. From the description of toric equivariant torsion-free sheaves above, the fixed locus decomposes as a disjoint union:
\be\label{NHST}
\mathcal{N}_S^J(N,c_1,c_2)^T \,=\,
\cup_{v_{a,\ell},\delta_{a,b,\ell,\ell'}} \, \mathcal{D}^s_{(v,\delta)}/SL(N,\mathbb{C})~,
\ee
where  the positive 
integers $v_{a,\ell}$ take values in the set:
\be\label{C_rank_r}
C\,:= \,\{ (v_{a, \ell})\,:\,N | -\, f_\ell \,+\, 
\sum_{a=1}^{N-1} a\, (v_{a,1}\, \xi_\ell\, +\, v_{a,2}\, \eta_\ell\, +\, v_{a,\ell} )~,
\quad \forall \ell\,=\,3,\dots, \chi \} \,\subset\, \IZ_{\geq 0}^{(N-1)\chi}~,
\ee
and $\delta_{a,b,\ell,\ell'}$ take values in the range $\{0,1,\dots, \min({a},{b})\}$. Moreover, 
$\mathcal{D}^s_{(v,\delta)}$ denotes the subset 
 inside the product of flag varieties $\prod_{\ell} {\rm Flag}(v_{1,\ell},\dots, v_{N-1,\ell})$
 corresponding to torsion-free sheaves with incidence data $\delta_{a,b,\ell,\ell'}$, furthermore satisfying 
the stability condition~\cite[Equation (19)]{Knutson:1997yt}:
 \be
\frac{1}{\dim F}\sum_{\ell=1}^{\chi} \sum_{i\in \IZ}  i\, \dim\left( \frac{ F\cap E_i^{(\ell)}}{F\cap E_{i-1}^{(\ell)}}\right) \int_S w_\ell \wedge J \,<\, 
\frac{1}{N} \sum_{\ell=1}^{\chi} 
\sum_{i\in \IZ}i\, \dim\left( \frac{ E_i^{(\ell)}}{E_{i-1}^{(\ell)}}\right) \int_S w_\ell \wedge J ~,
\ee
for any proper subspace $0\neq F\subsetneq \IC^N$. In practice, this implies the following linear inequalities
 on the integers $v_{a,\ell}$~\cite[Theorem 3.20]{kool2011fixed},\cite[Proposition 2.3.37]{vanBree2024}:
 \be\label{K-stab}
 \frac{1}{\dim F} \,\sum_{\ell=1}^{\chi} \, \sum_{a=1}^{N-1}\, 
 \dim(F\cap p_{a,\ell})\, v_{a,\ell} \,(D_\ell\,\cdot\, J)  \,<\, \frac{1}{N}\, 
 \sum_{\ell=1}^{\chi} \, \sum_{a=1}^{N-1}\, a \, v_{a,\ell}\, (D_\ell\,\cdot\, J)~. 
 \ee
 
The Euler number of the moduli space $\mathcal{M}_S^J(N,c_1,c_2)$ of slope-stable torsion-free sheaves of rank $r$ and fixed Chern classes 
$(c_1,c_2)$ is then obtained by including contributions of point-like instantons.
In terms of generating series, this amounts to dividing the generating series of 
Euler numbers of $\mathcal{N}_S^J(N,c_1,c_2)$ by a power of the Dedekind eta function~\cite[Theorem 3.5]{Kool2015}: 
\be
\label{KoolThm35}
\sum_{c_2} \,e(\mathcal{M}_{S}^{J}(N,c_1,c_2)) \,\sfq^{c_2}\, =\, 
\frac{1}{\overline{\eta}(\sfq)^{N\chi}} \,\sum_{v,\delta} 
\,e\left( \mathcal{D}^s_{(v,\delta))}/SL(N,\mathbb{C})\right) \, \sfq^{c_2(v,\delta)}~,
\ee
where $c_2(v,\delta)$ is given by the r.h.s. of~\eqref{c2Klyachko}. Recall that the normalized Dedekind eta function is defined as follows:
\begin{eqnarray}\label{eta bar}
      \overline{\eta} (\mathsf{q})\,\equiv\,\sfq^{-\frac{1}{24}}\,\eta(\sfq)\,=\, \prod_{n\geq 1}(1-\mathsf{q}^n)~.
\end{eqnarray}

\medskip
\noindent
\textbf{Specializing to the rank $2$ case.}
For rank $N=2$, the general prescription outlined above
 simplifies drastically since the indices $a,b$ only take a single value.
 The integers $v_\ell:=v_{1,\ell}$ then take values in the set~\eqref{C_rank_r}:
\be\label{C_rank_2}
C = \{ (v_\ell)\,:\, 2 \,| \,-\,f_\ell \,+\, v_\ell \,+\, v_1 \,\xi_\ell \,+\, v_2\, \eta_\ell~, \quad \forall \ell\,=\,3,\dots, \chi\}~,
\ee
while the integers $u_\ell$ are determined by the first Chern class~\eqref{c1Klyachko}:
\be
c_1(\cE)\, =\, -\, \sum_{\ell=1}^\chi\, \left( 2 \,u_\ell \,+\, v_\ell \right)\, D_\ell ~.
\ee
The second Chern class is furthermore given by~\eqref{c2Klyachko}:
\be
\label{c2Etoric}
\begin{split}
c_2(\cE) 
\,=\,&\frac14 \,\left( \sum_{\ell=3}^\chi\, f_\ell \,D_\ell \right)^2\,
-\, \frac14 \,\left( \sum_{\ell=3}^\chi \,(v_\ell \,+\, \xi_\ell\, v_1 \,+\, \eta_\ell\, v_2)\, D_\ell \right)^2
\,+\, \sum_{\ell=1}^\chi \,v_\ell \,v_{\ell+1} \,( 1\,-\,\delta_{\ell,\ell+1} )~,
\end{split}
\ee
where $\delta_{\ell,\ell'}=\dim(p_\ell \cap p_{\ell'})\in\{0,1\}$. 
Finally, $\mathcal{D}^s_{(v,\delta)}$ is either empty or equal to the configuration space of $n$ points on $\IP^1$, where $n$ is 
equal to $\chi$ minus the number of coincidence relations resulting from the incidence data\footnote{Namely,
$n$ is the number of connected components of the graph 
made of abstract points $\{e_\ell\}_{\ell=1,\dots,\chi}$,
connected by a link $\ell-\ell'$ whenever $\delta_{\ell,\ell'}=1$.} $\delta_{\ell,\ell'}$, depending on whether the inequalities:
\be\label{K-stable rank-2}
\sum_{\ell=1}^{\chi}  \,
 2\,\dim(F\cap p_{\ell})\, v_{\ell}\, (D_\ell\,\cdot\, J) \, <\, 
 \sum_{\ell=1}^{\chi} \, v_{\ell} \,(D_\ell\,\cdot\, J)~, 
 \ee
are satisfied for any line $F\subset \IC^2$. By acting with $SL(2,\IC)$, one can fix the position of 3 points
to $0,1$ and $\infty$. The quotient $\mathcal{D}^s_{(v,\delta)}/SL(2,\mathbb{C})$
is then the configuration space of $n-3$ points on $\IC\backslash \{0,1\}$, with Euler number: 
\be
e\left( \mathcal{D}^s_{(v,\delta)}/SL(2,\mathbb{C})\right) \,=\, (-1)^{n-3} \,(n-3)!~.
\ee

\section{Supersymmetric localization in 5D \texorpdfstring{$\cN=1^*$}{N=1} SYM}\label{sec_susyloc}

In this section we discuss the supersymmetric localization approach to $\mathcal{N}=1^*$ gauge theory on $S\times \mathbb{S}^1_{\boldsymbol{\beta}}$, where $S$ is a compact toric K\"ahler surface and 
$\mathbb{S}_{\boldsymbol{\beta}}^1$ a circle of radius $\boldsymbol{\beta}$. In this work, we will take the gauge group to be $U(N)$ with special focus on the $N=2$ case.
The $\mathcal{N}=1^*$ SYM theory is a mass deformation of the maximally supersymmetric YM theory in five dimensions. 
To preserve part of the supersymmetry, one has to perform a partial topological twist along $S$. The resulting twisted theory is topological\footnote{Actually, when $b_2^+(S)=1$, which is the case of interest in this paper, the topological observables are only piecewise constant with respect to the metric on $S$ and exhibit wall-crossing phenomena~\protect\cite{KotschickMorgan1994,gottsche1996modular,Moore:1997pc}.} along $S$ but retains dependence on the metric on $\mathbb{S}_{\boldsymbol{\beta}}^1$. As it is well known, there exist several topological twists along the four-dimensional manifold $S$~\cite{Yamron:1988qc}. For the purpose of this paper, we choose the Vafa--Witten twist~\cite{Vafa:1994tf}, which is equivalent to the Donaldson--Witten twist for complex K\"ahler surfaces because of the isomorphism between Spin and Dolbeault bundles $\mathcal{S}^+\cong \Lambda^{(0,{\rm even})} T^*S$, $\mathcal{S}^-\cong \Lambda^{(0,{\rm odd})} T^*S$~\protect\cite{Baulieu_2006}. 

\subsection{Topological Vafa--Witten partial twist}
Let us briefly describe the field content and equivariant twisted supersymmetry transformations of the 
$\mathcal{N}=1^*$ SYM on $M\times\mathbb{S}_{\boldsymbol{\beta}}^1$, starting with a Riemannian four-manifold $M$ for the sake of generality. The partial twisting along $M$ amounts to embedding the holonomy group $SO(4)$ of $M$ into the $Spin(5)$ R-symmetry group of the maximally supersymmetric $\mathcal{N}=2$ YM theory in five-dimensions. In the Vafa--Witten twist, the embedding corresponds to the group homomorphism under which the spinor representation $\boldsymbol{4}$ of $Spin(5)$ splits into the direct sum $\boldsymbol{2}\oplus\boldsymbol{2}$ chiral spinor representations of $Spin(4)$. This embedding commutes with a $SU(2)_R\subset Spin(5)$ symmetry group under which the two scalar supersymmetries arising from the topological twist transform as a doublet~\cite{witten2011fivebranesknots,Anderson_2013}.
The mass deformation to $\mathcal{N}=1^*$ is obtained by turning on the equivariant parameter for the $U(1)_{\madj}$ Cartan subgroup of $SU(2)_R$\footnote{We recall that equivariant parameters of the five-dimensional theory on the circle can be turned on as background vector fields along $\mathbb{S}_\beta^1$ of the corresponding global symmetry, see e.g.~\protect\cite{Kim_2011}.}. This splits the $\mathcal{N}=2$ supermultiplet (invariant under 16 supercharges) into an $\mathcal{N}=1$ vector multiplet (invariant under 8 supercharges) plus a massive $\mathcal{N}=1$ hypermultiplet in the adjoint representation of the gauge group. The field content reorganizes in the partially twisted $\mathcal{N}=1$ vector multiplet $(A_\mu,\psi_\mu,\eta,\chi^{\mu\nu\, +},\phi$) with
$\phi=\sigma + i A_t$, $\sigma$ being the real scalar of the vector multiplet and $A_t$ the component of the gauge field along $\mathbb{S}^1_{\boldsymbol{\beta}}$~\cite{Nekrasov_1998,Qiu_2017}. The $\mu,\nu$ indices refer to coordinates along $M$. In particular, $\chi^{\mu\nu\, +}$, $\psi_\mu$, and $\eta$ are respectively a fermionic self-dual two-form, a one-form, and a scalar on $M$. These fields are pulled back on $M\times \mathbb{S}^1_{\boldsymbol{\beta}}$ via the projection map $\pi:M\times \mathbb{S}_{\boldsymbol{\beta}}^1 \rightarrow M$. 
The equivariant twisted scalar supersymmetry $Q_v$ in presence of $\Omega$-background acts as follows~\cite{Hosseini:2018uzp,Bonelli_2026}:
\begin{eqnarray}
   && Q_v \,A \,=\, \psi ~ ,\qquad\quad  \, Q_v \,\psi\, =\, \I\, \iota_v\, F \, +\, D\,\phi ~, \nonumber \\
   && Q_v\, \phi\, = \,\I \, \iota_v\,\psi ~ ,~~\quad \, Q_v\, \overline\phi \,=\, \eta ~ , \qquad \, Q_v \,\eta \,= \mathcal{L}_v^A \, \bar\phi + \,[\phi\,,\,\overline\phi] ~, \nonumber\\
&&    Q_v\, \chi^{+} \,=\, H^{+} ~ , \,
~\,\,\quad Q_v\, H^{+} \,=\, \mathcal{L}_v^A \,\chi^{+ } \,+\, [\phi\,,\,\chi^+]~.
\label{QOmega1}
\end{eqnarray}
where $D$ is the covariant derivative of the gauge bundle.
The twisted adjoint hypermultiplet has components $(B_{\mu\nu}^+, C, \widetilde \psi_\mu,\widetilde \chi_{\mu\nu}^+,\widetilde \eta)$ and transforms under equivariant twisted supersymmetry as follows:
\begin{eqnarray}
   && Q_v \,B^+ \,=\, \widetilde\chi^+ ~ ,\qquad\qquad  Q_v\, \widetilde\chi^+ \,=\, \mathcal{L}_v^A \, B^+ \,+\, [\phi\,,\,B^+] \,+\, \madj\, \mathrm{J} B^+ ~, \nonumber \\
   && Q_v \,\widetilde\psi \,=\, H ~ ,\qquad \qquad\quad \, Q_v\, H\, =\, \mathcal{L}_v^A\, \widetilde\psi \,+\, [\phi,\widetilde\psi] \,+\,\madj\widetilde \psi~ , \nonumber\\
   && Q_v \,C \,= \,\widetilde \eta
   ~, \qquad\qquad\quad\,~ Q_v \,\widetilde\eta \,=\,\mathcal{L}_v^A \, C \,+\, [\phi\,,\, C] \,+\,
   \madj\, C~, 
   \label{QOmega2}
\end{eqnarray}
where $\madj$ is the mass of the adjoint hypermultiplet\footnote{We denote by $\mathrm{J}$ the generator of the Cartan subgroup of $SU(2)_R$ along which we turn equivariance. The self-dual two form field $B^+_{\mu\nu}=B^+_i\eta^i_{\mu\nu}$ transforms as a triplet under $SU(2)_R$, with $\mathrm{J}$-charges $q=-1,0,1$ for $i=1,2,3$ respectively, $\eta^i_{\mu\nu}$ being the 't Hooft symbols, see for example \cite{lozano2000dualitytopologicalquantumfield}. On K\"ahler manifolds, $B^+_2\equiv B$ is the real scalar component of $B^+$ along the K\"ahler form.} and  $\iota_v$ denotes the contraction with the vector field:
\be
v\,:=\,\partial_t \,+\,  \epsilon_1\, z_1\, \partial_{z_1}\,+\, \epsilon_2\, z_2 \,\partial_{z_2}\, +\, 
\overline{\epsilon_1\, z_1 \,\partial_{z_1}\,+\, \epsilon_2 \,z_2 \,\partial_{z_2}}~,
\ee
generating the rotation along the circle, combined with
an element in the $U(1)\times U(1)$ subgroup of the holonomy group. 

In~\eqref{QOmega1} and~\eqref{QOmega2},  $(H^+,H)$ are auxiliary fields enforcing the 
Haydys--Vafa--Witten equations~\cite{Haydys:2010dv,witten2011fivebranesknots,Anderson_2013}:
\be
\label{VWHaydysEq}
\begin{split}
& F_{\mu\nu}^{+}\,
-\,\frac14\, (B^+\,\times\, B^+)_{\mu\nu}\,
-\,\frac12\,
D_t
B^+_{\mu\nu}
\,=\,0~,\\
& F_{\mu t}\,+\,D^\nu\, B^+_{\nu\mu}\,
=\,0~,
\end{split}
\ee
where $\times$ denotes the cross product of self-dual two-forms \cite{witten2011fivebranesknots}:
\begin{equation}
(B^+\,\times\, B^+)_{\mu\nu}\,
=\,
\sum_\tau \left[
B^+_{\mu\tau},B^+_{\nu\tau}
\right]~.
\end{equation}

Before specializing to K\"ahler surfaces, we would like to remark that an analogous twist, without $U(1)_{\madj}$-equivariance, was studied for maximally 
supersymmetric YM on $M\times \mathbb{R}$ for a general Riemannian 4-manifold $M$~\cite{witten2011fivebranesknots,Anderson_2013}. It would be interesting to investigate if the resulting equations can play the r\^ole of gradient flow lines for a Morse theory on the moduli space of solutions to \eqref{VWHaydysEq}. This, of course, does not apply to the case of compact toric K\"ahler surfaces that we are interested in in the present paper, for which the Morse function is perfect, being the moment map of the toric action\footnote{In this case, the strong Morse inequalities are saturated and the Poincar\'e polynomial can be computed in terms of the index of the Morse function at the critical points~\protect\cite{Witten:1982im}.}.

\medskip
\noindent
\textbf{On complex K\"ahler surfaces.}
When $M$ is a complex K\"ahler surface, which we now denote by $S$, the self-dual two-form $B^+$ decomposes into a direct sum
$B^+ = \Omega^{(2,0)}\oplus \Omega^{(1,1)} \oplus\Omega^{(0,2)}$ where $\Omega^{(0,2)}$ is the complex conjugate
of $\Omega^{(2,0)}$ and $\Omega^{(1,1)}=B \omega$ is proportional to the K\"ahler form $\omega$. The real scalar field $B$
combines with $C$ to form a complex scalar which we denote by $\varphi=B+iC$. 
The fermionic superpartners $(\widetilde\chi^+, \widetilde{\eta})$ undergo a similar splitting.
Analogously, the doublet $(\chi^+,H^+)$
of the vector multiplet splits into $(\chi^{(0,2)},H^{(0,2)})$ and the real scalars
$(\chi,h)$. 
We thus recover the Donaldson--Witten twist by virtue of the isomorphism $\mathcal{S}^+\cong\Lambda^{(0,{\rm even)} }T^*S$.
When $b_2^+(S)=1$, the case of interest in this work, there are no holomorphic $(2,0)$-forms, and vanishing theorems forbid the existence of solutions 
to~\eqref{VWHaydysEq} with non-trivial $B^+$~\cite{Vafa:1994tf}. 
This implies that the so-called monopole branch (or vertical branch)
does not contribute, and Vafa-Witten invariants are entirely determined by the instanton (or horizontal) branch. Indeed, when the Chern vector $(N,c_1,c_2)$ is primitive,
they are given by the signed Euler number $(-1)^{\dim \mathcal{M}} e(\mathcal{M})$ of the moduli space $\mathcal{M}\equiv\mathcal{M}^J_S(N,c_1,c_2)$ of Gieseker-semi-stable 
coherent sheaves, which is a smooth projective variety of expected complex dimension:
\be
\label{expecteddim}
\dim_{\IC}\, \mathcal{M}^J_S(N,c_1,c_2) \,=\, c_1^2 \,-\, 2\, N \,\ch_2 \,-\, N^2\,+\,1~,
\ee
where we recall that $\ch_2=\frac12 c_1^2-c_2$.
Here, the polarization $J$ is equal to the K\"ahler form on $S$. In particular, $\mathcal{M}^J_S(N,c_1,c_2)$ is empty if~\eqref{expecteddim} is negative, in which case the VW invariant vanishes. 
When the Chern vector $(N,c_1,c_2)$ is not primitive, the space of slope-stable coherent sheaves is no longer smooth and projective, due to the existence of reducible connections (or, physically, 
to the existence of marginal bound states), and the definition of VW invariants requires more advanced techniques from DT theory. In either case, VW invariants are locally independent of the polarization $J$, but exhibit jumps when some stable coherent sheaves become strictly semi-stable.

\subsection{Twisted partition function and equivariant \texorpdfstring{$\chi_{y^2}$}{chiy}-genus}
Upon performing the partial topological twist, the partition function of $\cN=1^*$ SYM on $S\times \mathbb{S}_{\boldsymbol{\beta}}^1$ reduces to the partition function of supersymmetric quantum mechanics on the moduli space of instantons, and as such it computes the Witten index for the relevant twisted supersymmetric charge. In particular, the mass term for the adjoint hypermultiplet of the gauge theory induces a mass for the related chiral fermionic zero-modes on the instanton moduli space~\cite{Dorey:2002ik, Bruzzo:2002xf}. Since the latter can be identified with a basis for anti-holomorphic forms on the instanton moduli space, their mass term induces a twisting by the anti-holomorphic cotangent bundle~\cite{Hollowood_2003} such that the resulting supersymmetric index $\mathcal{Z}[S]$ computes the $\chi_{y^2}$-genus of the instanton moduli space:
\begin{align}
\mathcal{Z}[S]\,:=\,Z^{\mathcal{N}=1^*}_{S\times \mathbb{S}^1_{\boldsymbol{\beta}}}  \,=\,\sum_{c_2}\, \sfq^{c_2}\, \left( \sum_{p=0}^d \,\yadj^p\, \mathrm{Ind}
\left(\bar{\partial}\,\otimes\,\Lambda^{(0,p)} T^*\mathcal{M}^J_{S} (N,c_1,c_2) \right)\right)~, 
\end{align}
where $d$ is the expected dimension \eqref{expecteddim} and $\yadj\equiv e^{-\boldsymbol{\beta}\,\madj} $, $\madj$ being the mass of the adjoint hypermultiplet.

In the path integral formalism, the calculation of the index reduces to the integration of a suitable local density, providing a supersymmetric quantum mechanical version of classical index theorems~\cite{Alvarez-Gaume:1983zxc}.
This is a formidable task for general gauge theories, but
on toric surfaces, we can take advantage of the existence of a $U(1)\times U(1)$ isometry which is naturally lifted to $\mathcal{M}^J_S(N,c_1,c_2)$, and evaluate the index via equivariant localization.
While for smooth compact moduli spaces the dependence on the equivariant weights $y_i\equiv e^{-\boldsymbol{\beta}\epsilon_i}, i=1,2$ must drop out after summing over all fixed loci, the situation is subtler for singular and/or non-compact moduli spaces. 

Let us remark that analogous equivariant supersymmetric methods can also be used to evaluate the monopole branch contribution to the Vafa--Witten
invariants in the case of surfaces with $\chi(\cO_S)>1$. Indeed, this branch admits a local description in terms of a nested Hilbert scheme of points, and more generally a flag of sheaves, on $\mathbb{C}^2$~\cite{Tanaka:2017jom,Tanaka:2017bcw}. See~\cite{Arbesfeld:2026nxw} for the evaluation of the monopole branch contribution from the $\chi_{y^2}$-genus of the nested Hilbert scheme of points.  Equivariant localization approach for general flag of sheaves on $\mathbb{C}^2$, including the calculation of their $\chi_{y^2}$-genus from $\mathcal{N}=1^*$ gauge theory, can be found in~\cite{Bonelli:2019lal, Bonelli:2019het}.

In order to evaluate the path integral of the partially twisted topological $\mathcal{N}=1^*$ theory on $S\times \mathbb{S}^1_{\boldsymbol{\beta}}$, we will adapt the method proposed for four-dimensional theories in~\cite{nekrasov2006localizing} and further developed in~\cite{Bershtein:2015xfa,Bershtein:2016mxz,Bonelli:2020xps}. This was extended in~\cite{Kim:2025fpz} to pure $\cN=1$ SYM theory on $S \times \mathbb{S}^1_{\boldsymbol{\beta}}$, which computes K-theoretic equivariant Donaldson invariants of $S$.\footnote{One can consider a more general setup where the circle fibration is non-trivial over the toric surface $S$, see e.g.~\protect\cite{Closset:2022vjj}.}
Schematically, via supersymmetric localization, the evaluation of the partition function localizes on point-like instantons (or ideal sheaves) inserted at the fixed loci of the torus action. Their contribution is given in terms of a residue integral, with poles arising from each patch of the toric manifold. These poles turn out to be labeled by fluxes for the 
Cartan gauge field along the equivariant divisors of $S$. The choice of integration contour and the structure of the poles ensure that only fluxes associated to 
slope-semi-stable equivariant sheaves contribute.
A crucial ingredient for this calculation is a recurrence relation for the Nekrasov partition function on $\IC^2$~\cite{Poghossian:2009mk}, which follows from the standard
Zamolodchikov's recurrence relations for Virasoro conformal blocks via the AGT correspondence~\cite{Alday_2010}. First found for gauge group $U(2)$ in 4 dimensions, 
this recurrence relation was 
subsequently generalized to $U(N)$ in 4 dimensions (corresponding to $W_N$ conformal blocks on the punctured sphere) in~\cite{Sysoeva:2022syp} and to $U(2)$ in 5 dimensions 
(corresponding to $\sfq$-Virasoro  conformal blocks on the punctured sphere) in~\cite{Yanagida:2010vz,Kim:2025fpz}. One of our main new results is a generalization of this recurrence relation to 
$\sfq$-Virasoro  conformal blocks on the one-punctured torus, which is required for the five-dimensional $\mathcal{N}=1^* $ supersymmetric localization method.

\subsection{Supersymmetric localization for the rank \texorpdfstring{$2$}{2} theory}

Let us now provide a brief outline of our method for $U(2)$ gauge group\footnote{We refer to this as the `rank $2$' theory, having in mind that 
it encodes the  partition function of $U(2)$ SYM theory after averaging over all values of the first Chern class modulo 2.}, 
postponing detailed formulae and computations for specific toric manifolds to later sections. 
As in the previous Section, we denote by $(\epsilon_{1,\ell},\epsilon_{2,\ell}),\,\ell=1,\ldots,\chi$ the weights of the $\mathbb{C}^\times\times\mathbb{C}^\times$ torus action in each affine patch (see~\eqref{defn toric weights}). We further denote by $\sfp_\ell,\,\ell=1,\ldots,\chi$ the fluxes of the Cartan field strength $F$ along the toric divisors:
\begin{equation}\label{equiv-flux-defn}
F \,=\, \sum_{\ell=1}^{\chi} \sfp_\ell\, w_\ell~,\qquad \sfp_\ell\, =\, \int_{D_\ell} F~,
\end{equation}
where $w_\ell$ are  equivariant two-forms Poincar\'e dual to the equivariant divisors $D_\ell$. 
By supersymmetric localization,  the K-theoretic $U(2)$ partition function on $S\times \mathbb{S}^1_{\boldsymbol{\beta}}$ evaluates to an integral of the product of $\chi$ copies of $\mathbb{C}^2$ Nekrasov partition functions, with patch-dependent equivariant parameters: 
\begin{equation}\label{Zfull SU2 }
    \mathcal{Z}[S]\,=\,\sum_{\vec\sfp\in\IZ^\chi} \,\int\, \de a\, \prod_{\ell=1}^{\chi}\, Z_{\rm full}^{\,U(2)}[\mathbb{C}^2](a_{\ell},\epsilon_{1,\ell},\epsilon_{2,\ell})~,
\end{equation}
with: 
\be
a_\ell \,:=\, a \,+\, \frac12\,\sfp_\ell \,\epsilon_{1,\ell}\, +\, \frac12 \,\sfp_{\ell+1} \,\epsilon_{2,\ell}~.
\ee
Here $a$ is the v.e.v. of the complex scalar $\phi$  in the vector multiplet along the Cartan subgroup of the $U(2)$ gauge group. Unlike in the case of a non-compact surface such as
$\mathbb{C}^2$, it is now a normalizable zero mode which needs to be integrated over~\cite{nekrasov2006localizing}. Actually, on a compact K\"ahler surface,   besides $a$ there is a whole supermultiplet $(\eta,\chi,h)$ of scalar zero modes along the Cartan direction, whose careful treatment in the path integral is needed to determine the integration contour over $a$. Let us briefly recall this procedure, which was described in~\cite{Bonelli:2020xps}. The relevant measure for the integration over the Cartan zero modes is provided by the gauge-fixing term of the $\mathcal{N}=1$ vector multiplet $L=Q_v \mathcal{V}$, with: 
 \bea\label{vector-gf}
\mathcal{V}=\text{Tr}\,\big[
\I\,\chi^{+}\,\wedge\, F^{+}\,
+\,\frac{g_{\rm YM}^2}{4}\, \chi^{+}\,\wedge\, H^{+}  \,+\,
         \psi \, \wedge\,\star\,(Q_v\,\psi)^\dagger\, +\,\eta\,\wedge\,\star\,(Q_v\,\eta)^\dagger\,
          \big]~,
\eea
with $\star$ being the Hodge star operator on $S$ and $g_{\rm YM}$ is the gauge coupling.
It is easy to see that the above action, when specialized to the relevant zero modes, does not contain terms depending on $(\eta,\chi)$. Then, we add an extra $Q_v$-exact term:
\be\label{gf02}
{Q_v} \,\mathcal{V}' \,=\, 2\,\I\, s\,    {Q_v} \,   \Tr[ {\rm Im}(\phi) \, \chi ^+  \,\wedge \,\omega]~.
\ee
Here, $\omega$ denotes the K\"ahler form on $S$ and $s\in \mathbb{R}^+$ is a gauge-fixing parameter\footnote{Indeed, one can check that the final result of the computation does not depend on $s$, which provides a consistency check on our calculations.}. Computing 
$Q_v \left( \mathcal{V} + \mathcal{V}'\right)$
restricted to the zero modes, one gets:
 \be
S_{\rm zero\,modes}\,=\,
s\, \eta\,\chi^+ + \,2\,\I\,s\,{\rm Im}({a})\,h
\,+\,\frac{g_{\rm YM}^2}{4}\, h^2 \,+\, 2\,\pi\, \I  \, h \,\vec{\beta}\,\cdot\,\vec{\sfp}~,
\ee
with $\vec{\beta}\cdot\vec{\sfp}:=\sum_{\ell=1}^\chi\beta_\ell \,\sfp_\ell$, where $\beta_\ell$ are the coefficients of the K\"ahler form along the basis of equivariant two-forms:
\begin{equation}\label{vec-beta}
    \beta_\ell \,:=\, \frac{1}{2\pi}\, \int_S\, \omega\, \wedge\, w_\ell ~ .
\end{equation}
The path integral over $(\eta,\chi,h)$ then reads:
\bea
 \int   \de\chi\, \de\eta\,\de h\, 
e^{\left[ \frac{s}{2}\, \eta\,\chi\, -\,\frac{g_{\rm YM}^2}{8} \, h^2 \,-\,2\,\pi\,{\I }\, h \, \vec{\beta}\,\cdot\,\vec{\sfp}\,
-\,  {\I}\,s\,{\rm Im}(a)\,h \right]}
\,=\, \frac{s \,\sqrt{2\,\pi }}{g}\,  e^{-\,\frac{2}{g_{\rm YM}^2}\,
\left(s\,{\rm Im} (a)\, +\,2\,\pi\,\vec{\beta}\,\cdot\, \vec{\sfp} \right)^2  }~,
\label{zzm}
\eea
leading to:
\be
\mathcal{Z}[S]\,=\,
\frac{1}{4 \,\pi^2} \,\sum_{\vec\sfp\in\IZ^\chi}\,
  \frac{s\,\sqrt{2\,\pi}}{g}
\,\int_{{\mathbb C}}\,
\de a\, \wedge\, \de \bar a \,
e^{-\,\frac{2}{g_{\rm YM}^2}\,
\left(s\,{\rm Im} (a) \,+\,2\,\pi\,\vec{\beta}\,\cdot\, \vec{\sfp} \right)^2}
Z_{\rm full}^{\,U(2)}[S](a, \vec{\sfp})
\label{zero}\ee
where 
$Z_{\rm full}^{\,U(2)}[S] := \prod_{\ell=1}^\chi Z_{\rm full}^{\,U(2)}[\mathbb{C}^2_\ell]$.
The complex integral over the remaining zero modes $(a,\bar a)$ gives then rise in the  limit $g_{\rm YM}\to 0$ to~\cite{Bonelli:2020xps}:
 \be\label{sign}
\mathcal{Z}[S] \,=\,
\frac{1}{4} 
\,\sum_{\vec{\sfp} \in {\IZ^\chi}}\, \sgn(\vec{\beta}\,\cdot\,\vec{\sfp}) \, \operatorname*{Res}_{a=0} \, Z_{\rm full}^{\,U(2)}[S]~. 
\ee

In order to compare with the results arising from the classification of equivariant slope-stable sheaves in the previous Section, it is convenient to collect terms in the above sum into orbits generated by flipping the signs of the components of a flux vector $\vec{\sfp}$ with positive entries:
\begin{equation}\label{Z[S]}
    \mathcal{Z}[S]\,=\,\sum_{{\vec\sfp}\in\mathbb{Z}_{\geq 0}^\chi}\,Z_{\rm orbit}[S]\mid_{\vec\sfp}~,
\end{equation}
where:
\begin{equation}\label{Zorbit}
    {Z}_{\rm orbit}[S]\mid_{\vec\sfp}\,:=\,\frac{1}{4}\,
\sum_{\eta\in\IZ_2^\chi}\,\sgn(\vec{\beta}\,\cdot\,\vec{\sfp}_\eta)\,\operatorname*{Res}_{a=0} \, Z_{{\rm full}}^{\,U(2)}[S]\mid_{\vec{\sfp}_\eta}~.
\end{equation}
The sum on the r.h.s. is taken over $\eta\in\{{\rm diag}(\pm 1, \pm 1, \dots, \pm 1)\}$ and $\vec{\sfp}_\eta\equiv \eta\cdot\vec{\sfp}$, subject to the condition
that $\eta_\ell=+1$ whenever $\sfp_\ell=0$ in order to avoid double counting. This prescription of grouping the different fluxes together is motivated by an `abstruse duality' connecting the residues of the different elements of an orbit \cite{Bonelli:2020xps,Kim:2025fpz}. In particular, it was shown in 
\cite[Appendix C]{Bonelli:2020xps}, in the context of 4D $\cN=2$ SYM on a toric surface $S$, that for fluxes 
satisfying the strict inequality
\be
\exists \, \ell:   \qquad   2 \,\beta_\ell \, \sfp_\ell\,  >\, \vec{\beta}\,\cdot\,\vec{\sfp}~,    \label{comb-st}
\ee
the contributions in the orbit cancel out, leaving only contributions
from fluxes which satisfy the inequalities 
\be
\forall \,\ell :  \qquad    2 \,\beta_\ell \, \sfp_\ell  \,\leq \, \vec{\beta}\,\cdot\,\vec{\sfp} ~,\label{st} 
\ee
Moreover, whenever one of the inequalities in \eqref{st} is saturated, it turns out that some contributions in the orbit
vanish due to $\sgn(\vec{\beta}\cdot \vec{\sfp}_\eta)=\sgn(0)=0$, leading to an 
overall factor $1/2$. It is worth noting that the inequalities
\eqref{st} are identical to~\eqref{K-stable rank-2} upon identifying $v_{\ell}=\sfp_\ell$, where we recall that the latter is the positive representative in the Weyl orbit. 
In the next section, we will generalize the abstruse duality for the 5D $\mathcal{N}=1^*$ theories relevant for this work, which ensures that these cancellations continue to hold in the case of 
5D $\cN=1^*$ SYM. There is therefore a precise matching between supersymmetric fixed points contributing to the partition function and fixed loci in the moduli space of equivariant semi-stable
coherent sheaves. In particular simple poles of $Z^{\,U(2)}_{\rm full}[S]$ with $\vec{\sfp}$ satisfying the (semi-)stability condition correspond to isolated fixed points in Klyachko's classification, while higher order poles of $Z^{\,U(2)}_{\rm full}[S]$ correspond to fixed loci of strictly positive dimension. This will be illustrated at length in the case of $S=\IP^2$ and $S=\IF_n$ in the remainder of this work. 

Finally, let us remark that a higher rank generalization of supersymmetric localization is possible by using the Jeffrey--Kirwan (JK) localization. The key point is to apply this method directly on Klyachko's space of filtrations rather than on the moduli space of coherent sheaves~\cite{BonelliToAppear}. In this way one obtains from the JK method a general prescription for $U(N)$ theories that reduces in the $U(2)$ case to the one
found in~\cite{Bonelli:2020xps} and just recalled above. In this approach, the puzzle raised in~\cite[Section 9.6]{Kim:2025fpz} is avoided.

%%%%%%%%%%%%%%%%%%%%%%%%%%%%%%%%%%%%%%%%%
%%%%%%%%%%%%%%%%%%%%%%%%%%%%%%%%%%%%%%%%%
%%%%%%%%%%%%%%%%%%%%%%%%%%%%%%%%%%%%%%%%%
\section{The partition function of 5D \texorpdfstring{$\mathcal{N}=1^*$}{N=1} SYM}
\label{sec_5Dpart}

In this section, we construct the partition function of 5D $\mathcal{N}=1^*$ $U(N)$ SYM theory on $S\times \mathbb{S}_{\boldsymbol{\beta}}^1$ with $S$ being a compact toric surface with Euler number $\chi\equiv \chi(S)$.  We start by reviewing the form of the partition function on a local $\mathbb{C}^2\times\mathbb{S}_{\boldsymbol{\beta}}^1$ patch, and then work out the form for $S\times\mathbb{S}_{\boldsymbol{\beta}}^1$ by gluing along the local affine patches. After spelling out the details of the integrand for the $U(N)$ case, we specialize to the rank $2$ case, where we discuss the corresponding analytic structure near the origin of the Coulomb branch $a=0$. In particular, to analyze the poles of the instanton partition function, we argue that it satisfies a Zamolodchikov-type recurrence relation and `abstruse duality' generalizing the approach in \cite{Bershtein:2015xfa,Bonelli:2020xps}. 

%%%%%%%%%%%%%%%%%%%%%%%%%%%%%%%%%%%%%%%%%
%%%%%%%%%%%%%%%%%%%%%%%%%%%%%%%%%%%%%%%%%
\subsection{For \texorpdfstring{$\mathcal{N}=1^*$}{N1} \texorpdfstring{$U(N)$}{UN} theory on \texorpdfstring{$\mathbb{C}^2\times\mathbb{S}_{\boldsymbol{\beta}}^1$}{C2S1b}}
The full partition function of the 5D $\mathcal{N}=1^*$ $U(N)$ SYM theory on $\mathbb{C}^2\times\mathbb{S}^1_{\boldsymbol{\beta}}$ is decomposed into three different contributions~\cite{Nakajima:2005fg,Gottsche:2006bm}:
\begin{equation}\label{ZSU(N) full}
     Z_{\rm full}^{\,U(N)}[\mathbb{C}^2] \,=\, Z_{\rm class}^{\,U(N)}[\mathbb{C}^2]\, Z_{\rm 1-loop
     }^{\,U(N)}[\mathbb{C}^2]\,Z_{\rm inst,\,adj}^{\,U(N)}[\mathbb{C}^2]~.
\end{equation}
Note that we are following a different notational convention from that in~\cite{Nakajima:2005fg, Gottsche:2006bm} for the perturbative part of the partition function. In Appendix \ref{app: perturbative part}, we show that what they call the perturbative part in fact includes both the classical and 1-loop contributions. But here, let us discuss the details of each one of the components appearing on the r.h.s. of the above equation.

\medskip
\noindent
\textbf{The classical contribution.} Starting with the classical contribution, we have~\eqref{Zwbos C2 simplified}:
\begin{equation} \label{logClassC2}
    \log Z_{\rm class}^{\,U(N)}[\mathbb{C}^2]\,:=\,\left(-\,\frac{\sum_{\alpha\neq\beta}\,a_{\alpha,\beta}^2}{8\,\eps_1\,\eps_2}\,-\,{N\,(N\,-\,1)}\,\frac{\eps_1^2\,+\,\eps_2^2\,+\,3\,\eps_1\,\eps_2}{48\,\eps_1\,\eps_2}\right)\,\log\sfq~.
\end{equation}
This expression can be simplified further when gluing the affine patches along a compact toric surface $S$. We will come back to this point momentarily.
%%%%%%%%%%%%%%%%%%%%%%%%%%%%%%%%%%%%%%%%%
\subsubsection{The 1-loop contribution}
Following the analysis performed in Appendix~\ref{app: perturbative part}, this contribution comes from the adjoint W-bosons in the gauge vector multiplet and from the massive adjoint hypermultiplet:\footnote{Note that, there is an extra term of the form $e^{\mathsf{F}_N[\mathbb{C}^2]}$ -- see~\protect\eqref{FNC2} for definition -- that we are ignoring here. As we argue in Appendix~\protect\ref{app: perturbative part}, this will give us an overall factor in the equivariant parameters, and it will not be of importance to us.}
\begin{equation}
    Z_{\rm 1-loop}^{\,U(N)}[\mathbb{C}^2]\,=\,Z_{\rm 1-loop,\,W}^{\,U(N)}[\mathbb{C}^2]\,\times\,Z_{\rm 1-loop,\,adj}^{\,U(N)}[\mathbb{C}^2]~. 
\end{equation}
The contribution of the W-bosons is given by:
\begin{equation}\label{Z Wbos C2 SUN}
    Z_{\rm 1-loop,\,W}^{\,U(N)}[\mathbb{C}^2]\,:=\,{\rm Exp}[-\,\chi_N^{\mathbb{C}^2}(x,y_1,y_2)]~,
\end{equation}
where the character $\chi_N^{\mathbb{C}^2}$ is defined by:
\begin{equation}\label{chiC2N0}
    \chi_N^{\mathbb{C}^2}(x,y_1,y_2)\,:=\,\sum_{\substack{\alpha,\beta=1\\\alpha\neq \beta}}^N\,\frac{x_{\alpha,\beta}}{(1\,-\,y_1^{-1})\,(1\,-\,y_2^{-1})}~,
\end{equation}
and the plethystic exponential is defined by 
\begin{equation}\label{pexp defn}
    {\rm Exp}[f(x,y_1,\dots)] \,:=\, \exp\left(\sum_{n\geq 1}\frac{f(x^n,y_1^n,\dots)}{n}\right)~.
\end{equation}
As in~\eqref{defy12l}, we introduced the parameters:
\begin{eqnarray}\label{K to coh}
   y_{1,2}\,\equiv\,e^{-\boldsymbol{\beta}\,\eps_{1,2}} ~, \qquad x_\alpha\,\equiv\,e^{-\boldsymbol{\beta}\,a_\alpha}~,\quad \forall \alpha\,=\,1, \cdots, N~.
\end{eqnarray}
As for the contribution of the massive adjoint hypermultiplet, it is given by:
\begin{equation}\label{pert hyper C2}
     Z_{\rm 1-loop,\,adj}^{\,U(N)}[\mathbb{C}^2]\,:=\,{\rm Exp}[\chi_N^{\mathbb{C}^2}(x\,\yadj,y_1,y_2)]\,=\,{\rm Exp}[\,\yadj\,\chi_N^{\mathbb{C}^2}(x,y_1,y_2)]~,
\end{equation}
such that: 
\begin{equation}
    Z_{\rm 1-loop,\,adj}^{\,U(N)}[\mathbb{C}^2]\,=
    {\rm Exp}\left[(\yadj-1)\,\chi_N^{\mathbb{C}^2}(x,y_1,y_2)\right]~.
\end{equation}

%%%%%%%%%%%%%%%%%%%%%%%%%%%%%%%%%%%%%%%%%
\begin{figure}[t]
    \centering
    \includegraphics[width=0.5\linewidth]{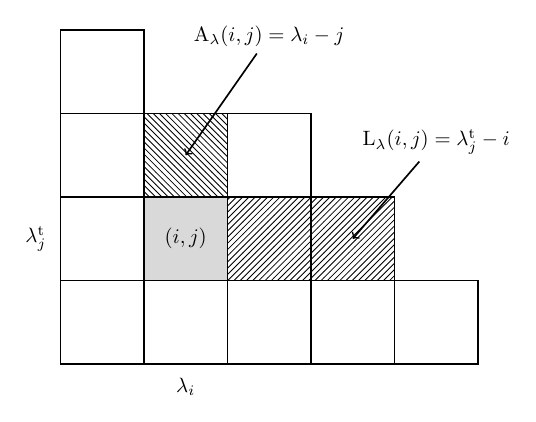}
    \caption{The arm and leg of an entry $\Box \equiv (i,j)\in\lambda$ in the Young tableau $\lambda = (\lambda_1, \cdots)$.}
    \label{fig:yngTabl}
\end{figure}

\subsubsection{The instanton partition function}\label{subsec:instantonC2}

On a local affine patch, the instanton partition function of the 5D $\mathcal{N}=1^*$ theory is given by:
\begin{equation}\label{adj K instanton U(N)}
    Z_{\rm inst,\,adj}^{\,U(N)}[\mathbb{C}^2] \,=\, \sum_{\vec{\lambda} = (\lambda_1,\cdots,\lambda_N)} \,\mathsf{q}^{|\vec{\lambda}|}\,Z^{\,U(N)}_{\vec{\lambda},\,{\rm adj}}~,
\end{equation}
with~\cite[theorem 2.11]{Nakajima:2003pg},~\cite[Equation (1.4)]{Nakajima:2005fg},\cite[Equation (2.4)]{Poghossian:2008ge}:
\begin{multline}\label{K inst adj U(N)}
    Z^{\,U(N)}_{\vec{\lambda},\,{\rm adj}} \,:=\, \prod_{\alpha,\beta=1}^N\,\Bigg[\,\prod_{\square\in \lambda_\alpha}\, \frac{\left(\,1\,-\,\frac{x_\beta}{x_\alpha}\,\yadj\,y_1^{-{\rm L}_\beta(\square)}\,y_2^{{\rm A}_\alpha(\square)+1}\,\right)}{\left(\,1\,-\,\frac{x_\beta}{x_\alpha}\,y_1^{-{\rm L}_\beta(\square)}\,y_2^{{\rm A}_\alpha(\square)+1}\,\right)}\,\\\times\prod_{\square\in \lambda_\beta}\,\frac{\left(\,1\,-\,\frac{x_\beta}{x_\alpha}\,\yadj\,y_1^{{\rm L}_\alpha(\square)+1}\,y_2^{-{\rm A}_\beta(\square)}\,\right)}{\left(\,1\,-\,\frac{x_\beta}{x_\alpha}\,y_1^{{\rm L}_\alpha(\square)+1}\,y_2^{-{\rm A}_\beta(\square)}\,\right)}\,\Bigg]~.
\end{multline}
The leg length and arm length of a box $\square\equiv (i,j)\in \lambda$ are defined as follows (see Figure \ref{fig:yngTabl}):
\begin{equation}\label{leg and arm defn}
    {\rm L}_{\lambda}(\square)\,:= \,\lambda^{\rm t}_j\,-\,i~, \qquad {\rm A}_{\lambda}(\square)\,:=\,\lambda_i\,-\,j~,
\end{equation}
where $\lambda^{\rm t}$ denotes the transpose of the partition $\lambda$.
It will be important to note that  under a simultaneous inversion
$(x_\alpha,\yadj, y_1, y_2)$, \eqref{K inst adj U(N)} picks up a factor of $\yadj^{-2N|\vec{\lambda}|}$, such that the instanton partition function
\eqref{adj K instanton U(N)} is invariant under:
\be
\label{yadjflip}
(x_\alpha,\, \yadj\,,\,y_1\,,\,y_2\,,\,\sfq)\,\longmapsto \,(1/x_\alpha, \, 1/\yadj\,,\,1/y_1\,,\,1/y_2\,,\,\sfq\, \yadj^{2N})~.
\ee
Moreover, in the limit $\yadj\rightarrow 0$ (corresponding to integrating out the adjoint hypermultiplet), \eqref{adj K instanton U(N)} reduces to the instanton partition function of the pure 5D $\mathcal{N}=1$ $U(N)$ SYM theory.\footnote{Integrating out a massive charged hypermultiplet in 5D gauge theory generates shifts in the 5D Chern--Simons (CS) terms~\protect\cite{Intriligator:1997pq} -- see also~\protect\cite[Appendix A]{Closset:2018bjz} for a review. This is analogous to what happens in 3d gauge theory -- see~\protect\cite[Subsection 2.2]{Closset:2023vos} for a recent review. For the case at hand, since the hypermultiplet is in the adjoint representation, the CS shift is trivial. We thank Cyril Closset for the discussion on this point.} Conversely, in the massless limit  $\yadj\rightarrow 1$, \eqref{adj K instanton U(N)} reduces to
the generating series of Euler numbers of the Hilbert scheme of $N$ points on $\IC^2$:
\begin{equation}\label{ZinstSU(N)->eta^-N}
    Z_{\rm inst,\,adj}^{\,U(N)}[\mathbb{C}^2]\mid_{\yadj\rightarrow 1}\,=\,\sum_{\vec{\lambda}=(\lambda_1, \cdots,\lambda_2)}\,\sfq^{|\vec{\lambda}|}\,=\,\overline{\eta}(\sfq)^{-N}~,
\end{equation}
where $\overline{\eta}$ is the normalized Dedekind eta function introduced in~\eqref{eta bar}.

In mathematical terms, the 5D $\mathcal{N}=1^*$ instanton partition function~\eqref{K inst adj U(N)} at order $\sfq^{k}$ with $k=|\vec{\lambda}|$ 
is understood as the equivariant virtual $\chi_{\yadj}$-genus of the ADHM moduli space of $k$ instantons on $\mathbb{C}^2$, 
or equivalently of rank $N$ framed torsion-free sheaves on
$\mathbb{P}^2$ with second Chern character equal to $k$ \footnote{For the evaluation of the Poincar\'e polynomial of the moduli space of framed sheaves on the stacky Hirzebruch surface, see~\protect\cite{Bruzzo_2011}.}. Invariance under \eqref{yadjflip} can be viewed as a generalization of~\cite[Corollary 4.9]{Fantechi_2010} at the equivariant level, 
with $2kN$ being the complex dimension of the moduli space of framed sheaves.
As explained in~\cite[Theorem 3.8]{Nakajima:2003uh}, the Poincar\'e polynomial of the modulus space of framed torsion-free sheaves can be obtained from the virtual equivariant $\chi_{\yadj}$ -genus~\eqref{K inst adj U(N)} by the limit $(y,x)\to 0^+$ of the equivariant weights of the torus action $y_1,y_2$ and $x_\alpha, \alpha=1,\ldots, N$ . Indeed, the moment map of the $U(1)^2 \times U(1)^N$ circle action is a perfect Morse function~\cite{AtiyahBott1984} on the ADHM moduli space (see~\cite[Chapter 5]{Nakajima1999}). Let us briefly recall this for completeness. The K\"ahler form on the ADHM moduli space is the restriction to the gauge orbits of the $(1,1)$ form:
\be
\varpi
=\frac{\I}{2}\operatorname{Tr}_V
\left(
\de B_1\,\wedge \,\de B_1^\dagger
\,+\,\de B_2\,\wedge\, \de B_2^\dagger
\,+\,\de I\,\wedge\, \de I^\dagger
\,+\,\de J^\dagger\,\wedge\, \de J
\right)~,
\ee
where $(B_1,B_2)\in {\rm End}(V)$ and $(I,J^{\dagger})\in {\rm Hom}(V,W)$, with $V\cong \mathbb{C}^k$ and $W\cong \mathbb{C}^N$.
For a generic torus element $\xi=(\epsilon_1,\epsilon_2,\bf{a})$, ${\bf a}=\operatorname{diag}(a_1,\ldots,a_N)$, the corresponding moment map reads:
\begin{align}
H_\xi
={}&
\epsilon_1\,\operatorname{Tr}_V(B_1B_1^\dagger)
+\epsilon_2\,\operatorname{Tr}_V(B_2B_2^\dagger)
-\operatorname{Tr}_V(I\,\mathbf{a}\,I^\dagger)
+\operatorname{Tr}_V\!\left(J^\dagger(\epsilon_1+\epsilon_2+\mathbf{a})J\right)~.
\end{align}
The character of the equivariant circle action is then naturally computed in terms of eigenvalues of the Hessian of the above Morse function at the critical points. 
It can be shown 
from~\eqref{K inst adj U(N)} that each negative eigenvalue of the toric action on the tangent space at the fixed point, which corresponds to a negative eigenvalue of the Hessian of the Morse function at the critical point, contributes a factor $\yadj$, while positive eigenvalues contribute $1$, recovering the homological weight $\yadj^{p}$ associated with the critical point, where $p$ is the number of negative Hessian eigenvalues.  

However, it should be pointed out that this prescription for extracting the Poincar\'e polynomial from the equivariant $\chi_{\yadj}$-genus does not directly apply
to the case of compact toric
surfaces of interest in this paper. Indeed, the proof of~\cite{Nakajima:2003uh} requires the weights of the toric action to be generic and assumes a special ordering 
$y_1\gg x_N>\ldots>x_2>x_1\gg y_2$
in order to simplify the proof. This choice in particular ensures the absence of zero eigenvalues.
However, when evaluating the supersymmetric partition function via contour integral in the equivariant weights $x$, 
we sit precisely at the poles for one or more factors~\eqref{K inst adj U(N)} in the integrand. We then get an extra factor $(1-\yadj)^p$ where $p$ is the order of the pole, e.g., $p=1$ for simple poles. Indeed, we will see the appearance of this factor in the explicit evaluations below. For the cases in which only simple poles contribute, this produces an overall factor $(1-\yadj)$ which needs to be divided out in order to get the Poincar\'e polynomial,
see for example \eqref{ZorbitP2} and \eqref{ZorbitFn}.

%%%%%%%%%%%%%%%%%%%%%%%%%%%%%%%%%%%%%%%%%
%%%%%%%%%%%%%%%%%%%%%%%%%%%%%%%%%%%%%%%%%
\subsection{For \texorpdfstring{$\mathcal{N}=1^*$}{N1} \texorpdfstring{$U(N)$}{UN} theory on \texorpdfstring{$S\times\mathbb{S}_{\boldsymbol{\beta}}^1$}{SS1b}}

Having reviewed the partition function on $\IC^2\times \mathbb{S}^1_{\boldsymbol{\beta}}$, we now obtain the partition of the $\mathcal{N}=1^*$ $U(N)$ gauge theory on the geometry $S\times \mathbb{S}^1_{\boldsymbol{\beta}}$ with $S$ being a compact toric surface with Euler number $\chi\equiv \chi(S)$. In this case, the partition function can be obtained by gluing together the contributions~\eqref{ZSU(N) full} associated to each of the  $\mathbb{C}^2_\ell$ affine patches~\cite{nekrasov2006localizing,Bershtein:2015xfa,Bershtein:2016mxz,Bonelli:2020xps}:
\begin{equation}\label{Zfull SUN glue}
    Z_{\rm full}^{\,U(N)}[S](a,\yadj,y_1,y_2)\,=\,\prod_{\ell=1}^{\chi}\, Z_{\rm full}^{\,U(N)}[\mathbb{C}_\ell^2](a_\ell,\yadj, y_{1,\ell}, y_{2,\ell}) ~.
\end{equation}
Splitting each factor into a classical, one-loop, and instanton part, we have, more explicitly:
\begin{equation}\label{ZFull SUN  S explicit}
    Z_{\rm full}^{\,U(N)}[S]\,=\,Z_{\rm class}^{\,U(N)}[S]\,Z_{\rm 1-loop,\,W}^{\,U(N)}[S]\,Z_{\rm 1-loop,\,adj}^{\,U(N)}[S]\,Z_{\rm inst,\,adj}^{\,U(N)}[S]~,
\end{equation}
where we now explain each of the factors in turn. Starting with the classical part, we have~\eqref{log Zclass S SUN}:
\begin{equation}\label{Zclass S SUN}
    Z_{\rm class}^{\,U(N)}[S]\,:=\,\sfq^{\,-\,\frac{N\,(N\,-\,1)}{4}\,-\,\sum_{\ell=1}^\chi\frac{\sum_{\alpha\neq\beta}\,(a^{(\ell)}_{\alpha,\beta})^2}{8\,\eps_{1,\ell}\,\eps_{2,\ell}}}~,
\end{equation}
with the Coulomb branch parameters associated with the $\ell$-th affine patch given by:
\begin{equation}\label{a ell alpha beta}
    a_{\alpha,\beta}^{(\ell)}\,\equiv\,a_{\alpha,\beta}\,+\,\frac{1}{2}\,\sfp_{\alpha, \beta}^{(\ell)}\,\eps_{1,\ell}\,+\,\frac{1}{2}\,\sfp^{(\ell+1)}_{\alpha,\beta}\,\eps_{2,\ell}~.
\end{equation}
Here $\sfp^{(\ell)}_{\alpha,\beta}\equiv \sfp_\alpha^{(\ell)}-\sfp_\beta^{(\ell)}$ are the equivariant magnetic fluxes for the $U(N)$ gauge group generalizing the $U(2)$ ones introduced in~\eqref{equiv-flux-defn}.

The contribution of the W-bosons appearing in~\eqref{ZFull SUN  S explicit} is of the form~\eqref{W-bos SUN S}:
\begin{equation}
    Z_{\rm 1-loop,\,W}^{\,U(N)}[S]\,=\,{\rm Exp}[-\,\chi^S_N(x,y_1,y_2)]~,
\end{equation}
with the character $\chi_N^S$ as defined in~\eqref{chiSN}. Similarly, the massive adjoint matter multiplet has the following contribution~\eqref{pert hyper C2}:
\begin{equation}
    Z_{\rm 1-loop,\,adj}^{\,U(N)}[S]\,=\,{\rm Exp}[\yadj\,\chi_{N}^S(x,y_1,y_2)]~.
\end{equation}

The instanton contribution in~\eqref{ZFull SUN  S explicit} is given by:
\begin{equation}
    Z_{\rm inst,\,adj}^{\,U(N)}[S]\,=\,\prod_{\ell=1}^\chi\, Z_{\rm inst,\,adj}^{\,U(N)}[\mathbb{C}^2_\ell]~,
\end{equation}
with each toric component contribution being of the form~\eqref{adj K instanton U(N)} upon substituting the corresponding toric weights and Coulomb branch parameters~\eqref{a ell alpha beta}.

\subsection{Specializing to rank \texorpdfstring{$2$}{2}}
\label{subsec:rank2 case}
In this subsection, we will specialize the earlier results we have had in this section to the rank $2$ gauge theory case. Starting with a local $\mathbb{C}^2\times \mathbb{S}^1_{\boldsymbol{\beta}}$,  we analyze the poles and zeros structure of the partition function~\eqref{ZSU(N) full}. Then, we generalize this discussion for the partition function~\eqref{ZFull SUN  S explicit} for any compact toric surface $S$.

Before we delve into the details of the analysis, let us first mention the convention that we will be following throughout this work. For the Coulomb branch parameters of the $U(2)$ theory, we will take:
\begin{equation}\label{su2 coulombs}
    a_1\,=\,-\,a_2\,\equiv\,a~.
\end{equation}
In particular, we have $a_{12} = -a_{21} = 2a$. The same applies for the corresponding 5D parameters; namely, $x_1 = x_2^{-1} \equiv x\equiv e^{-\boldsymbol{\beta}\,a}$. In particular, the classical contribution is~\eqref{logClassC2}:
\be
\label{Zclass S SU2}
  \log\, Z_{\rm class}^{\,U(2)}[\mathbb{C}^2] \,=\, 
   \left({-\frac{(2a)^2}{4\,\eps_1\,\eps_2}\, -\, \frac{\eps_1^2\,+\,\eps_2^2\,+\,3\,\eps_1\,\eps_2}{24\,\eps_1\,\eps_2}}\right)\,\log\,\sfq ~.
\ee

\subsubsection{Analytic structure of the \texorpdfstring{$1$}{1}-loop part on \texorpdfstring{$\mathbb{C}^2\times\mathbb{S}^1_{\boldsymbol{\beta}}$}{C2S1b}}
Let us start with the $1$-loop contribution of the W-bosons given in~\eqref{Z Wbos C2 SUN}:
\begin{equation}
     Z_{\rm 1-loop,\,W}^{\,U(2)}[\mathbb{C}^2]\,=\,{\rm Exp}\,\left[-\,\frac{x^2\,+\,x^{-2}}{(1\,-\,y_1^{-1})\,(1\,-\,y_2^{-1})}\right]~.
\end{equation}
Here we are following the convention in~\eqref{su2 coulombs}. Depending on the signs of the real parts of the equivariant parameters $\eps_1$ and $\eps_2$, the above expression could have zeros or poles. More explicitly, one can easily find~\cite[Appendix B]{Bonelli:2020xps},~\cite[Appendix K.1]{Kim:2025fpz}:
\begin{equation}\label{ZWbos C2 cases}
    Z_{\rm 1-loop,\,W}^{\,U(2)}[\mathbb{C}^2]\,=\,\begin{cases}
    \prod_{i,j\geq 0} \,\mathsf{G}_{i+1,j+1}^{(+)}\,\times\,\mathsf{G}_{i+1,j+1}^{(-)}~, ~ &{\rm Re}(\eps_1)\,,\,\text{Re}(\eps_2)\,>\,0~,\\
        % \prod_{i,j\geq 0} \,\mathsf{G}(i+1,j+1)~, ~ &{\rm Re}(\eps_1)\,,\,\text{Re}(\eps_2)\,>\,0~,\\
        % \prod_{i,j\geq 0}\,\mathsf{G}(-i,-j)~,~ &{\rm Re}(\eps_1)\,,\,\text{Re}(\eps_2)\,<\,0~,\\
        \prod_{i,j\geq 0}\,\mathsf{G}^{(+)}_{-i,-j}\,\times\,\mathsf{G}^{(-)}_{-i,-j}~,~ &{\rm Re}(\eps_1)\,,\,\text{Re}(\eps_2)\,<\,0~,\\
        % \prod_{i,j\geq 0}\,\mathsf{G}(i+1, -j)^{-1}~, ~ &{\rm Re}(\eps_1)\,>\,0\,,\,{\rm Re}(\eps_2)\,<\,0~,\\
        \prod_{i,j\geq 0}\,\left(\mathsf{G}^{(+)}_{i+1,-j}\,\times\,\mathsf{G}^{(-)}_{i+1,-j}\right)^{-1}~, ~ &{\rm Re}(\eps_1)\,>\,0\,,\,{\rm Re}(\eps_2)\,<\,0~,\\
        % \prod_{i,j\geq 0}\,\mathsf{G}(-i,j+1)^{-1}~, ~ &{\rm Re}(\eps_1)\,<\,0\,,\,{\rm Re}(\eps_2)\,>\,0~,\\
        \prod_{i,j\geq 0}\,\left(\mathsf{G}^{(+)}_{-i,j+1}\,\times\,\mathsf{G}^{(-)}_{-i,j+1}\right)^{-1}~, ~ &{\rm Re}(\eps_1)\,<\,0\,,\,{\rm Re}(\eps_2)\,>\,0~,\\
    \end{cases}
\end{equation}
where, for compactness, we introduced the function:
\begin{equation}\label{G+-}
    \mathsf{G}^{(\pm)}_{i,j}\,:=\,\left(\,1\,-\,x^{\pm 2}\,y_1^{i}\,y_2^{j}\,\right)~, \qquad \mathsf{G}_{i,j}\,:=\,\mathsf{G}^{(\pm)}_{i,j}\mid_{x=1}\,=\,\left(\,1\,-\,y_1^{i}\,y_2^{j}\,\right)~,
\end{equation}
for any two integers $i$ and $j$. The function $\mathsf{G}_{i,j}$ will be of importance for us later on.

From this, we see that the poles and zeros of the full $U(2)$ partition function~\eqref{ZFull SUN  S explicit} coming from the perturbative contribution of the W-bosons depend explicitly on the signs of Re($\eps_{1,2}$). More explicitly, the locations of these zeros or poles are of the following form:
\begin{equation}\label{pert zeros and poles}
    2\,a_{m,n,s}\,:=\,m\,\eps_1\,+\,n\,\eps_2\,+\,\frac{2\pi i \,s}{\boldsymbol{\beta}}~, \qquad m,\,n,\,s\,\in\,\mathbb{Z}~,
\end{equation}
such that:
\begin{equation}
    \begin{cases}
        m\,n\,>\,0~, \qquad &{\rm Re}(\eps_1)\,,\,\text{Re}(\eps_2)\,>\,0~,\\
        m\,n\,\geq\,0~,\qquad &{\rm Re}(\eps_1)\,,\,\text{Re}(\eps_2)\,<\,0~,\\
        (m\,<\,0\,\leq\,n)~\lor~(n\,\geq\,0\,<\,m)~,&{\rm Re}(\eps_1)\,>\,0\,,\,{\rm Re}(\eps_2)\,<\,0~,\\
        (n\,<\,0\,\leq\,m) ~\lor~(m\,\geq\,0\,<\,n)~, &{\rm Re}(\eps_1)\,<\,0\,,\,{\rm Re}(\eps_2)\,>\,0~.
    \end{cases}
\end{equation}

As for the perturbative contribution of the massive adjoint hypermultiplet~\eqref{pert hyper C2}, we do not get any zeros nor poles contributions.

\subsubsection{Instanton contribution of the theory on \texorpdfstring{$\mathbb{C}^2\times\mathbb{S}^1_{\boldsymbol{\beta}}$}{C2S1b}}

Let us now consider the instanton part of the partition function given in~\eqref{adj K instanton U(N)}, specializing to the rank $2$ case $N=2$. In order to study the structure of zeros and poles of this partition function, it will be crucial to establish a Zamolodchikov-type recurrence relation, generalizing the recursion found in~\cite[Equation (2.4)]{Yanagida:2010vz},~\cite[Equation (9.79)]{Kim:2025fpz} in the absence of the massive adjoint hypermultiplet. For this purpose, we shall 
split the instanton partition function~\eqref{adj K instanton U(N)} into a product:
\begin{equation}\label{ZU(2) decomp}
    Z_{\rm inst,\,adj}^{\,U(2)}[\mathbb{C}^2]\,=\,\mathsf{Z}^2\,H_{\rm adj}^{\rm K}(x,\yadj,y_1,y_2,\mathsf{q})~,
\end{equation}
where the prefactor $\mathsf{Z}$ is given by: 
\begin{equation}\label{prefactor for ZU(2)}
    \mathsf{Z}\,:=\,{\rm Exp}\left[\frac{\yadj\,\mathsf{q}\,(1\,-\,\yadj\,y_1)\,(1\,-\,\yadj\,y_2)}{(1\,-\,y_1)\,(1\,-\,y_2)\,(1\,-\,\yadj^2\,\mathsf{q})}\right]~.
\end{equation}
This prefactor will be recognized as the $U(1)$ instanton partition function~\eqref{Zinst K U(1) C2} in the next section, up to a shift $\mathsf{q}\,\to\,\yadj\,\mathsf{q}$. 
The details of deriving this prefactor from~\eqref{adj K instanton U(N)} are relegated to Appendix \ref{app:Z factor}. 
Moreover, it is straightforward to check that \eqref{prefactor for ZU(2)} is invariant under the inversion \eqref{yadjflip} of equivariant parameters,
provided $\sfq$ is rescaled into $\sfq\, \yadj^4$. 

We now claim that the function $H^{\rm K}_{\rm adj}$ satisfies the following recurrence relation:
\begin{equation}\label{HKadj recurrence}
    H^{\rm K}_{\rm adj}(x,\yadj,y_1,y_2)\,=\,1\,-\,\sum_{m,n=1}^\infty \,\frac{\mathsf{q}^{mn}\,T^{\rm adj}_{m,n}(y_1, y_2)\,H^{\rm K}_{\rm adj}(y_1^{\frac{m}{2}}\,y_2^{-\frac{n}{2}},\yadj,y_1,y_2)}{(y_1\,y_2)^{mn}\,(x^{-2}\,-\,y_1^m\,y_2^n)\,(1\,-\,x^2\,y_1^{-m}\,y_2^{-n})}~,
\end{equation}
where:
\begin{equation}\label{Tadjmn}
    T_{m,n}^{\rm adj}(\yadj,y_1, y_2)\,:=\, T_{m,n}(y_1,y_2)\,\prod_{i=-m+1}^m\,\prod_{j=-n+1}^n\,\left(\,1\,-\,y_1^i\,y_2^j\,\yadj\,\right)~.
\end{equation}
Here, $T_{m,n}$ is given by~\cite[Equation (2.4)]{Yanagida:2010vz},\cite[Equation (9.79)]{Kim:2025fpz}:
\begin{equation}\label{Tmn}
    T_{m,n}(y_1,y_2)\,:=\, \frac{1\,+\,y_1^m\,y_2^n}{(y_1\,y_2)^{mn}}\,\underbrace{\prod_{i=-m+1}^{m}\,\prod_{j=-n+1}^n}_{(i,j)\neq(0,0),(m,n)}\,\frac{y_1^i\,y_2^j}{1\,-\,y_1^i\,y_2^j}~.
\end{equation}
We have checked explicitly at low orders in $\sfq$ that this recurrence relation reproduces the sum over Young tableaux in~\eqref{adj K instanton U(N)} for $N=2$. Moreover, using:
\be
\begin{split}
  &T_{m,n}(y_1,y_2)\,=\,  T_{m,n}(1/y_1,1/y_2)~,\\
   &T_{m,n}^{\rm adj}( \yadj, y_1, y_2)
   \,=\,
  (y_1 \,y_2\, \yadj^2)^{-2mn}\, T_{m,n}^{\rm adj}( 1/\yadj, 1/y_1, 1/y_2)~,  
\end{split}
\ee
one easily checks that $H^{\rm K}_{\rm adj}(x)$ is invariant under a simultaneous
inversion of $(x,\yadj,y_1,y_2)$ combined with a rescaling $\sfq\mapsto \sfq \yadj^4$, as required by the symmetry under
\eqref{yadjflip}. Moreover, in the infinite mass limit $\yadj\rightarrow 0$, the prefactor $\mathsf{Z}$~\eqref{prefactor for ZU(2)} trivializes and the recurrence relation for $H^{\rm K}_{\rm adj}$ reduces to that given in~\cite[Theorem 4.1]{Yanagida:2010vz} and~\cite[Equation (9.78)]{Kim:2025fpz}.\footnote{When comparing with the conventions of~\protect\cite{Kim:2025fpz}, we have the identification $\mathsf{q}\equiv \mathcal{R}^4\,y_1\,y_2$. With this in mind, we have an extra $(y_1y_2)^{mn}$ appearing in the denominator in~\protect\eqref{HKadj recurrence}.} Another check for the above recurrence relation is to look at the 4D limit. That is, taking $q$ to be the 4D gauge parameter, we take~\eqref{K to coh} along with:\footnote{In the more general $U(N)$ case, the relation between the two gauge parameters is $\sfq = \boldsymbol{\beta}^{2N}\,q$.}
\begin{eqnarray}\label{qktoqcoh}
    \mathsf{q}\,=\, \boldsymbol{\beta}^{4}\, q~,
\end{eqnarray}
and extract the leading order of $\boldsymbol{\beta}$ in the small radius limit.  In this limit, the above recurrence relation reduces to the one studied by~\cite[Equation (5.3)]{Poghossian:2009mk} -- see also~\cite[Equation (4.6)]{Bershtein:2015xfa} and~\cite[Equation (4.50)]{Sysoeva:2022syp}. Taking $m_{\rm adj}\rightarrow \infty$ limit, the final recurrence relation becomes that of the instanton partition function of 4D $\mathcal{N}=2$ $U(2)$ gauge theory also considered in~\cite{Poghossian:2009mk,Bershtein:2015xfa,Sysoeva:2022syp}.

\medskip
\noindent
\textbf{Further comments on the structure of poles and zeros.} From the recurrence relation~\eqref{HKadj recurrence}, we can see that, regardless of the signs of Re$(\eps_{1,2})$, the instanton partition function of the $U(2)$ theory has only poles. These are simple poles at the following locations~\cite[Appendix K.1]{Kim:2025fpz},~\cite[Section 3]{Bonelli:2020xps}:
\begin{equation}\label{poles inst C2}
    2\,a_{m,n,s}\,:=\,m\,\eps_1\,+\,n\,\eps_2\,+\,\frac{2\pi i \,s}{\boldsymbol{\beta}}~, \qquad m,\,n,\,s\,\in\,\mathbb{Z}~, ~~m\,n>0~.
\end{equation}

Combining this observation with the analogous one made for the 1-loop part of the partition function~\eqref{pert zeros and poles}, we have the following poles/zeros structure of the full $U(2)$ partition function on $\mathbb{C}^2\times\mathbb{S}^1_{\boldsymbol{\beta}}$:
\begin{itemize}
    \item If ${\rm Re}(\eps_1)\,{\rm Re}(\eps_2)>0$, then the 1-loop part contributes zeros of degree 1 at the same locations of the simple poles coming from the instanton partition function. Therefore, the overall result is that the full partition function does not have poles or zeros in this regime.
    \item If ${\rm Re}(\eps_1)\,{\rm Re}(\eps_2)<0$, then both the 1-loop and the instanton partition function contribute simple poles.
\end{itemize}

\subsubsection{Abstruse duality}
As was pointed out in~\cite[Section 3.1]{Bonelli:2020xps} and further elaborated in~\cite{Sysoeva:2022syp}, the full partition function of the 4D $U(N)$ SYM theory on $\mathbb{C}^2$ satisfies an `abstruse duality' that relates its residues at different poles $a=m\epsilon_1\pm n\epsilon_2$. 
This duality was generalized for the pure 5D $\mathcal{N}=1$ $U(2)$ SYM theory on $\mathbb{C}^2\times\mathbb{S}^1_{\boldsymbol{\beta}}$ in~\cite[Appendix K.2]{Kim:2025fpz}. We claim that it  continues to hold in the 5D $\mathcal{N}=1^*$ $U(2)$ SYM theory, namely: 
\begin{align}\label{abstruce-duality-C2}
\begin{split}
        \frac{\lim_{x\,\rightarrow\, x_{m,n}}\,Z_{\rm full}^{\,U(2)}[\mathbb{C}^2]}{\lim_{x\,\rightarrow\, \hat{x}_{m,n}}\,Z_{\rm full}^{\,U(2)}[\mathbb{C}^2]}\,=\,-\,\sgn({\rm Re}(\eps_1))~,\qquad \frac{\lim_{x\,\rightarrow\, x_{m,n}}\,Z_{\rm full}^{\,U(2)}[\mathbb{C}^2]}{\lim_{x\,\rightarrow\, \hat{x}^{-1}_{m,n}}\,Z_{\rm full}^{\,U(2)}[\mathbb{C}^2]}\,=\,-\,\sgn({\rm Re}(\eps_2))~,
        \end{split}
\end{align}
where, $x_{m,n}\equiv y_1^my_2^n$ and $\hat{x}_{m,n}\equiv y_1^my_2^{-n}$.
This relation will play a key role in our calculation of the full partition function of the 5D theory on $S$. 

\medskip
\noindent
\textbf{Sketch of the proof.} The proof  follows the same steps as in~\cite[Appendix K.2]{Kim:2025fpz} for the case of pure $\cN=1$ 5D SYM. The only difference concerns the contribution of the massive adjoint matter multiplet. But, from our discussion above on the 1-loop and instanton partition functions, we saw that the presence of the massive adjoint hypermultiplet does not affect the analytic structure of the full partition function, hence,~\eqref{abstruce-duality-C2} should hold for the reasons as in~\cite{Kim:2025fpz}.

For instance, let us look at the case with $m,n>0$. For the 1-loop part of the ratio appearing on the r.h.s. of the first~\eqref{abstruce-duality-C2}, the adjoint multiplet adds a factor of the form:
\begin{equation}\label{adj-abstr-factor}
   \prod^{m}_{i=-m+1}\prod_{j=-n+1}^{n}\,\frac{1}{1\,-\,y_1^{i}\,y_2^{j}\,\yadj}~,
\end{equation}
to the r.h.s. of~\cite[Equation (K.37)]{Kim:2025fpz}\footnote{Note here that in~\protect\cite{Kim:2025fpz}, they denote the K-theoretic $\Omega$-background parameters by $t_{1,2}$. They are related to our conventions via: $y_i = t_i^{-1}$ for $i=1,2$.}. In particular, the new factor appears in the denominator due to the overall minus sign difference in the contribution of the W-bosons~\eqref{Z Wbos C2 SUN} and massive adjoint multiplet~\eqref{pert hyper C2}.
As for the instanton part of the ratio in~\eqref{abstruce-duality-C2}, we get the r.h.s. of~\cite[Equation (K.39)]{Kim:2025fpz} with $T_{m,n}$ replaced by $T_{m,n}^{\rm adj}$ defined in~\eqref{Tadjmn}. This amounts to the addition of the second factor appearing on the r.h.s. of~\eqref{Tadjmn}, which is the inverse of the term~\eqref{adj-abstr-factor}. This proves the first duality in~\eqref{abstruce-duality-C2} for $m,n>0$. One can use the same reasoning to prove the remaining cases.

\subsubsection{Full partition function for the theory on \texorpdfstring{$S\times \mathbb{S}^1_{\boldsymbol{\beta}}$}{SS1b}}
Let us now combine the analysis of poles and zeros done above in $\mathbb{C}^2\times\mathbb{S}^1_{\boldsymbol{\beta}}$ to calculate the degree of the zeros and poles of the full partition function of the $U(2)$ theory in $S\times\mathbb{S}^1_{\boldsymbol{\beta}}$~\eqref{ZFull SUN  S explicit}. As a starter, let us further simplify the form of the classical term given in~\eqref{Zclass S SUN}. In the $U(2)$ case, following the conventions in~\eqref{su2 coulombs}, we have only one independent Coulomb branch parameter $a=\frac12 a_{1,2}=-\frac12 a_{2,1}$. For each one of the affine patches, \eqref{a ell alpha beta} becomes:
\begin{equation}\label{2 a ell}
    a_\ell\,=\,a\,+\,\frac12\,\sfp_\ell\,\eps_{1,\ell}\,+\,\frac12\,\sfp_{\ell+1}\,\eps_{2,\ell}~, \qquad \ell \,=\, 1\,,\, \cdots\,,\, \chi~.
\end{equation}
where, to lighten the notation, we introduced $2\,a_\ell \equiv a^{(\ell)}_{1,2}$ and $\sfp_\ell \equiv \sfp^{(\ell)}_{1,2}$.
For the classical term, we get from~\eqref{Zclass S SUN}:
\begin{equation}\label{Zclass S SU22}
    Z_{\rm class}^{\,U(2)}[S]\,:=\,\sfq^{\,-\,\frac{1}{2}\,-\,\sum_{\ell=1}^\chi\frac{(2\,a_{\ell})^2}{4\,\eps_{1,\ell}\,\eps_{2,\ell}}}~.
\end{equation}
Substituting $a_\ell$ by~\eqref{2 a ell} and using the toric identities~\eqref{toric-weights-ident}, we end up with the following result:
\begin{equation}
    \sum_{\ell=1}^{\chi}\frac{(2\,a_\ell)^2}{\eps_{1,\ell}\,\eps_{2,\ell}} \,=\, -\sum_{\ell=1}^\chi \,h_\ell\,\mathsf{p}_\ell^2\,+\,2\,\sum_{\ell=1}^\chi\,\mathsf{p}_\ell\,\mathsf{p}_{\ell+1}~,
\end{equation}
where the intersection numbers $h_\ell$ were introduced in~\eqref{divisor-intersection}.

The full partition function of the $U(2)$ theory is therefore given by~\eqref{ZFull SUN  S explicit}:
\begin{equation}\label{rank 2 full partition}
     Z_{\rm full}^{\,U(2)}[S]\,=\,\sfq^{\Delta_{c_1,\vec{\sfp}}[S]}\,{\rm Exp}\left[(\yadj-1)\,\chi_{S}(x,y_1,y_2)\,\right]\,\prod_{\ell=1}^\chi Z_{\rm inst,\,adj}^{\,U(2)}[\mathbb{C}^2_\ell]~.
\end{equation}
Here, we further introduced the notation:
\begin{equation}\label{DeltaS}
    \Delta_{c_1,\vec{\sfp}}[S]\,:=\,\frac{1}{4}\,\sum_{\ell}\,h_\ell\,\mathsf{p}_\ell^2\,-\,\frac{1}{2}\,\sum_{\ell}\,\mathsf{p}_\ell\,\mathsf{p}_{\ell+1}\,+\,\frac{1}{4}\,c_1^2~,
\end{equation}
where $c_1 = \sum_{\ell=3}^\chi f_\ell\,D_\ell$~\eqref{c1E} is the first Chern class of the gauge bundle.
Moreover, we use the simpler notation $\chi_S(x,y_1,y_2)\equiv \chi_2^S(x,y_1,y_2)$ for the character of the toric surface $S$ in the rank $2$ case.

\medskip
\noindent
\textbf{The structure of poles and zeros.} The observations below~\eqref{poles inst C2} concerning the zeros and poles of the $U(2)$ partition function on $\mathbb{C}^2\times\mathbb{S}^1_{\boldsymbol{\beta}}$ can be carried over to any compact toric surface $S$ with Euler number $\chi$. Note that, for any such surface, there are only two affine patches with the property that ${\rm Re}(\eps_{1,\ell}){\rm Re}(\eps_{2,\ell})>0$. Therefore, the full partition function has poles of degree at most $\chi-2$~\cite[Remark 2]{Bonelli:2020xps}.

%%%%%%%%%%%%%%%%%%%%%%%%%%%%%%%%%%%%%%%%%
%%%%%%%%%%%%%%%%%%%%%%%%%%%%%%%%%%%%%%%%%
%%%%%%%%%%%%%%%%%%%%%%%%%%%%%%%%%%%%%%%%%
\section{Rank \texorpdfstring{$1$}{1} refined Vafa--Witten invariants for any toric surface}
\label{sec_U1part}
In this section, we specialize to the rank $1$, Abelian case. In this case, the integrand of the partition function after localization does not depend on the Coulomb branch parameter. This follows from the fact that the appearance of the Coulomb branch parameter in~\eqref{ZFull SUN  S explicit} comes from the adjoint matter multiplets, which are neutral under the $U(1)$ gauge group. Therefore, in the Abelian case we do not have to worry about taking any residues, and the final form of the 5D partition function can be worked out exactly for any toric surface $S$ and in any order of the gauge parameter $\sfq$.  

Our main result in this section is to derive G\"ottsche's formula for the Poincaré polynomial of the Hilbert scheme of points on a closed toric surface $S$, Hilb$[S] = \bigoplus_{k\geq 0} {\rm Hilb}^k[S]$~\cite[Equation (1b)]{GottscheBetti} -- see also~\cite[Equation (6.2)]{nakajima1999lectures}:\footnote{When comparing with \protect\cite{GottscheBetti,nakajima1999lectures}, keep in mind that, for compact toric surfaces, we have $b_1(S) = b_3(S) = 0$ and $b_0(S) = b_4(S)= 1$. Therefore, $b_2(S) = \chi - 2$.}
\begin{equation}\label{Gottsche for S}
    \sum_{k\geq 0} p({\rm Hilb}^k[S],\yadj) \,\mathsf{q}^n\,=\,\prod_{n\geq 1} \frac{1}{(1-\yadj^{n-1}\,\mathsf{q}^n)\,(1-\yadj^{n}\,\mathsf{q}^n)^{\chi-2}\,(1-\yadj^{n+1}\,\mathsf{q}^n)}~.
\end{equation}
Here, $\chi \equiv\chi(S)$ is the Euler number of the toric surface. In the unrefined limit, the above equation becomes:
\begin{align}\label{unrefined Gottsche}
    \sum_{n\geq 0} e({\rm Hilb}^n[S])\,\mathsf{q}^n \,=\, \overline{\eta}(\sfq)^{-\chi(S)} ~,
\end{align}
where the normalized eta function is defined in~\eqref{eta bar}.
%%%%%%%%%%%%%%%%%%%%%%%%%%%%%%%%%%%%%%%%%
%%%%%%%%%%%%%%%%%%%%%%%%%%%%%%%%%%%%%%%%%
\subsection{5D \texorpdfstring{$\mathcal{N}=1$}{N1} pure \texorpdfstring{$U(1)$}{U1} gauge theory on \texorpdfstring{$\mathbb{C}^2\times \mathbb{S}_{\boldsymbol{\beta}}^1$}{C2S1b}}
As a warm-up, let us consider pure $\mathcal{N}=1$ $U(1)$ gauge theory on the 5D geometry $\mathbb{C}^2\times\mathbb{S}_{\boldsymbol{\beta}}^1$. In this case, it is well known that the moduli space of instantons is none other than the Hilbert scheme of points on $\mathbb{C}^2$~\cite{nakajima1999lectures}:
\begin{equation}
    {\rm Hilb}[\mathbb{C}]\,:=\,\bigoplus_{k\geq 0} \,{\rm Hilb}^{k}[\mathbb{C}^2]~,
\end{equation}
with $k$ being the instanton number. From the 5D gauge theory perspective, the K-theoretic Nekrasov partition function is given by~\eqref{adj K instanton U(N)}:
\begin{equation}\label{U(1) inst par func}
    Z_{\rm inst}^{\,U(1)}[\mathbb{C}^2](y_1,y_2)\, =\,\sum_{\lambda} \,\mathsf{q}^{|\lambda|}\, Z_\lambda^{\,U(1)}(y_1,y_2)~,
\end{equation}
with, 
\begin{equation}
    Z^{\,U(1)}_\lambda (y_1,y_2)\,:=\, \prod_{\square \in \lambda}\, \left(\,1\,-\,y_1^{-{\rm L}_\lambda(\square)}\,y_2^{{\rm A}_\lambda(\square)+1}\,\right)^{-1}\,\left(\,1\,-\,y_1^{{\rm L}_\lambda(\square)+1}\,y_2^{-{\rm A}_\lambda(\square)}\,\right)^{-1}~.
\end{equation}
The leg and arm lengths of $\square\equiv (i,j)\in \lambda$ are defined in~\eqref{leg and arm defn}.

More compactly, one can write the instanton partition function~\eqref{U(1) inst par func} as follows~\cite[Equation (2.5)]{Poghossian:2008ge},~\cite[Equation (4.5)]{Nakajima:2003pg}:
\begin{equation}\label{Zinst first patch}
    Z_{\rm inst}^{\,U(1)}[\mathbb{C}^2](y_1,y_2)\, =\, {\rm Exp}\left[\, \frac{\mathsf{q}}{\,(1\,-\,y_1)\,(1\,-\,y_2)}\,\right]\, =\, \prod_{k_1, k_2\geq 0}\,\frac{1}{1\,-\,\mathsf{q}\,y_1^{k_1}\,y_2^{k_2}}~.
\end{equation}
One can indeed check that both expressions~\eqref{U(1) inst par func} and~\eqref{Zinst first patch} match order by order in $\mathsf{q}$. Expanding the above expression up to $\mathsf{q}^3$, we get:
\begin{multline}
       Z_{\rm inst}^{\,U(1)}[\mathbb{C}^2](y_1,y_2) \,=\,   1\,+\,\frac{\mathsf{q}}{\left(1\,-\,y_1\right)\, \left(1\,-\,y_2\right)}\,\\+\,\frac{ \left(1\,+\,y_1\, y_2\,\right)\,\mathsf{q}^2}{ \left(1\,+\,y_1\right)\,\left(1\,-\,y_1\right)^2\, 
   \left(1\,+\,y_2\right)\,\left(1\,-\,y_2\right)^2}\,+\,\cdots~.
\end{multline}

\medskip
\noindent
\textbf{The 4D limit.} Another limit that one can consider for the instanton partition function~\eqref{Zinst first patch} is the $\boldsymbol{\beta}\rightarrow 0$ one. Substituting the relations~\eqref{K to coh} and~\eqref{qktoqcoh} and looking at the leading order in the radius $\boldsymbol{\beta}$, we get~\cite[Equation (5.8)]{Pestun:2007rz},\cite[Equation (4.7)]{Nakajima:2003pg},\cite[Equation (3.2)]{Losev:2003py}:
\begin{equation}\label{4D U(1) instanton}
    Z^{{\rm 4D},\,U(1)}_{\rm inst}[\mathbb{C}^2](\eps_1,\eps_2) \,=\,\exp\left(\frac{q}{\eps_1\eps_2}\right) \,=\, 1 \,+\, \frac{q}{\eps_1\eps_2}\, + \,\frac{q^2}{2 \,\eps_1^2\,\eps_2^2}\,+\,\cdots~.
\end{equation}

%%%%%%%%%%%%%%%%%%%%%%%%%%%%%%%%%%%%%%%%%
%%%%%%%%%%%%%%%%%%%%%%%%%%%%%%%%%%%%%%%%%
\subsection{5D \texorpdfstring{$\mathcal{N}=1^*$}{N1} \texorpdfstring{$U(1)$}{U1} gauge theory on \texorpdfstring{$\mathbb{C}^2\times \mathbb{S}_{\boldsymbol{\beta}}^1$}{C2S1b}}
Let us now add in a massive adjoint hypermultiplet to the theory considered in the previous subsection. Recall that the adjoint hypermultiplet is neutral under $U(1)$. The full instanton partition function becomes of the form~\eqref{adj K instanton U(N)}:
\begin{equation}\label{K inst U(1)}
    Z_{\rm inst,\,adj}^{\,U(1)}[\mathbb{C}^2](\yadj,y_1,y_2) \,=\, \sum_\lambda \mathsf{q}^{|\lambda|} \,Z^{\,U(1)}_{\lambda,\,{\rm adj}} (\yadj,y_1,y_2)~,
\end{equation}
where, 
\begin{equation}
    Z^{\,U(1)}_{\lambda,\,{\rm adj}} (\yadj,y_1,y_2)\,:=\,\prod_{\square\in \lambda} \frac{ \left(\,1\,-\,\yadj\,y_1^{-{\rm L}_\lambda(\square)}\,y_2^{{\rm A}_\lambda(\square)+1}\,\right)}{ \left(\,1\,-\,y_1^{-{\rm L}_\lambda(\square)}\,y_2^{{\rm A}_\lambda(\square)+1}\,\right)} \,\frac{\left(\,1\,-\,\yadj\,y_1^{{\rm L}_\lambda(\square)+1}\,y_2^{-{\rm A}_\lambda(\square)}\,\right)}{\left(\,1\,-\,y_1^{{\rm L}_\lambda(\square)+1}\,y_2^{-{\rm A}_\lambda(\square)}\,\right)}~.
\end{equation}
Note that in the infinite mass limit $m_{\rm adj}\rightarrow\infty$, or equivalently $\yadj \rightarrow 0$, this reduces to 
the partition function  of the pure 5D $\mathcal{N}=1$ $U(1)$ theory given in~\eqref{U(1) inst par func}.
In fact, the K-theoretic partition function~\eqref{K inst U(1)} can be written more compactly as~\cite[Equation (2.10)]{Poghossian:2008ge},~\cite[Theorem 1.3]{rains2018nekrasov}:
\begin{equation}\label{Zinst K U(1) C2}
    \begin{split}
    Z_{\rm inst,\,adj}^{\,U(1)}[\mathbb{C}^2](\yadj,y_1,y_2) \,&=\, {\rm Exp}\left[\frac{\mathsf{q}\,(\,1\,-\,\yadj\,y_1\,)\,(\,1\,-\,\yadj\,y_2\,)}{\,(\,1\,-\,y_1\,)\,(\,1\,-\,y_2\,)\,(\,1\,-\,\yadj\,\mathsf{q}\,)\,}\right]~.
\end{split}
\end{equation}
It is straightforward to check that this matches with~\eqref{K inst U(1)} order by order in $\mathsf{q}$. Moreover, it coincides with \eqref{ZU(2) decomp}
up to a shift $\mathsf{q}\,\to\,\yadj\,\mathsf{q}$.

\medskip
\noindent
\textbf{The non-equivariant limit.} In the limit $y_1,y_2\rightarrow 0$, one finds the following expression for the instanton partition function~\cite[Equation (2.11)]{Poghossian:2008ge}:
\begin{equation}\label{Poincare HilbC^2}
\begin{split}
    Z_{\rm inst,\,adj}^{\,U(1)}[\mathbb{C}^2] \,&=\,\prod_{k\geq 1}\frac{1}{1\,-\,\yadj^{k-1}\,\mathsf{q}^{k}}\\
    &=\,1\,+\,\mathsf{q}\,+\,(1\,+\,\yadj)\,\mathsf{q}^2\, +\, \left(1\,+\,\yadj\,+\,y^2_{\rm adj}\right)\,\mathsf{q}^3\,\\
    &~~+\,\left(1+\yadj+2 \yadj^2+y^3_{\rm adj}\right)\mathsf{q}^4+\cdots ~.
    \end{split}
\end{equation}
This is recognized as the generating series of the Poincaré polynomial for the Hilbert scheme of points on $\mathbb{C}^2$~\cite[Corollary 5.9]{nakajima1999lectures},~\cite[Corollary 3.10]{Nakajima:2003uh}:\footnote{To match with these two references, one needs to identify $\yadj \equiv t^2$.}
\begin{equation}
    Z_{\rm inst,\,adj}^{\,U(1)}[\mathbb{C}^2]|_{y_1, y_2\rightarrow 0} \,=\, \sum_{k\geq 0}\, p({\rm Hilb}^k[\mathbb{C}^2], \yadj)\,\mathsf{q}^k~. 
\end{equation}

\medskip
\noindent
\textbf{The unrefined limit.} Another interesting limit is when the adjoint hypermultiplet becomes massless, i.e., $\yadj\rightarrow 1$. 
In this case, the $\cN=1$ supersymmetry of the 5D theory enhances to $\mathcal{N}=2$, and the 5D partition function computes the topological Euler numbers of these moduli spaces:
\begin{equation}\label{KVW C2}
    Z_{\rm inst, \, adj}^{ U(1)}[\mathbb{C}^2] |_{\yadj \rightarrow 1} \,=\, \sum_{k\geq 0} e({\rm Hilb}^k[\mathbb{C}^2])\,\mathsf{q}^k~.
    \end{equation}
Performing this limit on the r.h.s. of~\eqref{Poincare HilbC^2}, we find that the dependence on the equivariant parameters drops out and get~\eqref{ZinstSU(N)->eta^-N}:
\begin{equation}\label{dedkind eta defn}
    Z_{\rm inst, \, adj}^{ U(1)}[\mathbb{C}^2] |_{\yadj \rightarrow 1} \,=\,  \overline{\eta}(\mathsf{q})^{-1}\,=\,1\,+\,\mathsf{q}\,+\,2\, \mathsf{q}^2\,+\,3\, \mathsf{q}^3\,+\,5\,\mathsf{q}^4\,+\,\cdots~.
\end{equation}

%%%%%%%%%%%%%%%%%%%%%%%%%%%%%%%%%%%%%%%%%
%%%%%%%%%%%%%%%%%%%%%%%%%%%%%%%%%%%%%%%%%
\subsection{5D \texorpdfstring{$\mathcal{N}=1^*$}{N1} \texorpdfstring{$U(1)$}{U1} gauge theory on \texorpdfstring{$S\times \mathbb{S}_{\boldsymbol{\beta}}^1$}{SS1}}

Let us now consider the case of a general compact toric surface $S$  with Euler number $\chi \equiv\chi(S)$. Our aim is to show that the partition function of the 5D $\mathcal{N}=1^*$ $U(1)$ gauge theory on $S\times \mathbb{S}_{\boldsymbol{\beta}}^1$ computes the generating series of Poincaré polynomials of the Hilbert scheme of points on $S$, given by~\eqref{Gottsche for S}:
% The claim is that:
\begin{eqnarray}\label{main claim}
    Z_{\rm inst,\,adj}^{\,U(1)}[S]\,=\, \sum_{k\geq 0} \,p({\rm Hilb}^k[S], \yadj)\,\mathsf{q}^k~.
\end{eqnarray}

Recall from~\eqref{Zfull SUN glue} that the partition function of the 5D theory can be obtained simply by gluing together the partition functions associated to each of the local affine patches:\footnote{In the rank 1 case, the dependence on the Coulomb parameter is trivial, so there is no need to integrate over it. }
\begin{equation}\label{Zinst U(1) on S}
    Z_{\rm inst,\, adj}^{\,U(1)}[S]\,=\,\prod_{\ell=1}^\chi\,Z_{\rm inst,\,adj}^{\,U(1)}[\mathbb{C}^2_\ell] \,=\, {\rm Exp}\left[\sum_{\ell=1}^{\chi}\,\frac{\mathsf{q}\,(\,1\,-\,\yadj\,y_{1,\ell}\,)\,(\,1\,-\,\yadj\,y_{2,\ell}\,)}{(\,1\,-\,y_{1,\ell}\,)\,(\,1\,-\,y_{2,\ell}\,)\,(\,1\,-\,\yadj\,\mathsf{q}\,)}\right]~.
\end{equation}
In the last equality, we used~\eqref{Zinst first patch}. The r.h.s. can be simplified further by observing that:
\begin{equation}\label{claim}
    \sum_{\ell=1}^\chi\,\frac{\,(\,1\,-\,\yadj\,y_{1,\ell}\,)\,(\,1\,-\,\yadj\,y_{2,\ell}\,)}{(\,1\,-\,y_{1,\ell}\,)\,(\,1\,-\,y_{2,\ell}\,)}\,=\,1\,+\,(\chi\,-\,2)\,\yadj\,+\,\yadj^2~.
\end{equation}
This relation can be derived straightforwardly by expanding the terms on the r.h.s. in powers of $\yadj$ and simplifying the coefficient in each term using the identities~\eqref{Identity I,II}--\eqref{identity III}. Plugging~\eqref{claim} back into~\eqref{Zinst U(1) on S} and recalling the definition~\eqref{pexp defn}, we get the r.h.s. of G\"ottsche's formula~\eqref{Gottsche for S}; hence, proving~\eqref{main claim}.
Explicitly, setting $b_2=\chi-2$, we have: 
\begin{equation}\label{GottscheExplicit}
\begin{split}
    Z_{\rm inst,\,adj}^{\,U(1)}[S]  
     &= 1\,+\,  \left(\frac{1}{\yadj}\,+\,b_2+\,\yadj\right)\,(\yadj\,\mathsf{q})\\
     &+\, \left(\frac{1}{\yadj^2}\,+\, \frac{b_2+1}{\yadj}\,+\frac12 (b_2+1)(b_2+2) \,+\,\cdots\right)\,(\yadj\,\mathsf{q})^2\,\\
    &+\,\left(\frac{1}{\yadj^3}\,+\,\frac{b_2+1}{\yadj^2}\,+
    \frac{(b_2+1)(b_2+4)}{2\yadj}\,+ 2+\frac16 b_2(b_2+2)(b_2+7) 
    \, +\,\cdots\right)\,(\yadj\,\mathsf{q})^3\,\\&+\,\cdots~,\\
\end{split}
\end{equation}
where the dots inside the brackets denote terms required by invariance under
$\yadj\mapsto  1/\yadj$. In the unrefined limit, this becomes:
\begin{equation}
\label{unrefined U(1) S}
    Z_{\rm inst,\,adj}^{\,U(1)}[S]\mid_{\yadj\rightarrow 1}\,=
    \overline{\eta}(\sfq)^{-\chi} = 
    \,1\,+\chi\,\sfq\,+\frac12 \chi(\chi+3)\,\sfq^2\,+
    \frac16 \chi( \chi^2+9\chi+8)
    \,\sfq^3\,+\,\cdots~.
\end{equation}

\medskip
\noindent
\textbf{Modularity of the full K-theoretic partition function.} Let us now rewrite the full partition function of the 5D $\mathcal{N}=1^*$ theory on $S\times \mathbb{S}_{\boldsymbol{\beta}}^1$~\eqref{Gottsche for S} using Jacobi theta functions. Recall:
\begin{equation}
    \theta_1(m|\tau) := 2\,\mathsf{q}^{\frac{1}{8}}\,\sin(\pi m)\,\prod_{n\geq 1}\,(\,1\,-\,y^{-1}\,\mathsf{q}^n\,) \,(\,1\,-\,\mathsf{q}^n\,)\,(\,1\,-
    \,y\,\mathsf{q}^n)~,
\end{equation}
with $\mathsf{q} \equiv e^{2\pi \I \tau} $ and $y\equiv e^{2\pi \I m}$. Let us define the normalized Jacobi theta function $\overline{\theta}_1(m|\tau)$ to be:
\begin{equation}
    \overline{\theta}_1(m|\tau) \,:=\, \frac{1}{2}\, \mathsf{q}^{-\frac{1}{8}}\,\theta_1(m|\tau)~.
\end{equation}

Using this form of the theta function along with the normalized eta function~\eqref{eta bar}, we can rewrite the full partition function of the 5D theory~\eqref{Gottsche for S} as follows~\cite[Equation (2.8)]{Beaujard:2020sgs}:
\begin{equation}\label{ZU(1) = h1S}
{Z_{\rm inst,\,adj}^{\,U(1)}[S]\,\,=:\,h_1^{S}(\tau, m_{\rm adj})\,=\,\frac{\sin(\pi\, m_{\rm adj})}{\overline{\eta}(\yadj\,\mathsf{q})^{\chi-3}\,\overline{\theta}_1(m_{\rm adj}|\tau+m_{\rm adj})}~.}
\end{equation}
This shows that the generating series of $U(1)$ Vafa--Witten invariants is a Jacobi form of weight $-\frac12 b_2(S)$ and index $-2$. More generally, it is expected to be a 
vector-valued mock Jacobi form of weight $-\frac12 b_2(S)$, index $-\frac16 K_S^2 (N^3-N)-2N$ and depth $N-1$~\cite{Alexandrov:2019rth}, where the mock modularity follows from singularities in the instanton moduli space due to reducible connections~\cite{Vafa:1994tf,Dabholkar:2020fde} (or, in the algebraic language, due to strictly semi-stable sheaves).

%%%%%%%%%%%%%%%%%%%%%%%%%%%%%%%%%%%%%%%%%
%%%%%%%%%%%%%%%%%%%%%%%%%%%%%%%%%%%%%%%%%
%%%%%%%%%%%%%%%%%%%%%%%%%%%%%%%%%%%%%%%%%
\section{Rank \texorpdfstring{$2$}{2} refined Vafa--Witten invariants for \texorpdfstring{$\mathbb{P}^2$}{P2}}\label{sec:VW2P2}
In this section, we apply the formalism developed in Subsection \ref{subsec:rank2 case} to compute the partition function on 
$\cN=1^*$ SYM with gauge group $U(2)$ on the complex projective plane $\mathbb{P}^2$, and compare
to existing results in the literature for the rank $2$ refined Vafa--Witten invariants.

\medskip

\begin{figure}[t]
        \begin{subfigure}[a]{0.3\textwidth}
    \centering
    \includegraphics[scale=0.6]{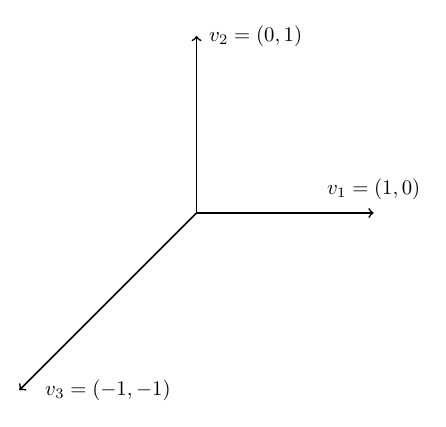}
    \end{subfigure}
    \begin{subfigure}[b]{0.5\textwidth}%
    \centering
\begin{equation*}
    \begin{array}{|c||c|c|c|}
    \hline
        ~\ell~ &~ (\epsilon_{1,\ell}\,,\, \epsilon_{2,\ell})~ &~(y_{1,\ell}\,\,y_{2,\ell})~ &~x^2_{\ell}~ \\
        \hline
        \hline
        ~ 1~&~(\epsilon_1\,,\, \epsilon_2)~&~(y_1\,,\,y_2)~&~x^2\,y_{1}^{\sfp_1}\,y_2^{\sfp_2}~\\
         \hline
         ~2~&~(\epsilon_2\,-\,\epsilon_1\,,\, -\,\epsilon_1)~&~(y_2\,y_1^{-1}\,,\,y_1^{-1})~&~x^2\,y_1^{-\sfp_2-\sfp_3}\,y_2^{\sfp_2}~\\
         \hline
         ~3~&~(-\epsilon_2\,,\, \epsilon_1\,-\,\epsilon_2)~&~(y_2^{-1}\,,\,y_1\,y_2^{-1})~&~x^2\,y_1^{\sfp_1}\,y_2^{-\sfp_1-\sfp_3}~\\
         \hline
    \end{array}
\end{equation*}
    \end{subfigure}%
  %%%%%%%%
    \caption{\textsc{Left:}  The toric fan for the projective surface $\mathbb{P}^2$.  \textsc{Right:} The toric weights and the shifted 5D Coulomb branch parameters for the $4$ patches of $\mathbb{P}^2$. The Coulomb branch parameters depend on the flux vector $\vec{\sfp} = (\sfp_1,\sfp_2,\sfp_3)$.}
    \label{tab:toricDataP2}
\end{figure}

Recall that the toric surface $\mathbb{P}^2$ is obtained by gluing together $\chi=3$ affine patches. The self-intersection numbers of the 3 toric divisors $\{D_\ell\}_{\ell=1,2,3}$ are $h_\ell=-1$.  In Table \ref{tab:toricDataP2}, we list the corresponding weights under the toric action and the equivariant 5D Coulomb branch parameters following the conventions in~\eqref{a ell alpha beta}.

%%

\subsection{Results on VW invariants from the literature}

The generating function for the Poincar\'e polynomials of the moduli space of Gieseker-stable rank $2$ sheaves on the complex projective plane
 was computed by Yoshioka in~\cite{Yoshioka1994} for the case $c_1 = 1$, and in~\cite{Yoshioka1995} for the case $c_1=0$, by counting methods over finite fields and using the Weil-conjecture proven by Deligne. Related results can be found in~\cite[Section 2]{Bringmann:2010sd} and~\cite[Appendix A.2]{Beaujard:2020sgs}.

\subsubsection{Generating function for \texorpdfstring{$c_1 = 1$}{c11}}\label{app:rank2VWP2c11}
For $c_1 = 1\mod 2$, the generating function for refined VW invariants is given by~\cite[Equation (2.11)]{Bringmann:2010sd}:
\begin{multline}\label{gen fun ref P2 c11}
    \sum_{n\geq 1}\, p(\mathcal{M}_{\mathbb{P}^2}^J(2,-1,n),s)\,t^n\,=\,\frac{\prod_{d\geq 1}\,\boldsymbol{Z}_{s^2}(\mathbb{P}^2, s^{4d-2}\,t^d)^2}{(s^2\,-\,1)\,\sum_{n\in\mathbb{Z}}s^{2n(2n-1)}t^{n^2}\,}\\\times\sum_{b\geq 0}\,\left(\frac{s^{2(b+1)(2b+1)}}{1\,-\,s^{8(b+1)}\,t^{2b+1}}\,-\,\frac{s^{2b(2b+5)}}{1\,-\,s^{8b}\,t^{2b+1}}\right)\,t^{(b+1)^2}~,
\end{multline}
where:
\begin{equation}
    \boldsymbol{Z}_s(\mathbb{P}^2,t)\,:=\,\frac{1}{(1\,-\,t)(1\,-\,s\,t)(1\,-\,s^2\,t)}~.
\end{equation}
Setting $\yadj\equiv s^2$ and $\sfq\equiv t$, the prefactor in \eqref{gen fun ref P2 c11} can be rewritten as: 
\begin{equation}
    \prod_{d\geq 1 } \,\boldsymbol{Z}_{s^2}(\mathbb{P}^2, s^{4d-2}t^d)^2 \,=\,h_1^{\mathbb{P}^2}(\tau+\madj, \madj)^2~,
\end{equation}
which is recognized as the square of the $U(1)$ partition function in equation~\eqref{ZU(1) = h1S}. 
With these identifications, we can rewrite the generating function~\eqref{gen fun ref P2 c11} as follows:
\begin{multline}
\label{Yoshiokac1=1}
    \sum_{n\geq 1}\, p(\mathcal{M}_{\mathbb{P}^2}^J(2,-1,n),\yadj)\,\sfq^n\,=\,\frac{\prod_{d\geq 1}\boldsymbol{Z}_{\yadj}(\mathbb{P}^2, \yadj^{2d-1}\,\sfq^d)^2}{(\yadj\,-\,1)\,\sum_{n\in\mathbb{Z}}\yadj^{n(2n-1)}\sfq^{n^2}\,}\\\times\sum_{b\geq 0}\,\left(\frac{\yadj^{(b+1)(2b+1)}}{1\,-\,\yadj^{4(b+1)}\,\sfq^{2b+1}}\,-\,\frac{\yadj^{b(2b+5)}}{1\,-\,\yadj^{4b}\,\sfq^{2b+1}}\right)\,\sfq^{(b+1)^2}~.
\end{multline}
Expanding the r.h.s. in powers of $\sfq$, we get:
\begin{multline}\label{Yoshioka expanded c1 = 1}
        \sum_{n\geq 1} p(\mathcal{M}_{\mathbb{P}^2}^J(2,-1,n),\yadj)\,\sfq^n\,=\,\sfq +\,\left(\frac{1}{\yadj^2}\,+\,\frac{2}{\yadj} \,+\,3\,+\,\cdots\right)\,(\yadj\,\sfq)^2\\
        \\+\,\yadj\,\left(\frac{1}{\yadj^4}\,+\,\frac{2}{\yadj^3}\,+\,\frac{6}{\yadj^2}\,+\,\frac{9}{\yadj}\,+\,12\,+\,\cdots\right)\,(\yadj\,\sfq)^3\,+\,\cdots~,
\end{multline}
which agrees with~\cite[Equation (A.38)]{Beaujard:2020sgs} upon identifying $\yadj=y^2$ and replacing $\sfq\to \sfq/y^4$. 

\medskip
\noindent
\textbf{The unrefined limit.} In the limit $\yadj\rightarrow 1$, the above generating function reduces to~\cite[page 19]{Kool2015}:
\begin{equation}
    \begin{split}
        \sum_{n\geq 1}\, e(\mathcal{M}_{\mathbb{P}^2}^J(2,-1,n))\,\sfq^n\,&=\,\frac{\overline{\eta}(\sfq)^{-6}}{2\,\sum_{m\in\mathbb{Z}} \sfq^{m^2}}\,\sum_{n\geq 0}\left(\frac{2\,-\,4\,n}{1\,-\,\sfq^{2n+1}}\,+\,\frac{8\,\sfq^{2n+1}}{(1\,-\,\sfq^{2n+1})^2}\right)\,\sfq^{(n+1)^2}\\
         &=\,\sfq\,+\,9\, \sfq^2\,+\,48\, \sfq^3\,+\,203\, \sfq^4\,+\,729\, \sfq^5\,+\,2346\, \sfq^6\,+\,\cdots~.
    \end{split}
\end{equation}
This reproduces the generating series of Euler numbers of the moduli space of strictly slope-stable rank $2$ torsion-free sheaves computed in~\cite[Corollary 4.1]{Kool2015}:
\be
\label{koolc11}
    \sum_{n\geq 1} e(\mathcal{M}_{\mathbb{P}^2}^{J}(2,1,n))\,\sfq^n \,=\,
       \,\frac{F_1}{\prod_{k\geq 1}(1\,-\,\sfq^k)^6}~,
\ee
where $F_1$ is the generating series of  Euler numbers of the moduli space of slope-stable locally-free sheaves, computed by the toric localization method reviewed in 
 Subsection~\ref{subsec:toric-localization}:
\be
\begin{split}
F_1= & \sum_{\substack{v_1,v_2,v_3>0 \\
v_1+v_2+v_2 \, {\rm odd}\\
v_1<v_2+v_3\\
v_2<v_3+v_1\\
v_3<v_1+v_2}} 
\sfq^{\frac14 + \frac12(v_1v_2+v_2v_3+v_3v_1)-\frac14(v_1^2+v_2^2+v_3^2)} \\
= & \sum_{m,n\geq1} \frac{\sfq^{mn}}{1-\sfq^{m+n-1}} 
=  \sfq+3\sfq^2+3\sfq^3+6\sfq^4+\dots ~.
\end{split}
\ee
It is worth noting that $F_1$ is proportional to the 
generating series of odd Hurwitz class numbers~\cite{zagier1975nombres,klyachko1991vector}, namely:
\be
\label{defF1P2}
F_1\,=\,3\, \sfq^{1/4}\, H_1(\tau)\ ,\qquad H_1(\tau)\,:=\,\sum_{n\geq 1} \,H(4n-1)\,\sfq^{n-\frac14}~,
\ee
where: 
\begin{equation}\label{Hurwitz numbers}
    \sum_{n\geq 0} \,H(n)\,\sfq^n\,=\,-\,\frac{1}{12}\,+\,\frac{1}{3}\,\sfq^3\,+\,\frac{1}{2}\,\sfq^4\,+\,\sfq^7\,+\,\sfq^8\,+\,\frac{4}{3}\,\sfq^9\,+\,\sfq^{11}\,+\,\frac{4}{3}\,\sfq^{12}\,+\,\sfq^{13}\,+\,2\,\sfq^{15}\,+\,\cdots~.
\end{equation}
The even Hurwitz class numbers $H(4n)$ will enter in the Euler numbers for $c_1=0$ in the next subsection, albeit in a more complicated fashion. 
It is also worth noting that the generating series of refined VW invariants~\eqref{Yoshiokac1=1} can be written as the ratio:
\be
  \sum_{n\geq 1}\, p(\mathcal{M}_{\mathbb{P}^2}^J(2,-1,n),\yadj)\,\sfq^n\,=\,\frac{3 \,H_1^{\rm ref}(\tau,y)}
  {\prod_{k\geq 1}\,(1\,-\,\sfq^k)^2\,(1\,-\,\sfq^k\, y^2)^2\,(1\,-\,\sfq^k/y^2)^2}~,
\ee
where $3H_1^{\rm ref}(\tau,y)$, denoted by $g_1(\tau,w)$ in~\cite{Bringmann:2010sd} (up to a factor of $\sfq^{1/4}/(y-1/y)$)
can be viewed as a refined version of the generating series of odd Hurwitz class numbers:
\be
\begin{split}
3 \,H_1^{\rm ref}(\tau,y)  \,& =\, \frac{1}{y\,-\,1/y}\, \frac{ \,y^3}
{\sum_{n\in \IZ} \sfq^{n^2}\, y^{2n}}
\sum_{n\in \IZ} \frac{\sfq^{n^2}\, y^{-2n}}{1\,-\,y^4 \sfq^{2n-1}}\,
\\
&=\, \sfq\,+\,\left(y^4\,+\,1\,+\,\dots\right) \,\sfq^2 \, +\,
   \left(y^8\,+\,y^4\,-\,y^2\,+\,1\,+\,\dots\right) \,\sfq^3\,+ \\ & +
\,   \left(y^{12} \,+\, y^8 \,-\, y^6\, + \,2 \,y^4 \,-\, y^2 \,+\, 2 \,+\, \dots \right)\, \sfq^4\,+\,\dots~,
\end{split}
\ee
where the dots denote monomials required by invariance under $y\mapsto 1/y$. In the limit $y\to 1$, this reduces to the
standard generating series of Hurwitz class numbers $F_1$ above. However, while the coefficients in the $\sfq$ expansion
are integral palindromic polynomials in $y$, they are not positive.

%%%%%%%%%%%%%%%%%%%%%%%%%%%%%%%%%%%%%%%%%
\subsubsection{Generating function for \texorpdfstring{$c_1 = 0$}{c10}}\label{subsubsec:P2c10literature}
Let us now turn to the case with even first Chern class $c_1=0\mod 2$. In this case, the refined rank $2$ VW invariants were computed by Yoshioka~\cite{Yoshioka1995}. More explicitly, we have~\cite[Equation (2.12)]{Bringmann:2010sd}:
\begin{multline}
\label{Yoshiokac1=0}
    \sum_{n=2}^\infty p(\mathcal{M}_{\mathbb{P}^2}^J(2,0,n),\yadj)\,\sfq^n\,=\,\frac{\prod_{d\geq 1}\,\boldsymbol{Z}_{\yadj}(\mathbb{P}^2,\yadj^{2d-1}\sfq^d)^2}{(1-\yadj)\sum_{n\in\mathbb{Z}}\yadj^{n(2n+1)}\sfq^{n(n+1)}}\\\times\left(\sum_{b\geq 0} -\left(\frac{\yadj^{(b+1)(2b+3)}}{1-\yadj^{4(b+1)}\sfq^{2b+1}}-\frac{\yadj^{b(2b+7)}}{1-\yadj^{4b}\sfq^{2b+1}}\right)\sfq^{b^2+3b+1}+\sum_{b\geq 0}\frac{\yadj^{(b+1)(2b+1)}-\yadj^{b(2b+1)}}{2\yadj}\sfq^{b(b+1)}\right)
\\+\frac{\prod_{d\geq 1}\boldsymbol{Z}_{\yadj^2}(\mathbb{P}^2,\yadj^{4d-4}\sfq^{2d})}{2\,\yadj\,(1\,+\,\yadj)}~.
\end{multline}
In the unrefined limit, this reduces to the generating series of integer DT invariants associated to the moduli space of Gieseker semi-stable rank $2$ sheaves on $\IP^2$: 
\begin{equation}\label{unrefined Yosh c10}
     \sum_{n=2}^\infty p(\mathcal{M}(2,0,n),1)\,\sfq^n \,=\,6\,\sfq^2\,+\,38\,\sfq^3\,+\,180\,\sfq^4\,+\,678\,\sfq^5\,+\,\cdots~.
\end{equation}
The relation to the Euler numbers of the moduli space of slope-stable sheaves, computable by toric localization, is, however, much more subtle than in the $c_1=1$ case, due to the existence of strictly semi-stable fixed points. The former is obtained by enumerating triplets satisfying strict triangular inequalities:
\begin{align}
\begin{split}
F_0&=\,  \sum_{\substack{v_1,v_2,v_3>0 \\
v_1+v_2+v_2 \, {\rm even}\\
v_1<v_2+v_3\\
v_2<v_3+v_1\\
v_3<v_1+v_2}} 
\sfq^{\frac12(v_1v_2+v_2v_3+v_3v_1)-\frac14(v_1^2+v_2^2+v_3^2)} \\
&= \, \sum_{m,n\geq1} \frac{\sfq^{mn+m+n}}{1-\sfq^{m+n}} 
\,=  \,0\,\sfq\,+\,0\,\sfq^2\,+\,\sfq^3\,+\,0\,\sfq^4\,+\,3\,\sfq^5 \,+\, 0 \,\sfq^6\, +\, 3 \,\sfq^7\,+\,\dots ~.
\end{split}
\end{align}
Keeping only these strictly-stable contributions and including point-like instantons would lead to:
\begin{equation}\label{Koolc10}
\frac{F_0}{\overline{\eta}(\sfq)^6}   \,=\, \sfq^3\,+\,6\,\sfq^4\,+\,30\,\sfq^5\,+\,116\,\sfq^6\,+\,399\,\sfq^7\,+\cdots~,
\end{equation}
which has no obvious relation to~\eqref{unrefined Yosh c10}. 
In order to reconcile the two, we need to
include strictly semi-stable fixed points,  corresponding to cases where one of the triangular inequalities is saturated:
\be
\begin{split}
F'_0 &=\, 3 \, \sum_{\substack{v_2,v_3>0 \\
v_1=v_2+v_3}}
\sfq^{\frac12(v_1v_2+v_2v_3+v_3v_1)-\frac14(v_1^2+v_2^2+v_3^2)} \\
&=\, 3\, \sfq \,+\, 6\, \sfq^2 \,+\, 6\, \sfq^3\,+\, 9\, \sfq^4\,+\,6\, \sfq^5 \,+\,\dots~.
\end{split}
\ee
Counting these strictly semi-stable contributions with a factor $1/2$, and including  a constant term\footnote{This constant term may be assigned to the contributions of triplets $(k,k,0),(k,0,k),(0,k,k)$ contributing to $c_2=0$, which evaluates
to $\frac{1}{2}\,\times\left( 1\,+ \,3\,\times \sum_{k=1}^{\infty}\,1\right)\,=\,\frac{1}{2}\,\times\,\left(1\,-\,\frac{3}{2}\right)\,=\, -\,\frac{1}{4}$
upon using Zeta function regularization. Unfortunately, this interpretation does not seem to extend to higher del Pezzo surfaces.\label{fookk0}}
$-1/4$, the sum:
\be
\label{F0F0p}
F_0 \,+\, \frac12\, F'_0 \, -\,\frac14 \,=\, -\,\frac14 \,+\, \sfq^3\,+\, 3\,\sfq^5\,+\,3\,\sfq^7\, 
+\, \frac12\,( 3\, \sfq \,+\, 6\, \sfq^2 \,+\, 6\, \sfq^3\,+\, 9\, \sfq^4\,+\,6\, \sfq^5 ) \, = \,3\, H_0(\tau)~, 
\ee
combines into the generating series of even Hurwitz class numbers. After including point-like instantons, one obtains:
\be
\label{r2ratDTgen}
\frac{3 H_0(\tau)}{{\overline \eta}(\tau)^6}  \,=\,
-\, \frac{1}{4}\,+\, 0\,\sfq \,+\, \frac{21}{4}\, \sfq^2\, +\, 38\, \sfq^3 \,+\, \frac{711}{4} \,\sfq^4 \,+\, 678\, \sfq^5 \,+\, \frac{4509}{2}\, \sfq^6\,+\, \dots~, 
\ee
whose terms with odd powers of $\sfq$ agree with the generating series ~\eqref{unrefined Yosh c10}.
We note that the constant term $-\frac14$ is crucial for obtaining the correct Euler numbers, even for odd $c_2$.
In order to find agreement for all powers of $\sfq$, one should use the multi-cover formula relating rational DT invariants (counted by~\eqref{r2ratDTgen}) and integer DT invariants, which amounts
to adding one quarter of the generating series of rank 1 VW invariants with $\tau$ replaced by $2\tau$: 
\be
\frac{3\, H_0(\tau)}{{\overline \eta}^6(\tau)}  \,+\,\frac{1}{4\,{\overline \eta}^3(2\tau)} \,=\, 
 0\,+\, 0\,\sfq \, + 6\, \sfq^2 \,+\, 38\, \sfq^3 \,+\, 180\, \sfq^4 \,+\, 678\, \sfq^5 \,+\, 2260\, \sfq^6\,+\, \dots ~.
\ee
 Unfortunately, we do not know how to derive the refined DT invariants from the toric localization method. In particular, while the last term in~\eqref{Yoshiokac1=0} manifestly comes from the refined multi-cover formula, it is unclear how to split the remainder into pieces associated with strictly-stable and strictly semi-stable fixed points. Nonetheless, by subtracting the last term
in~\eqref{Yoshiokac1=0}, we can anticipate that  at the refined level, the class number generating function 
$3H_0$ should be replaced by the generating series:
\be
\begin{split}
3 \,H_0^{\rm ref}(\tau,y) \,=\,&
-\tfrac{y}{2 (y^2+1)}\,+\,\tfrac{ y^4+y^2+1}{y^3+y}\,\sfq\,+\,\tfrac{
  y^{12}+y^{10}+y^8+y^4+y^2+1}{y^7+y^5} \,\sfq^2 \\&\,+\, \tfrac{
  y^{20}+y^{18}+y^{16}+y^{12}+y^8+y^4+y^2+1}{y^{11}+y^9}\,\sfq^3 \\
 & \, +\,\tfrac{
   y^{28}+y^{26}+y^{24}+y^{20}+y^{16}-y^{14}+y^{12}+y^8+y^4+y^2+1}{y^{15}
   +y^{13}} \,\sfq^4 \\
 & \, +\,\tfrac{y^{36}+y^{34}+y^{32}+y^{28}+y^{24}-y^{22}+2 y^{20}+2
   y^{16}-y^{14}+y^{12}+y^8+y^4+y^2+1}{y^{19}+y^{17}}\,\sfq^5\, +\,\dots~,
 \end{split}
\ee
whose unrefined limit $y\to 1$ reproduces~\eqref{F0F0p}. This function coincides with 
$g_0(\tau,w)$ in~\cite{Bringmann:2010sd}, up to a factor of $y-1/y$:
\be
3 \,H_0^{\rm ref}(\tau,y)  \,=\, \frac{1}{y\,-\,1/y}\, \left[ \frac12\, +\, \frac{\sfq^{-3/4} \,y^5}
{\sum_{n\in \IZ+\frac12} \sfq^{n^2}\, y^{2n}}
\sum_{n\in \IZ} \frac{\sfq^{n^2+n}\, y^{-2n}}{1\,-\,y^4 \sfq^{2n-1}}
\right]~.
\ee
After multiplying by the point-like instanton contributions 
$\prod_{k\geq 1}(1-\sfq^k)^2(1-\sfq^k y^2)^2(1-\sfq^k/y^2)^2$, the coefficients of odd powers of $\sfq$ turn out to be palindromic 
polynomials in $y$, as required by the compactness of the moduli space of 
coherent sheaves for odd $c_2$.

\subsection{Results from SUSY localization: generalities}\label{subsec:SUSYLocalP2}

Let us now consider the 5D $\mathcal{N}=1^*$ SYM with gauge group $U(2)$ on $\mathbb{P}^2\times\mathbb{S}^1_{\boldsymbol{\beta}}$. From our earlier discussion in Section \ref{sec_susyloc}, the final form of the partition function is given in terms of the following residue~\eqref{Z[S]}:
\begin{equation}\label{Zc1P2}
    \mathcal{Z}_{c_1}[\mathbb{P}^2]\,=\,\sum_{\vec{\sfp}\,\in\,\mathbb{Z}_{\geq 0}^{3}}\,Z_{\rm orbit}[\mathbb{P}^2]\mid_{\vec{\sfp}}~, 
\end{equation}
with the orbit partition function given by~\eqref{Zorbit}:
\begin{equation}\label{ZorbitP2}
    {Z}_{\rm orbit}[\mathbb{P}^2]\mid_{\vec\sfp}\,:=\,\frac{1}{4}\,\times\,\frac{1}{1\,-\,\yadj}\,\times\,
\sum_{\eta}\,\sgn(|\vec{\sfp}_\eta|)\,\operatorname*{Res}_{a=0} \, Z_{{\rm full}}^{\,U(2)}[\mathbb{P}^2]\mid_{\vec{\sfp}_\eta}~.
\end{equation}
Here, we are taking the K\"ahler form $\omega\in H^{1,1}(\mathbb{P}^2)$ to be of the form 
$\omega = w_1+w_2+w_3$ with  $w_\ell$ the Poincar\'e dual to the toric divisor $D_\ell$. With this in mind, it is straightforward to see that $\sgn(\vec{\beta}\cdot\vec{\sfp}) = \sgn(|\vec{\sfp}|)$\footnote{Here, we define the sign function such that $\sgn(0)=0$.} with $|\vec{\sfp}| := \sfp_1+\sfp_2+\sfp_3$. On the r.h.s. above, we are dividing by $(1-\yadj)$ following the discussion earlier in Subsection \ref{subsec:instantonC2}. We will revisit this point below.

The full partition function appearing in the residue integral on the r.h.s. of~\eqref{ZorbitP2} can be written explicitly as~\eqref{rank 2 full partition}:
\begin{equation}
    Z_{\rm full}^{\,U(2)}[\mathbb{P}^2]\mid_{\vec\sfp}\,=\,\sfq^{\Delta_{c_1,{\vec\sfp}}[\mathbb{P}^2]}\,{\rm Exp}[(\yadj\,-\,1)\,\chi_{\mathbb{P}^2}]\,Z_{\rm inst}^{\,U(2)}[\mathbb{P}^2]~.
\end{equation}
Here, $\Delta_{c_1,\vec{\sfp}}[\mathbb{P}^2]$ defined as~\eqref{DeltaS}:
\begin{equation}\label{DeltaP2}
    \Delta_{c_1, \vec{\sfp}}[\mathbb{P}^2]\,:=\,-\,\frac{1}{4}\,|\vec{\mathsf{p}}|^2 \,+\,\frac{c_1^2}{4}~,
\end{equation}
which is related to the discriminant of the gauge vector bundle on $\mathbb{P}^2$.

\medskip
\noindent
\textbf{The final contribution of each orbit.}
Let us start with $\vec{\sfp}\in\mathbb{Z}^3_{> 0}$\footnote{As we discuss in details in Appendix \protect\ref{app:ZfullRes}, for the cases where one of the flux components is zero, the full partition function does not develop any pole at $a=0$ and it trivializes under the residue calculation.} as the representative of the corresponding orbit under reflections. When computing the residue on the r.h.s. of~\eqref{Zorbit} it proves useful to note the following relation:
\begin{equation}\label{res=-resP2}
   \operatorname*{Res}_{a=0} \, Z_{{\rm full}}^{\,U(2)}[\mathbb{P}^2]\mid_{\vec{\sfp}_\eta} \,=\,-\,   \operatorname*{Res}_{a=0} \, Z_{{\rm full}}^{\,U(2)}[\mathbb{P}^2]\mid_{\vec{\sfp}}~,
\end{equation}
for reflections $\eta$ that flip the sign of a single component of $\vec{\sfp}$. The proof of this statement follows the same line of reasoning as in~\cite[Equation (K.43)]{Kim:2025fpz} where one uses the abstruse duality given in~\eqref{abstruce-duality-C2} and the fact that poles of the full partition function are simple. See Appendix \ref{app:ZfullRes} for more details on the latter point.  From this, we directly deduce the following identifications:
\begin{align}
    \begin{split}
        &\sgn(|\vec\sfp_{\eta'}|)\,   \operatorname*{Res}_{a=0} \, Z_{{\rm full}}^{\,U(2)}[\mathbb{P}^2]\mid_{\vec{\sfp}_{\eta'}} \,=\,-\,\sgn(|\vec\sfp|)\,   \operatorname*{Res}_{a=0} \, Z_{{\rm full}}^{\,U(2)}[\mathbb{P}^2]\mid_{\vec{\sfp}}~,\\
        &\sgn(|-\vec\sfp|)\,   \operatorname*{Res}_{a=0} \, Z_{{\rm full}}^{\,U(2)}[\mathbb{P}^2]\mid_{-\vec{\sfp}} \,=\,\sgn(|\vec\sfp|)\,   \operatorname*{Res}_{a=0} \, Z_{{\rm full}}^{\,U(2)}[\mathbb{P}^2]\mid_{\vec{\sfp}}~,\\
    \end{split}
\end{align}
for $\eta'\in\{{\rm diag}(\pm 1,\pm 1,\pm 1)\}-\{\pm{\rm diag}(1,1,1)\}$. 

With this observation in mind, let us now consider the following three possibilities for the flux vector $\vec{\sfp}$:
\begin{itemize}
    \item \underline{$\vec{\sfp}$ is strictly-stable:} this is the case where the three components of the flux vector satisfy the triangular inequalities~\eqref{st}: $\sfp_1+\sfp_2 >\sfp_3$ and its cyclic permutations. In this case, we get a total contribution of $-2+6  =4$ where $-2$ comes from $\pm\vec{\sfp}$ and the remaining $6$ come from all possible sign flips.
    \item \underline{$\vec{\sfp}$ is semi-stable:} this is the case where one of the triangular inequalities mentioned above is saturated. In this case, there will be two missing contributions from the $6$ mentioned above. Therefore, in total, we have $-2+4 = 2$.
    \item \underline {$\vec{\sfp}$ is unstable:} this is the case where at least one of the triangular inequalities above gets flipped. In this case, the reflections $\vec{\sfp}_{\eta}$ organize into pairs with canceling contributions. 
\end{itemize}
In Table \ref{tab:orbit-example-P2}, we give an explicit example for each one of these three cases.

\begin{table}[t!]
\centering
\begin{minipage}{0.3\textwidth}
\centering
\begin{tabular}{cc}
\hline
$\vec{\sfp}_\eta$& contribution  \\
\hline
$\pm(1,1,1)$ & $-\,1$\\
$\pm(1,1,-1)$ & $+\,1$\\
$\pm(1,-1,1)$ &$ +\,1$\\
$\pm(-1,1,1)$&$+\,1$\\
\hline
\end{tabular}
\end{minipage}
\hfill
\begin{minipage}{0.3\textwidth}
\centering
\begin{tabular}{cc}
\hline
$\vec{\sfp}_\eta$& contribution  \\
\hline
$\pm(2,1,1)$ &  $-\,\yadj^2$\\
$\pm(2,1,-1)$ &  $\yadj^2$\\
$\pm(2,-1,1)$ & $\yadj^2$\\
$\pm (-2,1,1)$ &$0$\\
\hline
\end{tabular}
\end{minipage}
\hfill
\begin{minipage}{0.3\textwidth}
\centering
\begin{tabular}{cc}
\hline
$\vec{\sfp}_\eta$& contribution \\
\hline
$\pm(1,1,3)$ & $-\,\yadj^{-3}$ \\
$\pm(1,1,-3)$ & $-\,\yadj^{-3}$ \\
$\pm(1,-1,3)$ & $\yadj^{-3}$\\
$\pm(-1,1,3)$ &$\yadj^{-3}$\\
\hline
\end{tabular}
\end{minipage}
\caption{In this table, we exhibit the final contributions of the different terms in the reflection orbit of a flux vector  $\vec{\sfp}$ to the term $\sfq$ as appearing on the r.h.s. of~\eqref{Zorbit}. Here, we are taking the flux vectors $\vec{\sfp}=(1,1,1),(2,1,1)$ and $(1,1,3)$ which, respectively, are examples of stable, semi-stable, and unstable vectors.}
\label{tab:orbit-example-P2}
\end{table}

\subsubsection{Extracting the residue}
Following the above observations, the final form of the partition function of the 5D theory on $\mathbb{P}^2$ with first Chern class $c_1$ can be written as follows~\eqref{Zc1P2}:
\begin{equation}\label{ZP2 residue}
    \mathcal{Z}_{c_1}[\mathbb{P}^2]\,:=\,-\,\sum_{\vec{\sfp}\in \IZ^3_{>0}}\,\Theta_{c_1}(\vec{\sfp})\,\mathop{\mathrm{Res}}\limits_{a=0}\, Z_{\rm full}^{\,U(2)}[\mathbb{P}^2]\mid_{\vec{\sfp}}~,
\end{equation}
where the sum above is over the strictly positive flux vectors $\vec{\sfp}$ such that $|\vec{\sfp}|=c_1\mod 2$, and they satisfy the slope-stability conditions~\eqref{st}:
\begin{equation}\label{stab cond for P2}
    \begin{split}
        \sfp_i\,+\,\sfp_j\,\leq\,\sfp_k~,
    \end{split}
\end{equation}
for the three possible cyclic permutations $(i,j,k)$ of $(1,2,3)$. As for the factor  $\Theta_{c_1}(\vec{\mathsf{p}})$, as explained above, it depends on the (semi)-stability property of the flux $\vec{\sfp}$~\cite[Equation (9.161)]{Kim:2025fpz},~\cite[Equation (3.70)]{Bershtein:2015xfa}:
\begin{equation}\label{Thetac1p}
    \Theta_{c_1}(\vec{\mathsf{p}})\,:=\,\begin{cases}
        1~, \qquad & \forall \ell~, ~~\mathsf{p}_\ell>0~, ~~\mathsf{p}_\ell\,+\,\mathsf{p}_{\ell+1}\,>\,\mathsf{p}_{\ell+2}~,\\
        &\text{and}~~|\vec{\mathsf{p}}| \,=\, c_1 \mod 2~,\\
        \frac{1}{2}~, \qquad &\forall \ell~, ~~\mathsf{p}_\ell>0~, ~~\mathsf{p}_\ell\,+\,\mathsf{p}_{\ell+1}\,\geq\,\mathsf{p}_{\ell+2}~,\\
        &\exists \ell~, ~~\mathsf{p}_\ell\,+\,\mathsf{p}_{\ell+1}\,=\,\mathsf{p}_{\ell+2}~,~~\text{and}~~|\vec{\mathsf{p}}| \,=\, c_1 \mod 2~,\\
        0~, &\text{otherwise}~.
    \end{cases}
\end{equation}

On the r.h.s. of~\eqref{ZP2 residue}, we have the full partition function with the explicit form given in~\eqref{rank 2 full partition}. As we argue in Appendix \ref{app:ZfullRes}, the 1-loop part contributes a zero of order $2$ at $a=0$, while the instanton part has a pole of degree $3$ at the same point. Therefore, the full partition function has a simple pole at $a=0$, and the evaluation of the residue in~\eqref{ZP2 residue} is straightforward. The end result is:\footnote{As we shall see below, for $c_1$ even the result below misses a
$\sfq$-independent term presumably coming from triplets mentioned in Footnote~\protect\ref{fookk0}. }
\begin{multline}\label{ZFullRes}
   \mathcal{Z}_{c_1}[\mathbb{P}^2]\,=\,\frac{h_1^{\mathbb{P}^2}(\tau+\madj,m_{\rm adj})^2}{1\,-\,\yadj}\,\sum_{\vec{\mathsf{p}}\in \IZ^3_{>0}} \Theta_{c_1}(\vec{\mathsf{p}}) \,\mathsf{q}^{c_2(\vec{\sfp})} \,\times P_{\rm W}[\mathbb{P}^2]\mid_{\vec{\mathsf{p}}}\,\times \,P_{\rm adj}[\mathbb{P}^2]\mid_{\vec{\mathsf{p}}}\,\times\\\times\,\prod_{\ell=1}^3\frac{\,T_{\mathsf{p}_\ell,\mathsf{p}_{\ell+1}}^{\rm adj}(y_{1,\ell},y_{2,\ell})\,H^{\rm K}(y_{1,\ell}^{\frac{\mathsf{p}_\ell}{2}}\,y_{2,\ell}^{-\frac{\mathsf{p}_{\ell+1}}{2}}, \yadj, y_{1,\ell},y_{2,\ell})}{(y_{1,\ell}\,y_{2,\ell})^{\mathsf{p}_\ell\mathsf{p}_{\ell+1}}\left(y_{1,\ell}^{-\mathsf{p}_\ell}\,y_{2,\ell}^{-\mathsf{p}_{\ell+1}}\,-\,y_{1,\ell}^{\mathsf{p}_\ell}\,y_{2,\ell}^{\mathsf{p}_{\ell+1}}\right)}~,
\end{multline}
where overall factor $h_1^{\mathbb{P}^2}$ is defined in~\eqref{ZU(1) = h1S}. The addition of the overall factor of $(1-\yadj)^{-1}$ will be explained shortly. Moreover, here we defined :
\begin{eqnarray}
    c_2(\vec{\sfp})\,:=\,\Delta_{c_1, \vec{\sfp}}[\mathbb{P}^2]\,+\,\sum_{\ell=1}^\chi\,\mathsf{p}_\ell\,\mathsf{p}_{\ell+1}~,
\end{eqnarray}
which coincides with the second Chern class of the gauge bundle.
Upon taking the residue of the 1-loop partition function~\eqref{Z pert P2}, we get $P_{\rm W}\times P_{\rm adj}$ which have the following explicit forms:
\begin{equation}
 \begin{split}
        &P_{\rm W}[\mathbb{P}^2]\,:=\,\prod_{\substack{(i,j)\in\mathbb{N}^2\setminus\{(\mathsf{p}_2,\mathsf{p}_1)\}\\i+j\leq|\vec{\mathsf{p}}|}}\mathsf{G}_{\sfp_1-j,\sfp_2-i}\,\times\,\prod_{\substack{(i,j)\in\mathbb{N}^2\setminus\{(\mathsf{p}_2-1,\mathsf{p}_1-1)\}\\i+j\leq |\vec{\mathsf{p}}|-3}}\,\mathsf{G}_{1+j-\sfp_1,1+i-\sfp_2}~,\\
        &P_{\rm adj}[\mathbb{P}^2]\,:=\,\prod_{\substack{(i,j)\in\mathbb{N}^2\\i+j\leq|\vec{\mathsf{p}}|}}\mathsf{G}^{\rm adj}_{\sfp_1-j,\sfp_2-i}\,\times\,\prod_{\substack{(i,j)\in\mathbb{N}^2\\i+j\leq |\vec{\mathsf{p}}|-3}}\,\mathsf{G}_{1+j-\sfp_1,1+i-\sfp_2}^{\rm adj}\, ,
 \end{split}   
\end{equation}
where $\mathsf{G}_{i,j}$ was introduced in~\eqref{G+-}, and $\mathsf{G}_{i,j}^{\rm adj}$ is defined in a similar fashion:
\begin{eqnarray}
    \mathsf{G}_{i,j}^{\rm adj} \,:=\,\left(1\,-\,y_1^{i}\,y_2^{j}\,\yadj\right)^{-1}~.
\end{eqnarray}
In the next two subsections, we will separately consider the two possible non-equivalent cases for $c_1 = 0,1\mod 2$. For each case, we will evaluate the formula~\eqref{ZFullRes} at low orders
in the instanton counting parameter $\sfq$ and extract the corresponding refined VW invariants matching known results in the literature.

\subsubsection{The unrefined limit}\label{subsubsec:unrefinedlimit}
Let us now take the unrefined limit of the 5D partition function calculation done above. That is, let us take $\yadj\rightarrow 1$. As pointed out earlier, this amounts to integrating in the massive adjoint hypermultiplet and enhancing the supersymmetry of the theory to be $\mathcal{N}=2$. To perform this limit on the r.h.s. of~\eqref{ZFullRes}, we first observe that:
\begin{equation}
\begin{split}\label{divide-1-yadj}
      &P_{\rm W}[\mathbb{P}^2]\mid_{\vec{\sfp}}\,\times\,P_{\rm adj}[\mathbb{P}^2]\mid_{\vec{\sfp}}\,=\,\frac{1}{(1\,-\,\yadj)^2}\,\times\,(\cdots)~,\\
       &\prod_{\ell=1}^3\frac{\,T_{\mathsf{p}_\ell,\mathsf{p}_{\ell+1}}^{\rm adj}(y_{1,\ell},y_{2,\ell})}{(y_{1,\ell}\,y_{2,\ell})^{\mathsf{p}_\ell\mathsf{p}_{\ell+1}}\left(y_{1,\ell}^{-\mathsf{p}_\ell}\,y_{2,\ell}^{-\mathsf{p}_{\ell+1}}\,-\,y_{1,\ell}^{\mathsf{p}_\ell}\,y_{2,\ell}^{\mathsf{p}_{\ell+1}}\right)}\,=\,(1\,-\,\yadj)^3\,\times\,(\cdots)~,
\end{split}
\end{equation}
where, in both cases, the $(\cdots)$ stand for the remaining parts which go to $1$ in the unrefined limit. The overall factor of $(1-\yadj)$ here explains the similar factor that we are dividing by in the r.h.s. of~\eqref{ZFullRes}. Note here that this is the same overall factor that we discussed at the end of Subsection~\ref{subsec:instantonC2}.

With these observations in mind, along with the one already made in~\eqref{unrefined U(1) S} for the limit of the $h_1^{\mathbb{P}^2}$ factor, we see that the unrefined limit of the rank $2$ partition function takes the form:
\begin{equation}\label{non refined ZP2}
    \lim_{\yadj\rightarrow 1} \,\mathcal{Z}_{c_1}[\mathbb{P}^2]\,=\,\overline{\eta}(\sfq)^{-6}\,\sum_{\vec{\sfp}\in \IZ^3_{>0}}\, \Theta_{c_1}(\vec{\sfp}) \,\sfq^{c_2(\vec{\sfp})}~.
\end{equation}
For $c_1=1$,  the sum over $\vec{\sfp}$ reproduces the numerator $F_1$ in~\eqref{koolc11},
counting the number of toric fixed points in the moduli space of slope-stable locally free sheaves,
giving a one-to-one map between poles in the SUSY localization method (classified into Weyl orbits of the flux vectors $\vec{\sfp}$), and fixed points
in the toric localization method. For $c_1=0$, the sum over $\vec{\sfp}$ reproduces the 
count $F_0+\frac12 F'_0$ of  toric fixed points in the moduli space of slope-semi-stable locally free sheaves, with strictly semi-stable sheaves counted with a factor $1/2$, but it misses the constant $-1/4$ which must be added by hand to produce the correct rational DT invariants.
In the next subsections, we examine the refined result in more details for the two cases. 

\subsection{Refined Vafa--Witten invariants for odd \texorpdfstring{$c_1$}{c1}}\label{subsec:ref P2 c11}
Let us start with the case where the first Chern class of the $U(2)$ bundle is odd -- that is $c_1 = 1\mod 2$. In this case, solving the  stability condition inequalities~\eqref{stab cond for P2}, we get the following list of fluxes:
\begin{equation}\label{p for c11 P2}
   \underbrace{(1,1,1)}_{\mathsf{q}^{1}}~,~\underbrace{(1,2,2)}_{\sfq^{2}}~,~
    \underbrace{(1,3,3)}_{\sfq^{3}}~,~
     \underbrace{(1,4,4)~,~(2,2,3)}_{\sfq^{4}}~,~\cdots~,
\end{equation}
where the `$\sfq^n$' under the brace indicates at which order in $\sfq$ this particular flux starts contributing in the full generating function.  Moreover, it is understood here that one needs to consider all possible permutations of the components of the flux. Since all flux vectors satisfy the strict inequalities, they all contribute with coefficient  $\Theta_1(\vec{\sfp})=1$. As an explicit example, \
we show the contribution of the flux $\vec{\sfp}=(1,1,1)$ to the first three terms in the $\sfq$ expansion of the full generating function:
\begin{multline}
    \sfq\,+\,(1\,+\,\yadj\,+\,2\,\yadj^2\,+\,\yadj^3\,+\,\yadj^4)\,\sfq^2\,\\+\,(1
\,+\,y_{\mathrm{adj}}
\,+\,4\,y_{\mathrm{adj}}^2
\,+\,5\,y_{\mathrm{adj}}^3
\,+\,6\,y_{\mathrm{adj}}^4
\,+\,4\,y_{\mathrm{adj}}^5
\,+\,4\,y_{\mathrm{adj}}^6
\,+\,y_{\mathrm{adj}}^7
\,+\,y_{\mathrm{adj}}^8)\,\sfq^3\,+\,\cdots~.
\end{multline}
The first term indicates that it originates from a single  isolated fixed point on the moduli space of stable locally free sheaves with $c_2=1$, which in this case has complex dimension~\eqref{expecteddim} zero (indeed, it corresponds to the exceptional bundle on $\mathbb{P}^2$ with $(N,c_1,c_2)=(2,1,1)$, which coincides with 
the tangent bundle tensored with $\mathcal{O}(-1)$). 
 Higher order terms in $\sfq$ arise by dressing this isolated bundle with point-like instantons.

\begin{table}[t!]
\renewcommand{\arraystretch}{2}
\centering
\begin{equation*}
\begin{array}{|c||c|c|c|}
\hline
{~n~}&{\rm Coeff}_{(\yadj^2\sfq)^n}\,\left[\yadj^{2}\,\mathcal{Z}_1[\mathbb{P}^2]\right] &{\rm Coeff}_{\sfq^n}\,\left[\mathcal{Z}_1[\mathbb{P}^2]\mid_{\yadj\rightarrow 1}\right] \\
\hline\hline
~1~ &1 &1\\
\hline
~2~ &\parbox{10cm}{\centering $\yadj^2\,+\,2\,\yadj\,+\,3\,+\,\cdots$}&9\\
\hline
~3~ &\parbox{10cm}{\centering $\yadj^4\,+\,2\,\yadj^3\,+\,6\,\yadj^2\,+\,9\,\yadj\,+\,12\,+\,\cdots$}&48\\
\hline
~4~&\parbox{10cm}{\centering $\yadj^6\,+\,2\,\yadj^5\,+\,6\,\yadj^4\,+\,13\,\yadj^3\,+\,24\,\yadj^2\,+\,35\,\yadj\,+\,41\,+\,\cdots$}&203\\
\hline
~5~&\parbox{10cm}{\centering $\yadj^8\,+\,2\,\yadj^7\,+\,6\,\yadj^6\,+\,13\,\yadj^5\,+\,29\,\yadj^4\,+\,51\,\yadj^3\,+\,85\,\yadj^2\,+\,113\,\yadj^1\,+\,129\,+\,\cdots$}&729\\
\hline
~6~&\parbox{10cm}{\centering $\yadj^{10}\,+\,2\,\yadj^9\,+\,6\,\yadj^8\,+\,13\,\yadj^7\,+\,29\,\yadj^6\,+\,57\,\yadj^5\,+\,106\,\yadj^4\,+\,175\,\yadj^3\,+\,262\,\yadj^2\,+\,337\,\yadj\,+\,370\,+\,\cdots$}&2346\\
\hline
\end{array}
\end{equation*}
\caption{The first six terms in the partition function of the 5D $\mathcal{N}=1^*$ $U(2)$ theory on $\mathbb{P}^2\times\mathbb{S}^1_{\boldsymbol{\beta}}$ upon performing the residue calculation. These give us the rank $2$ refined VW invariants for $\mathbb{P}^2$ as shown in the second column. In the third column, we take the unrefined limit where we end up with the corresponding number of the moduli space of slope-stable sheaves on $\mathbb{P}^2$ with $(c_1,c_2) = (1,n)$. The dots inside the brackets denote terms required by invariance under
$\yadj\mapsto  1/\yadj$.}
\label{tab:P2c11}
\end{table}

Working out the contributions coming from the other flux vectors listed in~\eqref{p for c11 P2}, we get the results exhibited in the second column in Table \ref{tab:P2c11} above, which match perfectly with Yoshioka's formula~\eqref{Yoshioka expanded c1 = 1}. Note that each individual contribution depends on the equivariant parameters $y_1,y_2$ explicitly; however, this dependence cancels in the final result.

\subsection{Refined Vafa--Witten invariants for even \texorpdfstring{$c_1$}{c1}}

\medskip
\noindent
\textbf{Contribution of the strictly-stable fluxes.} 
Let us now consider the second case for the first Chern class of the gauge vector bundle, namely, $c_1 = 0$. Solving the stability conditions~\eqref{stab cond for P2}, we find the following list of strictly slope-stable flux vectors:
\begin{equation}\label{p for P2 c10 I}
    \underbrace{(2,2,2)}_{\sfq^3}~,~\underbrace{(2,3,3)}_{\sfq^5}~,~\cdots~.
\end{equation}
In writing these fluxes, we are following the conventions mentioned below~\eqref{p for c11 P2}.

For the flux $\vec{\sfp}=(2,2,2)$, we find the following contribution to the first few terms in the $\sfq$ expansion of the generating function~\eqref{ZP2 residue}:
\begin{multline}\label{p222}
    -\,\mathop{\mathrm{Res}}\limits_{a=0}\, Z_{\rm full}^{\,U(2)}[\mathbb{P}^2]\mid_{(2,2,2)}\,=\,\yadj^4\,\sfq^3\,+\,(2\,y_{\mathrm{adj}}^6\,+\,2\,y_{\mathrm{adj}}^7\,+\,y_{\mathrm{adj}}^8\,+\,y_{\mathrm{adj}}^9)\,\sfq^4\\+\,(4\,y_{\mathrm{adj}}^7\,+\,5\,y_{\mathrm{adj}}^8\,+\,7\,y_{\mathrm{adj}}^9\,+\,5\,y_{\mathrm{adj}}^{10}\,+\,4\,y_{\mathrm{adj}}^{11}\,+\,y_{\mathrm{adj}}^{12}\,+\,y_{\mathrm{adj}}^{13})\,\sfq^5\,+\,\cdots~,
\end{multline}
which, in the unrefined limit, becomes $\sfq^3\,+\,6\,\sfq^4\,+\,27\,\sfq^5\,+\,\cdots$. Meanwhile, looking at the contribution from the flux $\vec{\sfp}=(2,3,3)$ and its two other possible permutations, we get:
\begin{equation}\label{p233perms}
    -\,\mathop{\mathrm{Res}}\limits_{a=0}\, Z_{\rm full}^{\,U(2)}[\mathbb{P}^2]\mid_{(2,3,3)+{\rm perm}}\,=\,(\yadj^8\,+\,\yadj^9\,+\,\yadj^{10})\,\sfq^5\,+\,\cdots~.
\end{equation}
In the unrefined limit, this gives $3\,\sfq^5+\cdots$, which, when summed up with the contribution we got from $\vec{\sfp}=(2,2,2)$ above, leads to $\sfq^3\,+\,6\,\sfq^4\,+\,30\,\sfq^5\,+\,\cdots$. This matches with~\eqref{Koolc10} counting the strictly-stable rank $2$ sheaves on $\mathbb{P}^2$ with even first Chern class $c_1$. The sum of~\eqref{p222} and~\eqref{p233perms} can be viewed as the refined counting of~\eqref{Koolc10}.

\medskip
\noindent
\textbf{Strictly semi-stable fluxes.}
Moreover, we find the following list of strictly semi-stable flux vectors:
\begin{equation}\label{p for P2 c10 II}
    \underbrace{(0,k,k)}_{\sfq^0}~,~\underbrace{(1,1,2)}_{\sfq^1}~,~\underbrace{(1,2,3)}_{\sfq^2}~,~\underbrace{(1,3,4)}_{\sfq^3}~,~\underbrace{(1,4,5)~,~(2,2,4)}_{\sfq^4}~,~\underbrace{(1,5,6)}_{\sfq^5}~,~\cdots~,
\end{equation}
along with all their possible permutations. In the first flux vector, $k$ can take any integer non-negative values, contributing to the same power of $\sfq^0$.

\begin{table}[t!]
\renewcommand{\arraystretch}{3}
\centering
\begin{equation*}
\begin{array}{|c||c|c|c|}
\hline
~{n}~&~2\,\times\,{\rm Coeff}_{\sfq^n}\,\left[\mathcal{Z}_{0}[\mathbb{P}^2]\right]~ &~{\rm Coeff}_{\sfq^n}\,\left[\mathcal{Z}_{0}[\mathbb{P}^2]\mid_{\yadj\rightarrow 1}\right]~&~{\rm Coeff}_{\sfq^n}\,\left[\mathcal{Z}'_{0}[\mathbb{P}^2]\mid_{\yadj\rightarrow 1}\right]~ \\
\hline\hline
~1~&~\frac{1}{\yadj}\,+\,1\,+\,{\yadj^2}~&~\frac{3}{2}~&~0~\\
\hline
~2~ &~\parbox{6cm}{\centering $\frac{1}{\yadj}\,+\,{3}\,+\,{5\,\yadj}\,+\,4\,\yadj^2\,+\,4\,\yadj^3\,+\,4\,\yadj^4\,+\,2\,\yadj^5\,+\,{\yadj^6}$}~&~12~&~\frac{21}{4}~\\
\hline
~3 ~&~\parbox{6cm}{\centering $\frac{1}{\yadj}\,+\,{3}\,+\,8\,\yadj\,+\,{15\,\yadj^2}\,+\,{19\,\yadj^3}\,+\,{21\,\yadj^4}\,+\,{19\,\yadj^5}\,+\,18\,\yadj^6\,+\,{11\,\yadj^7}\,+\,{7\,\yadj^8}\,+\,2\,\yadj^9\,+\,{\yadj^{10}}$}~&~\frac{125}{2}~&~38~\\
\hline
~4~&~\parbox{6cm}{\centering $\frac{1}{y_{\mathrm{adj}}}\,+\,{3}\,+\,8\,y_{\mathrm{adj}}\,+\,{19}\,y_{\mathrm{adj}}^{2}\,+\,{37}\,y_{\mathrm{adj}}^{3}\,+\,{55}\,y_{\mathrm{adj}}^{4}\,+\,72\,y_{\mathrm{adj}}^{5}\,+\,76\,y_{\mathrm{adj}}^{6}\,+\,74\,y_{\mathrm{adj}}^{7}\,+\,{65}\,y_{\mathrm{adj}}^{8}\,+\,48\,y_{\mathrm{adj}}^{9}\,+\,30\,y_{\mathrm{adj}}^{10}\,+\,{15}\,y_{\mathrm{adj}}^{11}\,+\,{7}\,y_{\mathrm{adj}}^{12}\,+\,2\,y_{\mathrm{adj}}^{13}\,+\,y_{\mathrm{adj}}^{14}$}~&~\frac{513}{2}~&~\frac{711}{4}~\\
\hline
~5~&~\parbox{6cm}{\centering $\frac{1}{y_{\mathrm{adj}}}\,+\,{3}\,
+\,8\,y_{\mathrm{adj}}
\,+\,{19}\,y_{\mathrm{adj}}^{2}
\,+\,42\,y_{\mathrm{adj}}^{3}
\,+\,80\,y_{\mathrm{adj}}^{4}
\,+\,130\,y_{\mathrm{adj}}^{5}
\,+\,{189}\,y_{\mathrm{adj}}^{6}
\,+\,230\,y_{\mathrm{adj}}^{7}
\,+\,{251}\,y_{\mathrm{adj}}^{8}
\,+\,242\,y_{\mathrm{adj}}^{9}
\,+\,{217}\,y_{\mathrm{adj}}^{10}
\,+\,{165}\,y_{\mathrm{adj}}^{11}
\,+\,114\,y_{\mathrm{adj}}^{12}
\,+\,64\,y_{\mathrm{adj}}^{13}
\,+\,{35}\,y_{\mathrm{adj}}^{14}
\,+\,{15}\,y_{\mathrm{adj}}^{15}
\,+\,{7}\,y_{\mathrm{adj}}^{16}
\,+\,2\,y_{\mathrm{adj}}^{17}
\,+\,y_{\mathrm{adj}}^{18}$}~&~\frac{1815}{2}~&~678~\\
\hline
\end{array}
\end{equation*}
\caption{The first five terms of the refined VW invariants for $\mathbb{P}^2$ with $c_1=0$ computed from the localization formula \eqref{ZFullRes}, in the
non-equivariant limit $y_1,y_2\to 0$. In the second column, we indicate their unrefined limit $\yadj\to 1$. The result differs from 
the rational VW invariants displayed in the third column by the constant term contribution $-1/(4\overline{\eta}^6)$, as discussed in \eqref{r2ratDTgen} and below \eqref{P2q1y10}.}
\label{tab:P2c10}
\end{table}

Working out the contribution of those semi-stable fluxes and adding it to those of the strictly-stable ones discussed above (with coefficient $1/2$), in Table \ref{tab:P2c10} 
we list the coefficients of  $\mathcal{Z}_0[\mathbb{P}^2]$ up to order $\sfq^5$, taking the
non-equivariant limit $y_1\to 0, y_2\to 0$ for simplicity\footnote{In view of the symmetry under exchanging $y_1\leftrightarrow y_2$, the order of the two limits is irrelevant; however, it is important to take 
$y_1\gg y_2$ or $y_2\gg y_1$, as otherwise the result would depend on the ratio $y_1/y_2$.}. Indeed, the final expressions depend in a complicated way
on the equivariant parameters $y_1$ and $y_2$,
as a consequence of the non-compactness of the moduli space of semi-stable sheaves. For example, for the order $\sfq^1$ term the dependence on the equivariant parameters at leading order in $y_1$ is given by:
\be
\label{P2q1y10}
\frac{y_{\text{adj}}^3+y_{\text{adj}}+1}{2 y_{\text{adj}}}
+\frac{\left(1-y_{\text{adj}}\right) \left(y_2+1\right)  \left(
y_{\text{adj}}-y_2-y_2 y_{\text{adj}}^3+y_2 y_{\text{adj}}^2+y_2^2
   y_{\text{adj}}\right)}{2 y_2^2 y_{\text{adj}}^2} y_1 +\cO(y_1^2)~.
\ee
In particular, it is \textit{not} invariant under $y_{\text{adj}}\to 1/y_{\text{adj}}$, as this symmetry would map $y_1=0$ to $y_1=\infty$.
Moreover, in the unrefined limit it reduces to $3/2$, which is the correct $\cO(\sfq)$ coefficient in the generating series of Hurwitz class
numbers \eqref{F0F0p}, but differs from the rational DT invariant at order $\sfq$, which is seen to vanish in~\eqref{r2ratDTgen}. As
explained there, the resolution is that there is an additional contribution $-1/(4\overline{\eta}^6)$, presumably originating from fluxes with $c_2(\vec\sfp)=0$,
which cancels the $\cO(\sfq)$ term from \eqref{P2q1y10} in the unrefined, non-equivariant limit. In the last column of Table~\ref{tab:P2c10}, we
display the coefficients of the full generating function $\mathcal{Z}'_0[\mathbb{P}^2]$ including this term, showing agreement with the rational
DT invariants~\eqref{r2ratDTgen} in the unrefined limit. Unfortunately, we do not know how to extend this mechanism at the refined, equivariant level.
In particular, if we postulate an additional contribution proportional to $(h_1^S)^2$ where $h_1^S$ is the $U(1)$ instanton partition function defined in \eqref{ZU(1) = h1S}
with a $\sfq$-independent coefficient $C(\yadj,y_1,y_2)$, we can determine
this coefficient by requiring that it cancels the $\cO(\sfq)$ term in the generating series, given that the refined rational VW invariant for $(c_1,c_2)=(1,0)$
is known from~\eqref{Yoshiokac1=0} to vanish.\footnote{This is an example where the moduli space of Gieseker-stable sheaves is empty, despite its expected dimension~\protect\eqref{expecteddim} being positive.} However, one can check that this prescription does not cancel the dependence on equivariant parameters 
at higher orders in $\sfq$, nor does it reduce to the correct refined VW invariants in the non-equivariant limit. We leave it as an important
open problem to understand the relation between the equivariant $\chi_{y^2}$-genus computed by supersymmetric localization and the refined Vafa--Witten invariants
in cases where the first Chern class is even.

%%%%%%%%%%%%%%%%%%%%%%%%%%%%%%%%%%%%%%%%%
%%%%%%%%%%%%%%%%%%%%%%%%%%%%%%%%%%%%%%%%%
%%%%%%%%%%%%%%%%%%%%%%%%%%%%%%%%%%%%%%%%%
\section{Rank \texorpdfstring{$2$}{2} refined Vafa--Witten invariants for \texorpdfstring{$\mathbb{F}_n$}{Fn}}\label{sec:VW2Fn}

\begin{figure}[t]
        \begin{subfigure}[a]{0.3\textwidth}
    \centering
    \includegraphics[scale=0.6]{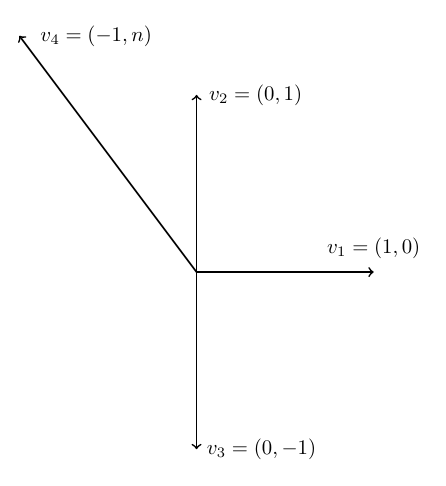}
    \end{subfigure}
    \begin{subfigure}[b]{0.5\textwidth}%
    \centering
\begin{equation*}
\begin{array}{|c||c|c|c|}
\hline
\ell& (\eps_{1,\ell}\,,\,\eps_{2,\ell})& (y_{1,\ell}\,,\, y_{2,\ell}) & x^2_\ell\\
\hline\hline
1 & (\eps_1\,,\, \eps_2) & (y_1\,,\, y_2) & x^2\,y_1^{\sfp_1}\,y_2^{\sfp_2}\\
\hline
2&(n\,\eps_1\,+\,\eps_2\,,\,-\eps_1) & (y_1^{n}\,y_2\,,\,y_1^{-1}) & x^2 \,y_1^{n\sfp_2-\sfp_3}\,y_2^{\sfp_2}\\
\hline
3 & (-\eps_1\, ,\,-\,n\,\eps_1\,-\,\eps_2) & (y_1^{-1}\,,\,y_1^{-n}\,y_2^{-1}) & x^2 \, y_1^{-\sfp_3-n\sfp_4}\,y_2^{-\sfp_4}\\
\hline
4&(-\,\eps_2\,,\,\eps_1)& (y_2^{-1}\,,\,y_1)& x^2\,y_1^{\sfp_1}\,y_2^{-\sfp_4}\\
\hline
\end{array}
\end{equation*}
    \end{subfigure}%
  %%%%%%%%
    \caption{\textsc{Left:}  The toric fan for the Hirzebruch surface $\mathbb{F}_n$.  \textsc{Right:} The toric weights and the shifted 5D Coulomb branch parameters for the $4$ patches of $\mathbb{F}_n$. The Coulomb branch parameters depend on the flux vector $\vec{\sfp} = (\sfp_1,\sfp_2,\sfp_3,\sfp_4)$.}
    \label{fig:Fn toric}
\end{figure}%%

In this section, we consider 5D $\mathcal{N}=1^*$ $U(2)$ SYM theory on the product $\mathbb{F}_n\times\mathbb{S}^1_{\boldsymbol{\beta}}$ 
of the $n$-th Hirzebruch surface times a circle. Recall that $\mathbb{F}_n$ is the projectivization of the rank $2$ bundle $\cO\oplus\cO(-n)$ over $\IP^1$, for any $n\geq 0$.
For $n=0$, it is just the product $\IP^1\times \IP^1$. For $n=1$, it is birational to the blow-up of $\IP^2$ in one point. 
The Euler number of $\mathbb{F}_n$ is $4$, and the corresponding toric fan is depicted on the l.h.s. of Figure~\ref{fig:Fn toric}. The toric weights of the local $\mathbb{C}^2$ patches are given in the Table on the r.h.s. of the same figure. The intersection numbers of the toric divisors $D_\ell$ defined in~\eqref{divisor-intersection} are given by:
\begin{equation}
    (h_1\,,\, h_2\,,\, h_3\,,\, h_4) \,=\,(0\,,\,n\,,\,0\,,\,-\,n)~.
\end{equation}
Since the
surfaces $\IF_n$ and $\IF_{n+2}$ are diffeomorphic as smooth manifolds, there is no loss of generality in restricting to $n\in\{0,1\}$.

In the first subsection, we will review results from the literature on the final form of the $\chi_{y^2}$-genus of the moduli space of slope (semi-)stable rank $2$ sheaves on $\mathbb{F}_n$~\cite{Beaujard:2020sgs}. These results were obtained via the wall-crossing formula. Later, we will match these results in the unrefined limit with those obtained via the toric localization method that we reviewed in Subsection~\ref{subsec:toric-localization}. In the last two subsections, we compute the partition function of the 5D theory
on $\IF_0\times \mathbb{S}^1_{\boldsymbol{\beta}}$ and $\IF_1\times \mathbb{S}^1_{\boldsymbol{\beta}} $ and match our results with the $\chi_{y^2}$-genus, where, as in the $\mathbb{P}^2$ case above, the refinement parameter $y^2$ is identified with the exponential of (minus) the 5D mass of the adjoint hypermultiplet $\yadj$, up to some power that we will specify as well.

\subsection{Results from wall-crossing}

Following~\cite[Appendix A.1]{Beaujard:2020sgs}, 
we denote by $C$ and $F$ the base and fiber of the projective bundle 
$\IF_n\to \IP^1$,\footnote{The base and fiber divisors $C$ and $F$ are related to $D_1,\cdots, D_4$ as follows: $D_1 = D_3 = F$, $D_2=C$ and $D_4 = C+nF$.} such that  $C^2=-n, F^2=0, C\cdot F=1$. We denote by $\mu=\beta C-\alpha F$ the first Chern class, and  parametrize the K\"ahler form by  $J_{m_1,m_2}=m_1(C+nF)+m_2 F$ with $m_1,m_2\geq 0$. For $m_1=0$, $J_{0,1}$ lies at the  boundary of the K\"ahler 
 cone, and the generating functions $H_{N,\mu}^{\IF_n,J_{0,1}}(\tau,w)$ of 
 the refined stacky invariants of the moduli spaces of rank $N$ slope-semi-stable sheaves
 are given by~\cite[Conjecture 4.1]{Manschot:2011ym} and proven in~\cite{Mozgovoy:2013zqx}:
\be 
\label{defHN} 
H_{N,\mu}^{\IF_n,J_{0,1}}(\tau,w)\,=\, 
H_{N}(\tau,w) \,:= \,\frac{\I \,(-1)^{N-1} \,\eta(\tau)^{2N-3}}
{\theta_1(\tau,2Nw)\, \prod_{k=1}^{N-1} \,\theta_1(\tau,2kw)^2}~,
\ee
for $\mu \cdot F=0 \mod N$, or 0 otherwise,
 independent of $n$. Note that  $H_N(\tau,w)$ 
is a Jacobi form of weight $-1$ and index $-\frac13(4N^3+2N)$. For rank $N=1$, this reduces to G\"ottsche's result~\eqref{ZU(1) = h1S}, with $y\equiv e^{2\pi \I w}$:
\be
H_1(\tau,w)\,=\,\frac{\I}{\eta(\tau)\, \theta_1(\tau,2w)} ~, \quad 
H_2(\tau,w)\,=\,\frac{-\,\I}{\eta(\tau)\, \theta_1(\tau,4w) \,\theta_1(\tau,2w)^2}~.
\ee

The stacky invariants for $m_1=\eps>0, m_2=1$ are obtained by wall-crossing. Restricting to $N=2$ for simplicity, we get:
\be 
H^{\IF_n,J_{\eps,1}}_{2,\beta C-\alpha F}(\tau,w) \,=\, 
\begin{cases} 
0~, \quad &\mbox{if} \quad \beta\,=\,1\mod 2~,
\\
 H_2  \,+\, \frac{y^2}{1\,-\,y^4}\, H_1^2~,
 \quad &\mbox{if} \quad (\alpha,\beta)=(1,0) \mod 2~,
 \\
 H_2 \,+\, \frac{1}{1\,-\,y^4}\,   H_1^2~,
 \quad &\mbox{if} \quad (\alpha,\beta)=(0,0) \mod 2~.
\end{cases}
\ee
Going further away from the boundary of the K\"ahler cone, we get, more generally,~\cite[Equation (5.14)]{Manschot:2016gsx}:
 \bea
\label{VWFm2wc}
&&H^{\IF_n,J_{m_1,m_2}}_{2,\beta C-\alpha F}   =
H^{\IF_n,J_{\epsilon,1}}_{2,\beta C-\alpha F}
\\
\quad && +\frac{1}{2} (h^{\IF_n}_{1})^2 
\sum_{\substack{a\in\IZ+\frac{1}{2}\alpha \\ b\in\IZ+\frac{1}{2}\beta}} 
\left[ \sgn(2b\, m_2-2 m_1 a + v) - \sgn(2b-2a\epsilon+v) \right]\, 
y^{-2(n-2)b-4a} \, \sfq^{n b^2 +2a b}~.\nn
\eea
for 
 $0<v\ll \epsilon \ll 1$. Here $\sfq\equiv e^{2\pi \I \tau}$. For $(m_1,m_2)$ generic, i.e., away from any wall, the generating function of rational Donaldson--Thomas (DT) invariants with respect to Gieseker stability is then obtained from the generating function of stacky invariants for slope stability
via:
\be
\label{htoH}
h^{\IF_n,J}_{N,\mu}(\tau,w) \,=\, \sum_{d=d_1+\dots +d_\ell}\,
 \frac{(-1)^{\ell-1}}{\ell}\,  \prod_{i=1}^\ell\, H^{\IF_n,J}_{d_i N_0, d_i \mu_0}(\tau,w)~,
\ee
where $d$ is the largest integer such that $(N_0,\mu_0):=(N/d,\mu/d)$ is primitive, and the sum runs over all ordered decompositions of $d$ with $d_i>0$.

\subsubsection{Specializing to the rank \texorpdfstring{$2$}{2} case}
For $N=2$, this reduces to:
\be
\label{htoH2}
h^{\IF_n,J}_{2,\mu}(\tau,w) \,=\, H^{\IF_n,J}_{2,\mu}(\tau,w) \,-\,\frac12 \,\delta_2(\mu)\, 
H_{1}(\tau,w)^2~,
\ee
where $\delta_2(\mu)=1$ if $\mu=0\mod 2$, or 0 otherwise. 
We further define:
\be
\label{deffrakh}
\mathfrak{h}^{\IF_n,J}_{N,\mu}(\tau,w) \,=\,
\sfq^{\frac{N\,-\,1}{2\,N}\, \mu^2 \,+\, \frac{N\,\chi(S)}{24}}\, (y\,-\,1/y) 
\left[\sum_{d|(N,\mu)}\, \frac{\mu(d)}{d} \,h^{\IF_n,J}_{N/d,\mu/d}(d\,\tau,d\,w) \right]~,
\ee
where $\mu(d)$ is the Möbius multiplicative function, such that  the coefficient 
of $\sfq^{c_2}$ in~\eqref{deffrakh} is the integer DT invariant of the moduli space of Gieseker semi-stable sheaves of rank $N$, first Chern class $c_1=\mu$, and second Chern class $c_2$:
\be
\mathfrak{h}^{\IF_n,J}_{N,\mu}(\tau,w) \,=\, \sum_{c_2\geq 0} \,
p(\cM_{\IF_n}^J(N,\mu,c_2),y)\, \sfq^{c_2}~.
\ee
A non-trivial check is that the coefficients $p(y)$ are monic Laurent polynomials with integer coefficients, invariant under $y\to 1/y$, with degree given by the expected dimension of the moduli space~\eqref{expecteddim}, unless they vanish identically.

\medskip
\noindent
\underline{\textbf{For $n=0$.}}
For $\IF_0$ in the canonical chamber $J \propto c_1(\IF_0)=2C+2F$, we find 
the first few terms in the series
\cite[Table 6]{Manschot:2016gsx}\cite[Equation (4.36)]{Haghighat:2012bm}:
\bea
\label{F0mod2} 
\mathfrak{h}^{\IF_0}_{2,(0,0)}&=& 
(y^5+2 y^3+3 y+\dots) \sfq^2 
   + (y^9+3 y^7+8 y^5+16 y^3+20y+\dots) \sfq^3 \nn  \\
&&   + ( y^{13}+3 y^{11}+10 y^9+24 y^7+51 y^5+83 y^3+104
   y +\dots) \sfq^4 +\dots~, \nn \\
\mathfrak{h}^{\IF_0}_{2,(0,1)}
&=& \underline{(y+1/y)}\, \sfq + \underline{(y^5+3 y^3+7y+\dots)}\sfq^2
\nn\\ && + (y^9+3 y^7 +10 y^5 +22 y^3 + 37 y+\dots) \sfq^3 
\nn\\ &&
+ ( y^{13} + 3 y^{11} + 10 y^9 + 26 y^7 + 60 y^5 + 110 y^3 + 161 y+\dots) \sfq^4 
+\dots~,
\nn\\
\mathfrak{h}^{\IF_0}_{2,(1,1)}
&=&
 \underline{(y^3+y+\dots)} \sfq^2 + (y^7+3y^5+7y^3+9y+\dots) \sfq^3 \nn\\
 && +
 ( y^{11} + 3 y^9 + 10 y^7 + 22 y^5 + 40 y^3 + 50 y + \dots ) \sfq^4 +\dots~.
\eea
Here we underlined the cases considered in~\cite{vanBree2023} (possibly for a different polarization), which are such that toric fixed loci are isolated points. 
In the unrefined limit, these become:
\bea
\label{F0mod2unrefined} 
\mathfrak{h}^{\IF_0}_{2,(0,0)}\mid_{y\rightarrow 1}&=& 
12 \sfq^2 
   + 96 \sfq^3 + 552 \sfq^4 + 2464 \sfq^5 + \dots~, \nn\\
   &=&
(-\tfrac14+\tfrac14)+ 0\sfq + (11+1)\sfq^2 + 96 \sfq^3 + \tfrac{1097+7}{2} \sfq^4+ 2496 \sfq^5+\dots  ~,
   \nn \\
\mathfrak{h}^{\IF_0}_{2,(0,1)}\mid_{y\rightarrow 1}
&=& 2\, \sfq + 22\sfq^2 + 146 \sfq^3+ 742 q^4 + \dots~,
\nn\\
\mathfrak{h}^{\IF_0}_{2,(1,1)}\mid_{y\rightarrow 1}
&=&
 4\sfq^2 + 40\sfq^3 + 252 \sfq^4 +\dots~,
\eea
where in the second line, we exhibit the contribution from the rational DT invariants.
At the refined level, this becomes:
\be
\begin{split}
\mathfrak{h}^{\IF_0}_{2,(0,0)} \,&=\, \frac{-\,y\,+\,y}{2\,(1\,+\,y^2)} \,+\, 0 \,\sfq\,\\&+\, \left( 
\frac{2 y^{12}+5 y^{10}+10 y^8+10 y^6+10 y^4+5 y^2+2}{2 \left(y^7+y^5\right)}
+ \frac{\left(y^4+1\right)^2}{2 y^3 \left(y^2+1\right)}\right) \sfq^2+ \dots~.
\end{split}
\ee

\medskip
\noindent
\underline{\textbf{For $n=1$.}}
For  $\IF_1$  in the canonical chamber $J \propto c_1(\IF_1)=2C+3F$  we find in the basis $\mu=(f_3,f_4)$,
related to the basis in~\cite{Beaujard:2020sgs} by $(d_H,d_C)=(f_3+f_4,-f_3)$:
\bea
\label{F1mod2}
\mathfrak{h}^{\IF_1}_{2,(0,0)} 
&=& 
(y^5+2 y^3+3   y+\dots) \sfq^2  + (y^9+3 y^7+8 y^5+16 y^3+21y+\dots) \sfq^3\nn\\
&&   + ( y^{13}+3 y^{11}+10 y^9+24 y^7+51 y^5+84 y^3+109  y+\dots ) \sfq^4 +\dots,
\nn\\
\mathfrak{h}^{\IF_1}_{2,(1,0)} &=& 
\underline{(y+1/y)} \sfq + (y^5+3y^3+7y+\dots) \sfq^2 
\nn\\&& + 
 \left(  y^9+  3 y^7 + 10 y^5 +22 y^3 + 36 y +\dots\right) \sfq^3 + \dots~,
\nn\\ 
 \mathfrak{h}^{\IF_1}_{2,(0,1)} &=& 
\underline{1} \sfq + \left( y^4 + 3 y^2 + 5 + \dots\right)\sfq^2 + \left(y^8+3y^6+10y^4+19y^2+27+\dots\right) \sfq^3 + \dots~,
\nn\\
\mathfrak{h}^{\IF_1}_{2,(1,1)} &=& 
\underline{(y^2+1+1/y^2)} \sfq^2 + 
(y^6 + 3 y^4 + 7 y^2 + 9+ \dots) \sfq^3  \nn \\
&& + (y^{10}+3
   y^8+10 y^6+22 y^4+40
   y^2+47+\dots) \sfq^4\nn \\
&&   + (y^{14}+3
   y^{12}+10 y^{10}+26 y^8+60 y^6+114 y^4+177
   y^2+205+\dots ) \sfq^5 +\dots~,
\nn\\&& 
\eea 
The underlined cases are those considered in~\cite{vanBree2023}, with $\Delta=Z,F,Z+F$, where toric fixed loci are isolated points. In the unrefined limit, these become:
\bea
\label{F1mod2unrefined} 
\mathfrak{h}^{\IF_1}_{2,(0,0)}\mid_{y\rightarrow 1}&=& 
12 \sfq^2 
   + 98 \sfq^3 + 564 \sfq^4 + 2558 \sfq^5 + \dots \nn\\
   &=&
(-\tfrac14+\tfrac14)+ 0\sfq + (11+1)\sfq^2 + 98 \sfq^3 + \tfrac{1121+7}{2} \sfq^4+
2558 \sfq^5 + \dots  ~,
   \nn \\
\mathfrak{h}^{\IF_1}_{2,(1,0)}\mid_{y\rightarrow 1}
&=& 2 \sfq + 22\sfq^2 + 144 \sfq^3+  \dots~,
\nn\\
\mathfrak{h}^{\IF_1}_{2,(0,1)}\mid_{y\rightarrow 1}
&=&
 \sfq+ 13 \sfq^2 + 93 \sfq^3 +497 \sfq^4 + \dots~,
    \nn \\
\mathfrak{h}^{\IF_1}_{2,(1,1)}\mid_{y\rightarrow 1}
&=& 3 \sfq^2 + 31 \sfq^3 + 199 \sfq^4 + 987 \sfq^5  + \dots~,
\eea
where in the second line, we exhibit the contribution from the rational DT invariants.
At the refined level, this is identical to the $\IF_0$ case, up to order $\sfq^2$:
\be
\begin{split}
\mathfrak{h}^{\IF_1}_{2,(0,0)} \,&=\, \frac{-\,y\,+\,y}{2\,(1\,+\,y^2)}\,\sfq^0 \,+\, 0\, \sfq\\
&\,+\, \left( 
\frac{2\, y^{12}\,+\,5 \,y^{10}\,+\,10\, y^8\,+\,10\, y^6\,+\,10\, y^4\,+\,5\, y^2\,+\,2}{2 \,\left(y^7\,+\,y^5\right)}
\,+\, \frac{\left(y^4\,+\,1\right)^2}{2\, y^3 \,\left(y^2\,+\,1\right)}\right)\, \sfq^2\,+\, \dots~.
\end{split}
\ee

\subsection{Results from toric localization}
For rank $2$ sheaves on the Hirzebruch surface $\IF_n$, the localization formula~\eqref{KoolThm35} leads to the following generating function
\cite[Corollary 4.2]{Kool2015}:\footnote{Here we quote the formula (10) appearing in the proof of the corollary. We are grateful to M. Kool for his help in spelling out the `similar terms' omitted in that formula.}
\be
\label{KoolProp10}
\overline{\eta}(\sfq)^8\, \sum_{c_2}\,e(\mathcal{M}_{\mathbb{F}_n}^{J}(2,(f_3,f_4),c_2))\, 
\sfq^{c_2} \,=\, \sfq^{\frac{1}{2}f_3 f_4 + \frac{n}{4}f_4^2}\, \sum_{i=1}^{11}\, F_{i}~,
\ee
with the following 11 contributions, each of which is subject to the condition:
\be
2\, \mid\, -\,f_3 \,+\, v_1 \,-\, n\,v_2 \,+\, v_3~, \quad 2 \,\mid\, -\,f_4\, +\, v_2\, +\,v_4
\ee
(denoting $\lambda=\frac{a}{b}, \lambda'=\lambda-n$): 
\begin{enumerate}
\item all $v_i>0$, all $\delta_i=0$, all $p_i$ distinct:
\begin{equation}\label{type1}
    F_1\,\equiv\,-\,\sum_{\substack{v_1,v_2,v_3,v_4 > 0 \\[2pt]
 v_1\, <\, \lambda'\, v_2\, + \, v_3 \,+\, \lambda\, v_4 \\
\lambda'\, v_2 \,<\,  v_1\, +\,  v_3 \,+\, \lambda \,v_4 \\
 v_3 \,<\,  v_1\, +\, \lambda'\, v_2\, + \,\lambda\, v_4 \\
\lambda\, v_4\, <\,  v_1 \,+\, \lambda'\, v_2\, +\,  v_3}}\,
\sfq^{ \frac{1}{2}(v_2+v_4)(v_1 + \frac{n}{2}v_2 + v_3 - \frac{n}{2}v_4)}~.
\end{equation}

\item all $v_i>0$, all $\delta_i=0$, $p_1=p_3$:
\begin{equation}\label{type2}
    F_2\,\equiv\,\sum_{\substack{v_1,v_2,v_3,v_4 \,>\, 0 \\[2pt]
 v_1 \,+\,  v_3 \,<\, \lambda' \,v_2 \,+\, \lambda \,v_4 \\
\lambda' \,v_2\, <\,  v_1 \,+\,  v_3 \,+\, \lambda\, v_4 \\
\lambda \,v_4\, <\,  v_1 \,+\, \lambda'\, v_2\, +\,  v_3}}\,
\sfq^{\frac{1}{2}(v_2+v_4)(v_1 + \frac{n}{2}v_2 + v_3 - \frac{n}{2}v_4)}~.
\end{equation}

\item all $v_i>0$, all $\delta_i=0$, $p_2=p_4$:
\begin{equation}\label{type3}
    F_3\,\equiv\,\sum_{\substack{v_1,v_2,v_3,v_4 \,>\, 0 \\[2pt]
\lambda'\, v_2\, + \,\lambda\, v_4 \,<\,  v_1 \,+\,  v_3 \\
 v_1\, <\, \lambda' \,v_2 \,+\,  v_3 \,+\, \lambda\, v_4 \\
 v_3 \,<\,  v_1 \,+\, \lambda'\, v_2\, +\, \lambda\, v_4}}\,
\sfq^{\frac{1}{2}(v_2+v_4)(v_1 + \frac{n}{2}v_2 + v_3 - \frac{n}{2}v_4)}~.
\end{equation}

\item all $v_i>0$, $\delta_i=(1,0,0,0)$ i.e. $p_1=p_2$:
\begin{equation}\label{type4}
    F_4\,\equiv\,\sum_{\substack{v_1,v_2,v_3,v_4 \,>\, 0 \\[2pt]
 v_1 \,+\, \lambda' \,v_2\, <\,  v_3 \,+\, \lambda\, v_4 \\
 v_3\, <\,  v_1 \,+\, \lambda' \,v_2 \,+\, \lambda \,v_4 \\
\lambda \,v_4 \,<\,  v_1 \,+\, \lambda'\, v_2 \,+\,  v_3}}\,
\sfq^{- \frac{1}{2}(v_2+v_4)(v_1 - \frac{n}{2}v_2 + v_3 + \frac{n}{2}v_4) + v_2 v_3 + v_3 v_4 + v_4 v_1}~.
\end{equation}

\item all $v_i>0$, $\delta_i=(0,1,0,0)$ i.e. $p_2=p_3$:
\begin{equation}\label{type5}
    F_5\,\equiv\,\sum_{\substack{v_1,v_2,v_3,v_4 \,>\, 0 \\[2pt]
\lambda'\, v_2 \,+\,  v_3 \,<\,  v_1 \,+\, \lambda\, v_4 \\
 v_1\, <\, \lambda' \,v_2 \,+\,  v_3\, +\, \lambda\, v_4 \\
\lambda\, v_4\, <\,  v_1\, +\, \lambda' \,v_2 \,+\,  v_3}}\,
\sfq^{- \frac{1}{2}(v_2+v_4)(v_1 - \frac{n}{2}v_2 + v_3 + \frac{n}{2}v_4) + v_1 v_2 + v_3 v_4 + v_4 v_1}~.
\end{equation}

\item all $v_i>0$, $\delta_i=(0,0,1,0)$ i.e., $p_3=p_4$:
\begin{equation}\label{type6}
    F_6\,\equiv\,\sum_{\substack{v_1,v_2,v_3,v_4 \,>\, 0 \\[2pt]
 v_3 \,+\, \lambda \,v_4 \,<\,  v_1 \,+\, \lambda'\, v_2 \\
 v_1\, <\, \lambda' \,v_2 \,+\,  v_3 \,+\, \lambda \,v_4 \\
\lambda' \,v_2 \,<\,  v_1 \,+\,  v_3\, +\, \lambda \,v_4}}\,
\sfq^{- \frac{1}{2}(v_2+v_4)(v_1 - \frac{n}{2}v_2 + v_3 + \frac{n}{2}v_4) + v_1 v_2 + v_2 v_3 + v_4 v_1}~.
\end{equation}

\item all $v_i>0$, $\delta_i=(0,0,0,1)$ i.e. $p_4=p_1$:
\begin{equation}\label{type7}
    F_7\,\equiv\,\sum_{\substack{v_1,v_2,v_3,v_4 \,>\, 0 \\[2pt]
\lambda\, v_4\, + \, v_1 \,  <\, \lambda'\, v_2\, +\,  v_3 \\
\lambda'\, v_2 \,<\,  v_1 \,+\,  v_3 \,+\, \lambda\, v_4 \\
 v_3 \,< \, v_1\, +\, \lambda' \,v_2 \,+\, \lambda \,v_4}}\,
\sfq^{- \frac{1}{2}(v_2+v_4)(v_1 - \frac{n}{2}v_2 + v_3 + \frac{n}{2}v_4) + v_2 v_3 + v_3 v_4 + v_1 v_2}~.
\end{equation}

\item $v_1=0$, other $v_i>0$:
\begin{equation}\label{type8}
    F_8\,\equiv\,\sum_{\substack{v_2,v_3,v_4 \,>\, 0 \\[2pt]
\lambda'\, v_2 \,<\,  v_3\, +\, \lambda\, v_4 \\
 v_3\, <\, \lambda' \,v_2 \,+\, \lambda \,v_4 \\
\lambda \,v_4\, < \,\lambda'\, v_2 \,+\,  v_3}}\,
\sfq^{\frac{1}{2}(v_2+v_4)(\frac{n}{2}v_2 + v_3 - \frac{n}{2}v_4)} ~.
\end{equation}

\item $v_2=0$, other $v_i>0$:
\begin{equation}\label{type9}
    F_9\,\equiv\,\sum_{\substack{v_1,v_3,v_4 \,>\, 0 \\[2pt]
 v_1 \,<\,  v_3 \,+\, \lambda\, v_4 \\
 v_3\, <\,  v_1 \,+\, \lambda \,v_4 \\
\lambda \,v_4\, < \, v_1 \,+\,  v_3}}\,
\sfq^{ \frac{1}{2}v_4(v_1 + v_3 - \frac{n}{2}v_4)}~.
\end{equation}

\item $v_3=0$, other $v_i>0$:
\begin{equation}\label{type10}
F_{10}\,\equiv\,\sum_{\substack{v_1,v_2,v_4 \,>\, 0 \\[2pt]
 v_1 \,<\, \lambda' \,v_2\, + \,\lambda\, v_4 \\
\lambda' \,v_2\, <\,  v_1 \,+\, \lambda\, v_4 \\
\lambda\, v_4\, <\,  v_1 \,+\, \lambda' \,v_2}}\,
\sfq^{\frac{1}{2}(v_2+v_4)(v_1 + \frac{n}{2}v_2 - \frac{n}{2}v_4)} ~.
\end{equation}

\item $v_4=0$, other $v_i>0$:
\begin{equation}\label{type11}
    F_{11}\,\equiv\,\sum_{\substack{v_1,v_2,v_3 \,>\, 0 \\[2pt]
 v_1 \,<\, \lambda'\, v_2\, +\,  v_3 \\
\lambda'\, v_2\, <\,  v_1 \,+\,  v_3 \\
 v_3 \,<\,  v_1 \,+\, \lambda'\, v_2}}\,
\sfq^{\frac{1}{2}v_2(v_1 + \frac{n}{2}v_2 + v_3)}~.
\end{equation}

\end{enumerate}
Note that the contributions of type (2,3) are mutually exclusive, and similarly those of type (4,6)  and (5,7). Moreover, whenever a type 1  contribution exists, there is a corresponding type 2  or type 3 contribution, such that the sum $F_1+F_2+F_3$ vanishes, as long as $\lambda$ is irrational\footnote{Indeed, it reduces to  $C_1$ in~\cite[Corollary 4.2]{Kool2015}. 
Similarly, $F_4+F_5+F_6+F_7$
reproduce  $C_2+C_3$, while $F_8+F_9+F_{10}+F_{11}$ add up to $C_4+C_5+C_6$ in the same corollary.}. Note also that quadruplets $(v_1,v_2,v_3,v_4)$ contributing to either $F_4+F_5+F_6+F_7$ necessarily contribute to $F_1$ as well.

% %%%%%%%%%%%%%%

\begin{table}[t]
\centering
\begin{tabular}{|c|c|p{11cm}|}
\hline
$c_1$ & $\sfq^n$ & Contributing $(v_1,v_2,v_3,v_4)$ \\
\hline
\multirow{6}{*}{$(0,0)$}
& $\sfq^2$
& $(1,1,1,1)_{2-1},\; (0,1,2,1)_8,\; (2,1,0,1)_{10}$\\
\cline{2-3}

&$ \sfq^3$
& $(1,1,3,1)_{4+7},\; (3,1,1,1)_{5+6}$\\
\cline{2-3}

& $\sfq^4$
& $(1,1,3,1)_{3-1},\; (1,2,1,2)_{2-1},\;
(2,1,2,1)_{3-1},\; (3,1,1,1)_{3-1},$\\

&
& $(2,1,4,1)_{4+7},\; (4,1,2,1)_{5+6},\;
(0,2,2,2)_8,\; (1,0,3,2)_9,\;
(1,2,3,0)_{11},$\\

&
& $(2,0,2,2)_9,\; (2,2,0,2)_{10},\;
(2,2,2,0)_{11},\; (3,0,1,2)_9,\;
(3,2,1,0)_{11}$\\
\hline

\multirow{7}{*}{$(1,0)$}
& $\sfq$
& $(0,1,1,1)_8,\; (1,1,0,1)_{10}$\\
\cline{2-3}

&$ \sfq^2$
& $(1,1,2,1)_{4+7},\; (2,1,1,1)_{5+6},\;
(0,2,1,2)_8,\; (1,2,0,2)_{10}$\\
\cline{2-3}

& $\sfq^3$
& $(1,1,2,1)_{3-1},\; (2,1,1,1)_{3-1},\;
(2,1,3,1)_{4+7},\; (3,1,2,1)_{5+6},$\\

&
& $(0,3,1,3)_8,\; (1,0,2,2)_9,\;
(1,2,2,0)_{11},\; (1,3,0,3)_{10},\;
(2,0,1,2)_9,$\\

&
& $(2,2,1,0)_{11}$\\
\cline{2-3}

& $\sfq^4$
& $(1,1,2,3)_5,\; (1,2,2,2)_{4+7},\;
(1,3,2,1)_6,\; (2,1,1,3)_4,\;
(2,2,1,2)_{5+6},$\\

&
& $(2,3,1,1)_7,\; (3,1,4,1)_{4+7},\;
(4,1,3,1)_{5+6},\;
(0,4,1,4)_8,\; (1,4,0,4)_{10}$\\
\hline

\multirow{7}{*}{$(0,1)$}
& $\sfq$
& $(1,0,1,1)_9,\; (1,1,1,0)_{11}$\\
\cline{2-3}

& $\sfq^2$
& $(1,1,1,2)_{4+5},\; (1,2,1,1)_{6+7},\;
(2,0,2,1)_9,\; (2,1,2,0)_{11}$\\
\cline{2-3}

& $\sfq^3$
& $(1,1,1,2)_{2-1},\; (1,2,1,1)_{2-1},\;
(1,2,1,3)_{4+5},\; (1,3,1,2)_{6+7},$\\

&
& $(0,1,2,2)_8,\; (0,2,2,1)_8,\;
(2,1,0,2)_{10},\; (2,2,0,1)_{10},\;
(3,0,3,1)_9,$\\

&
& $(3,1,3,0)_{11}$\\
\cline{2-3}

& $\sfq^4$
& $(1,1,3,2)_7,\; (1,2,3,1)_4,\;
(1,3,1,4)_{4+5},\; (1,4,1,3)_{6+7},\;
(2,1,2,2)_{4+5},$\\

&
& $(2,2,2,1)_{6+7},\; (3,1,1,2)_6,\;
(3,2,1,1)_5,\;
(4,0,4,1)_9,\; (4,1,4,0)_{11}$\\
\hline

\multirow{6}{*}{$(1,1)$}
& $\sfq^2$
& $(1,0,2,1)_9,\; (1,1,2,0)_{11},\;
(2,0,1,1)_9,\; (2,1,1,0)_{11}$\\
\cline{2-3}

&$ \sfq^3$
& $(1,1,2,2)_5,\; (1,2,2,1)_6,\;
(2,1,1,2)_4,\; (2,2,1,1)_7,\;
(2,0,3,1)_9,$\\

&
& $(2,1,3,0)_{11},\; (3,0,2,1)_9,\;
(3,1,2,0)_{11}$\\
\cline{2-3}

& $\sfq^4$
& $(1,1,2,2)_4,\; (1,2,2,1)_7,\;
(2,1,1,2)_5,\; (2,2,1,1)_6,\;
(1,2,2,3)_5,$\\

&
& $(1,3,2,2)_6,\; (2,2,1,3)_4,\;
(2,3,1,2)_7,\; (3,0,4,1)_9,\;
(3,1,4,0)_{11},$\\

&
& $(4,0,3,1)_9,\; (4,1,3,0)_{11}$\\
\hline
\end{tabular}
\caption{The contributing $4$-tuples $(v_1,v_2,v_3,v_4)$ contributing to the r.h.s. of~\eqref{KoolProp10} for $\mathbb{F}_0$ and for all four possible choices of the first Chern class $c_1$. The subscript indicates which type out of~\eqref{type1}--\eqref{type11} this $4$-tuple belongs to. For compactness, we write $(v_1,v_2,v_3,v_4)_{i\pm j}$ to indicate that this particular $4$-tuples contributes to types $i$ and $j$. Note that the subscript $i-j$ occurs only for the cases with $j = 1$.}
\label{tab:F0_tuples}
\end{table}

%%%%%%%%%%%
\subsubsection{For \texorpdfstring{$\mathbb{F}_0$}{F0}}
For $\IF_0$, in Table \ref{tab:F0_tuples} we  list all the $4$-tuples $(v_1,v_2,v_3,v_4)$ contributing to the r.h.s. of~\eqref{KoolProp10}. Counting these $4$-tuples, we find:
\begin{eqnarray}
   \sum_{c_2}\,e(\mathcal{N}_{\mathbb{F}_0}^{J}(2,(0,0),c_2))\,\sfq^{c_2}\,&=& 2\, \sfq^2 \,+ \,4\,\sfq^3\,+\,12\,\sfq^4\,+\,\cdots~,\\
   \sum_{c_2}\,e(\mathcal{N}_{\mathbb{F}_0}^{J}(2,(0,1),c_2))\,\sfq^{c_2}\,&=&\,2\,\sfq\,+\,6\,\sfq^2\,+\,10\,\sfq^3\,+\,14\,\sfq^4\,+\,\cdots~,
\\
   \sum_{c_2}\,e(\mathcal{N}_{\mathbb{F}_0}^{J}(2,(1,1),c_2))\,\sfq^{c_2}&=& 4\, \sfq^2 + \,8\,\sfq^3\,+\,12 \sfq^4\,+\, \cdots~.
\end{eqnarray}
Recall here that, as introduced above~\eqref{NHST}, $\mathcal{N}_{\mathbb{F}_0}^J$ denotes the moduli space of slope-stable locally-free sheaves on $\mathbb{F}_0$. Dividing by the contribution $\overline{\eta}(\sfq)^8$ of point-like instantons, we arrive at the Euler numbers of the moduli space of slope-stable torsion-free sheaves: 
\begin{eqnarray}
\label{Kool F0 c100}
   \sum_{c_2}\,e(\mathcal{M}_{\mathbb{F}_0}^{J}(2,(0,0),c_2))\,\sfq^{c_2}\,&=& 2\, \sfq^2 \,+ \,20\,\sfq^3\,+\,132\,\sfq^4\,+\,664\,\sfq^5\,+\,\cdots~,\\
   \label{Kool F0 c101}
   \sum_{c_2}\,e(\mathcal{M}_{\mathbb{F}_0}^{J}(2,(0,1),c_2))\,\sfq^{c_2}\,&=&\,2\,\sfq\,+\,22\,\sfq^2\,+\,146\,\sfq^3\,+\,742\,\sfq^4\,+\,\cdots~,
\\
\label{Kool F0 c111}
   \sum_{c_2}\,e(\mathcal{M}_{\mathbb{F}_0}^{J}(2,(1,1),c_2))\,\sfq^{c_2}&=& 4\, \sfq^2 + \,40\,\sfq^3\,+\,252 \sfq^4\,+1232\, \sfq^5 + \cdots~.
\end{eqnarray}

\medskip
\noindent
\textbf{A compact formula for the case $c_1 = (0,1)$.}
For completeness, we note that for $c_1= (0,1)$, the number of fixed points in the moduli space of strictly slope-stable sheaves was worked out for arbitrary $c_2$ in~\cite[Page 22]{Kool2015}:
\begin{multline}
\overline{\eta}(\sfq)^8\,\sum_{c_2}\,e(\mathcal{M}_{\mathbb{F}_0}^{J}(2,(0,1),c_2))\,\sfq^{c_2}\,=\,
    \sum_{m=1}^\infty \sum_{n=1}^{2m}\frac{4\,\sfq^{(2m+3)m-2m n+1}(\sfq^{(2m+1)n}\,-\,\sfq^{n^2})}{(1\,-\,\sfq^n)(\sfq^{2m+1}\,-\,\sfq^{n})}\,\\+\,\sum_{m=1}^\infty \frac{2\,(2\,m\,-\,1)\,\sfq^{(2m-1)m}}{1\,-\,\sfq^{2m-1}}\,+\,\sum_{m=1}^\infty \frac{4\,m\,\sfq^{(2m+1)m}}{1\,-\,\sfq^{2m}}\,\\+\,\sum_{m=1}^\infty\sum_{n=1}^\infty\sum_{p=1}^{2m-1}\frac{4\,\sfq^{(2m+1)m-2mp+1}(\sfq^{p(n+p-1)}\,-\,\sfq^{2m(n+p-1)})}{\sfq\,-\,\sfq^{n+p}}~.
\end{multline}
which reproduces and extends the formula~\eqref{Kool F0 c101} above.

\paragraph{Strictly semistable fixed points for $c_1=(0,0)$.}

Recall that the contribution from slope-stable fixed points is given by~\eqref{Kool F0 c100},
which we denote by $F_0$, in analogy with the $\IP^2$ case discussed in Subsection~\ref{subsubsec:P2c10literature}. Assuming that semi-stable fixed points 
are counted by an analogue of~\eqref{KoolProp10} where the strict inequalities $<$ 
are replaced by $\leq$, and extracting the contributions saturating these inequalities, we 
find that strictly semi-stable fixed points contribute:
\be
\label{F0sss}
 F'_0=  4 \,\sfq \,+\, 8 \,\sfq^2 \,+\, 8\, \sfq^3  \,+\, \dots  ~,
\ee
corresponding to the quadruplets:
\be
\label{semistabF0tuples}
\begin{split}
\sfq \,\left[(1,1,1,1)_{4+5+6+7} \right]  
&+\,\sfq^2 \, \left[(1,2,1,2)_{4+5+6+7} \,+\,(2,1,2,1)_{4+5+6+7}  \right]\\
 &  + \,\sfq^3\,
   \left[(1,3,1,3)_{4+5+6+7} \,+\,(3,1,3,1)_{4+5+6+7} \right] \\
   &\,+\, \dots~.
\end{split} 
\ee
Note that these contributing 4-tuples already appeared in Table~\ref{tab:F0_tuples}, albeit with a different power of $\sfq$, 
indexing strictly slope-stable sheaves of types 1--3 \eqref{type1}--\eqref{type3}.
Counting the strictly semi-stable fixed points with coefficient 1/2,  adding in the same 
constant term\footnote{A naive attempt to compute this constant term by Zeta function regularization of the sum over semi-stable fluxes
of the form 
$(n,0,n,0)$ and $(0,n,0,n)$ with $n\geq 0$, which all have vanishing $c_2(\vec\sfp)=0$, 
unfortunately produces $\frac12(1+2\sum_{n\geq 1} 1)=0$ 
rather than $-1/4$.
} $-1/4$ as for $\IP_2$ and  dividing by the pointlike instanton contribution $\overline{\eta}(\sfq)^8$, 
we reproduce the rational invariants listed in the second line of~\eqref{F0mod2unrefined}: 
\be
\frac{F_0\,+\,\frac12\,F'_0\,-\,\frac14}{\overline{\eta}(\sfq)^8} \,=\, 
-\,\frac14 \,+\, 0\, \sfq \,+\, 11 \,\sfq^2 \,+\, 96\, \sfq^3 \,+\, \frac{1097}{2}\, \sfq^4 \,+\, 2496\, \sfq^5\, +\, \dots~.
\ee

\begin{table}[t!]
\centering
\renewcommand{\arraystretch}{1.2}
\begin{equation*}
\begin{array}{|c||c|p{9cm}|}
\hline
c_1 & \sfq^n & \text{Contributing} $(v_1,v_2,v_3,v_4)$\\
\hline\hline
\multirow{3}{*}{$(0,0)$}
& \sfq^2
& $(1,1,2,1)_4,\; (2,1,1,1)_5$\\
\cline{2-3}

& \sfq^3
& $(1,1,2,1)_{3-1},\; (2,1,1,1)_{3-1},\;
  (2,1,3,1)_4,\; (3,1,2,1)_5,$\\

&
& $(1,0,3,2)_9,\; (1,2,1,0)_{11},\;
  (2,0,2,2)_9,\; (3,0,1,2)_9$\\
\hline\hline

\multirow{5}{*}{$(1,0)$}
& \sfq
& $(1,1,1,1)_{4+5}$\\
\cline{2-3}

& \sfq^2
& $(1,1,1,1)_{3-1},\; (1,0,2,2)_9,\;
  (1,2,2,2)_5,\; (2,0,1,2)_9,$\\

&
& $(2,1,2,1)_{4+5},\; (2,2,1,2)_4$\\
\cline{2-3}

& \sfq^3
& $(1,1,3,3)_5,\; (1,3,1,1)_{4+5},\;
  (1,3,3,3)_5,\; (3,1,1,3)_4,$\\

&
& $(3,1,3,1)_{4+5},\; (3,3,1,3)_4$\\
\hline\hline
\multirow{4}{*}{$(0,1)$}
& \sfq
& $(1,0,1,1)_9$\\
\cline{2-3}

& \sfq^2
& $(1,1,2,2)_5,\; (1,2,1,1)_{4+5},\;
  (2,0,2,1)_9,\; (2,1,1,2)_4$\\
\cline{2-3}

& \sfq^3
& $(1,1,2,2)_4,\; (1,2,3,3)_5,\;
  (1,3,2,2)_5,\; (2,1,1,2)_5,$\\

&
& $(2,2,2,1)_{4+5},\; (2,3,1,2)_4,\;
  (3,0,3,1)_9,\; (3,2,1,3)_4$\\
\hline\hline

\multirow{3}{*}{$(1,1)$}
& \sfq^2
& $(1,0,2,1)_9,\; (1,1,1,0)_{11},\;
  (2,0,1,1)_9$\\
\cline{2-3}

& \sfq^3
& $(3,1,1,2)_4,\; (1,1,3,2)_5,\;
  (0,2,1,1)_8,\; (1,2,0,1)_{10},$\\

&
& $(2,1,2,0)_{11},\; (3,0,2,1)_9,\;
  (2,0,3,1)_9$\\
\hline
\end{array}
\end{equation*}
\caption{The contributing $4$-tuples $(v_1,v_2,v_3,v_4)$ contributing to the r.h.s. of~\eqref{KoolProp10} for $\mathbb{F}_1$ and for all four possible choices of the first Chern class $c_1$. The convention for the subscript is explained in the caption of Table \ref{tab:F0_tuples}.}
\label{tab:F1_tuples}
\end{table}

\subsubsection{For \texorpdfstring{$\mathbb{F}_1$}{F1}}\label{subsec:F1Literature}

For $\IF_1$, we list the contributing $(v_1,v_2,v_3,v_4)$ in Table \ref{tab:F1_tuples} for the four possibilities of the first Chern class $c_1$. Counting the number of these $4$-tuples, we find:
\begin{eqnarray}
   \sum_{c_2}\,e(\mathcal{N}_{\mathbb{F}_1}^{J}(2,(0,0),c_2))\,\sfq^{c_2}\,&=& 2\, \sfq^2 \,+ \,6\,\sfq^3\,+\,\cdots~,\\
   \sum_{c_2}\,e(\mathcal{N}_{\mathbb{F}_1}^{J}(2,(1,0),c_2))\,\sfq^{c_2} &=&\,2\,\sfq\,+\,6\,\sfq^2\,+\,8\,\sfq^3\,+\,\cdots~,
\\
   \sum_{c_2}\,e(\mathcal{N}_{\mathbb{F}_1}^{J}(2,(0,1),c_2))\,\sfq^{c_2} &=& \sfq\,+\,5\,\sfq^2\,+\,9\,\sfq^3\,+\,\cdots~,
\\
   \sum_{c_2}\,e(\mathcal{N}_{\mathbb{F}_1}^{J}(2,(1,1),c_2))\,\sfq^{c_2}&=& 3\,\sfq^2\,+ \,7\, \sfq^3\,+\,\cdots~.
\end{eqnarray}
Dividing by the contribution $\overline{\eta}(\sfq)^8$ of point-like instantons, we arrive at:
\begin{eqnarray}
\label{Kool F1 c100}
   \sum_{c_2}\,e(\mathcal{M}_{\mathbb{F}_1}^{J}(2,(0,0),c_2))\,\sfq^{c_2}\,&=& 2\, \sfq^2 + \,22\,\sfq^3\,+\,144 \sfq^4\,+\,726\,\sfq^5\,+\,\cdots~,\\
   \label{Kool F1 c110}
      \sum_{c_2}\,e(\mathcal{M}_{\mathbb{F}_1}^{J}(2,(1,0),c_2))\,\sfq^{c_2} &=&\,2\,\sfq\,+\,22\,\sfq^2\,+\,144\,\sfq^3\,+\,730\,\sfq^4\,+\,\cdots~,
\\
\label{Kool F1 c101}
   \sum_{c_2}\,e(\mathcal{M}_{\mathbb{F}_1}^{J}(2,(0,1),c_2))\,\sfq^{c_2} &=& \sfq\,+\,13\,\sfq^2\,+\,93\,\sfq^3\,+\,496\,\sfq^4\,+\,\cdots~,
\\
\label{Kool F1 c111}
   \sum_{c_2}\,e(\mathcal{M}_{\mathbb{F}_1}^{J}(2,(1,1),c_2))\,\sfq^{c_2}&=& 3\,\sfq^2\,+\, 31\, \sfq^3\, +\, 199\,\sfq^4\,+\,987 \sfq^5\,+\,\cdots~.
\end{eqnarray}

\paragraph{Strictly semistable sheaves for $c_1=(0,0)$.}

Recall that the contribution from strictly-stable fixed points is given by \eqref{Kool F1 c100},
which we denote by $F_0$, in analogy with the $\IP^2$ case discussed in Subsection \ref{subsubsec:P2c10literature}. Assuming again that semi-stable fixed points 
are counted by an analogue of  \eqref{KoolProp10} where the strict inequalities $<$ 
are replaced by $\leq$, and extracting the contributions saturating these inequalities, we 
find that strictly semi-stable fixed points contribute:
\be
\label{F0sssF1}
 F'_0\,=\,  4 \,\sfq \,+\, 8 \,\sfq^2 \,+\, 8\, \sfq^3  \,+\, \dots  ~,
\ee
corresponding to the quadruplets:
\begin{align}\label{semistabF1c100tuples}
\begin{split}
\sfq\, &[(0,1,1,1)_8\,+\,(1,1,0,1)_{10}\,+\,(1,1,2,1)_5\,+\,(2,1,1,1)_4]\\
+\,\sfq^2 \,
&[(1,1,2,1)_7\,+\,(1,2,1,2)_{4+5} \,+\,(1,2,3,2)_5\,+\,(2,1,1,1)_6\,+\,(2,1,3,1
   )_5\,\\&+\,(3,1,2,1)_4\,+\,(3,2,1,2)_4]\\
   +\,\sfq^3\, &
   [(1,3,2,3)_5\,+\,(1,3,4,3)_5\,+\,(2,1,3,1)_7\,+\,(2,3,1,3)_4\,+\,(3,1,2,1)_6 \\
 &  +\,(3,1,4,1)_5\,+\,(4,1,3,1)_4\,+\,(4,3,1,3)_4] \,+\, \dots~.
 \end{split}
\end{align}
Counting the strictly semi-stable sheaves with coefficient 1/2,  adding in the same 
constant term\footnote{A naive attempt to compute this constant term by Zeta function regularization of the sum over semi-stable fluxes
of the form 
$(n,0,n,0)$  with $n\geq 0$, which all have vanishing $c_2(\vec\sfp)=0$, 
unfortunately produces $\frac12\times(1+\sum_{n\geq 1} 1)=1/4$ rather than $-1/4$.
} $-1/4$ as for $\IP_2$ and  dividing by $\overline{\eta}(\sfq)^8$, we reproduce
the rational invariants listed in the second line of \eqref{F1mod2unrefined}:
\be
\frac{F_0\,+\,\frac12 \,F'_0\,-\,\frac14}{\overline{\eta}(\sfq)^8} \,=\, 
-\,\frac14 \,+\, 0\, \sfq \,+\, 11\, \sfq^2 \,+\, 98\, \sfq^3 \,+\, \frac{1121}{2} \,\sfq^4 \,+\, 2558 \,\sfq^5 \,+\, \dots~.
\ee

\subsection{Supersymmetric localization for \texorpdfstring{$\mathbb{F}_n$}{Fn}: generalities}

Let us now compute the partition function of the 5D theory placed on $\mathbb{F}_n\times\mathbb{S}^1_{\boldsymbol{\beta}}$. For a fixed choice of the first Chern class $c_1$ of the rank $2$ gauge bundle and polarization  $J = (\boldsymbol{a},\boldsymbol{b})\equiv \boldsymbol{a} \,w_1\,+\,\boldsymbol{b}\,w_2$,\footnote{Here, we assume that $\boldsymbol{b}>0$ and $\boldsymbol{a}'\equiv \boldsymbol{a}-n\boldsymbol{b}>0$. These are the constraints for the divisor to be ample. See, for instance, Example 6.1.16 of~\protect\cite{cox2024toric}.} the partition function of the 5D $\mathcal{N}=1^*$ $U(2)$ SYM theory is given as~\eqref{Z[S]}:
\begin{eqnarray}\label{Zc1Fn=SumOrbits}
    \mathcal{Z}_{c_1}[\mathbb{F}_n]\,=\,\sum_{\vec\sfp\in\mathbb{Z}^4_{\geq0}}\,Z_{\rm orbit}[\mathbb{F}_n]\mid_{\vec\sfp}~.
\end{eqnarray}
The sum here is over positive flux vectors $\vec{\sfp}$ satisfying the stability conditions~\eqref{st}:
\begin{equation}\label{stab cond Fn}
    \begin{split}
        \boldsymbol{b}\,\sfp_1\,&\leq\,\boldsymbol{a}'\,\sfp_2\,+\,\boldsymbol{b}\,\sfp_3\,+\,\boldsymbol{a}\,\sfp_4~,\\
        \boldsymbol{a}'\,\sfp_2\,&\leq\,\boldsymbol{b}\,\sfp_1\,+\,\boldsymbol{b}\,\sfp_3\,+\,\boldsymbol{a}\,\sfp_4~,\\
        \boldsymbol{b}\,\sfp_3\,&\leq\,\boldsymbol{b}\,\sfp_1\,+\,\boldsymbol{a}'\,\sfp_2\,+\,\boldsymbol{a}\,\sfp_4~,\\
        \boldsymbol{a}\,\sfp_4\,&\leq\,\boldsymbol{b}\,\sfp_1\,+\,\boldsymbol{a}'\,\sfp_2\,+\,\boldsymbol{b}\,\sfp_3~,
    \end{split}
\end{equation}
along with the conditions \eqref{f3f4 defn} fixing the first Chern class.
In \eqref{Zc1Fn=SumOrbits}, the summand is given by the contribution of the orbit of $\vec{\sfp}$ under reflections~\eqref{Zorbit}: 
 \begin{equation}\label{ZorbitFn}
    \mathcal{Z}_{\rm orbit}[\mathbb{F}_n]\mid_{\vec\sfp}\,:\,=\frac{1}{4}\,\times\,\frac{1}{1-\,\yadj}\,\times\,\,\sum_{\eta}\,\sgn(\vec{\beta}\,\cdot\,\vec{\sfp}_\eta)\,\operatorname*{Res}_{a=0} \, Z_{{\rm full},\,c_1}^{\,U(2)}[\mathbb{F}_n]\mid_{\vec{\sfp}_\eta}~,
\end{equation}
where $\vec{\beta}=(\boldsymbol{b},\boldsymbol{a}-n\boldsymbol{b},\boldsymbol{b},\boldsymbol{a})$ for a particular choice of polarization $J = (\boldsymbol{a},\boldsymbol{b})$.\footnote{To stay away from walls of marginal stability, we shall choose the ratio $a/b$ to be irrational, or at least rational with a large enough denominator. in practice we choose $(\boldsymbol{a},\boldsymbol{b}) = (1.1,1)$.} As introduced around \eqref{Zorbit}, $\vec{\sfp}_\eta$ is the vector obtained from $\vec{\sfp}$ after applying the reflection $ \eta\in\left\{{\rm diag}(\pm 1, \pm1, \pm 1, \pm 1)\right\}$ with $\eta_\ell=+1$ whenever $\sfp_\ell=0$.

The residue at $a=0$ on the r.h.s. of~\eqref{ZorbitFn} is performed on the full partition function including the 1-loop and instanton contributions, evaluated at the reflected flux vector $\vec{\sfp}_{\eta}$. In the remainder of this subsection, we analyze its structure and the corresponding degree of the pole at $a=0$ depending on the explicit form of the flux vector $\vec{\sfp}$.

\subsubsection{Analytic structure of the full partition function}\label{subsubsec:analyzFullFuncFn}
The integrand $Z_{{\rm full},\,c_1}^{\,U(2)}[\mathbb{F}_n]$  decomposes as follows~\eqref{rank 2 full partition}:
    \begin{equation}\label{ZfullFn}
        Z_{{\rm full},\,c_1}^{\,U(2)}[\mathbb{F}_n]\mid_{\vec{\sfp}}\,=\, \sfq^{\Delta_{c_1,\vec\sfp}[\mathbb{F}_n]}\,Z_{{\rm 1-loop}}^{\,U(2)}[\mathbb{F}_n]\mid_{\vec\sfp}\,Z_{{\rm inst,\,adj}}^{\,U(2)}[\mathbb{F}_n]\mid_{\vec\sfp}\,~.
    \end{equation}
The leading power of $\sfq$, coming from the classical contribution~\eqref{DeltaS}, 
 can be written as:
     \begin{eqnarray}\label{DeltaFn}
        \Delta_{c_1,\vec{\sfp}}[\mathbb{F}_n] \,:=\,-\,\frac{1}{4}(2\,r(\vec{\sfp})\,r'(\vec{\sfp})\,+\,n\,r'(\vec{\sfp})^2)\,+\,\frac{1}{4}(2\,f_3\,f_4\,+\,n\,f_4^2)~,
    \end{eqnarray}
 where, following~\protect\cite[Equations (6.7)--(6.8)]{Bonelli:2020xps}\footnote{Note that there is a missing sign typo in~\cite[Equation (6.8)]{Bonelli:2020xps} in their definition of $\Delta_k$.}, we define:
  \begin{equation}\label{rr' defn}
        r(\vec{\sfp})\,:=\,\sfp_1\,-\,n\,\sfp_2\,+\sfp_3~, \qquad r'(\vec{\sfp})\,:=\,\sfp_2\,+\,\sfp_4~.
 \end{equation}
 These integers are constrained to satisfy:
    \begin{equation}\label{f3f4 defn}
    f_3=\,r(\vec{\sfp})\mod 2~, \quad \text{and}\quad f_4=\,r'(\vec{\sfp})\mod 2~,
\end{equation}
where $(f_3,f_4)$ parametrizes the first Chern class 
$c_1= f_3\,w_3\,+\,f_4\,w_4$ of the rank two sheaf. 
We further define:  
\begin{equation}\label{c2pFn}
    c_2(\vec{\sfp})\,:=\,\Delta_{c_1, {\vec\sfp}}[\mathbb{F}_n]\,+\,\sum_{\ell=1}^4\sfp_\ell\,\sfp_{\ell+1}~,
\end{equation}
which coincides with the second Chern class \eqref{c2Etoric} of a rank $2$ sheaf 
in the toric localization computation, upon identifying $v_\ell=\sfp_\ell$ and setting
$\delta_{\ell,\ell+1}=0$. As we shall see momentarily,  the flux vector $\vec\sfp$ actually
contributes to an infinite set of powers of $\sfq$, which may differ from
\eqref{c2pFn} by terms of the form $n-\sum_\ell \delta_{\ell,\ell+1} \sfp_\ell \sfp_{\ell+1}$
where $n$ is a positive integer.

\begin{table}[t!]
\renewcommand{\arraystretch}{2.5}
\centering
\begin{equation*}
\begin{array}{|c||c|c|c|}
\hline
{~\vec{\sfp}~}&~c_2({\vec\sfp})~&~{{\rm Exp}\left[-\,\chi_{\mathbb{F}_0}(x,y_1,y_2)\,\right]\mid_{\vec{\sfp}}}~ \\
\hline\hline
~(1,0,1,1)~&~1~&~\parbox{12cm}{\centering$(1\,-\,x^2)\, (1\,-\,\frac{x^2}{y_1})\, (1\,-\,x^2\, y_1)\, (1\,-\,\frac{x^2}{y_2})\,
   (1\,-\,\frac{x^2}{y_1 \,y_2})\, (1\,-\,\frac{x^2\, y_1}{y_2})$}~\\
\hline
~(1,1,1,0)~ &~1~&~\parbox{12cm}{\centering$(1\,-\,x^2) \,(1\,-\,\frac{x^2}{y_1})\, (1\,-\,x^2\, y_1)\, (1\,-\,x^2\, y_2)\,
   (1\,-\,\frac{x^2 \,y_2}{y_1})\, (1\,-\,x^2\, y_1\, y_2)$}\\
\hline
~(1,1,1,2) ~&~3~&~\parbox{12cm}{\centering$(1\,-\,\frac{1}{x^2}) \,(1\,-\,x^2)\, (1\,-\,\frac{x^2}{y_1})\, (1\,-\,x^2\, y_1)\,
   (1\,-\,\frac{x^2}{y_2^2})\, (1\,-\,\frac{x^2}{y_1\, y_2^2})\, (1\,-\,\frac{x^2\,
   y_1}{y_2^2})\, (1\,-\,\frac{x^2}{y_2})\, (1\,-\,\frac{x^2}{y_1 \,y_2}) \,(1\,-\,\frac{x^2
   y_1}{y_2})\, (1\,-\,\frac{y_2}{x^2})\, (1-x^2\, y_2)\, (1\,-\,\frac{x^2 \,y_2}{y_1})
   \,(1\,-\,x^2\, y_1\, y_2)$}\\
\hline
~(1,2,1,1)~&~3~ &~\parbox{12cm}{\centering$(1\,-\,\frac{1}{x^2})\, (1\,-\,x^2)\, (1\,-\,\frac{x^2}{y_1})\, (1\,-\,x^2\, y_1)\,
   (1\,-\,\frac{x^2}{y_2^2})\, (1\,-\,\frac{x^2}{y_1\, y_2^2})\, (1\,-\,\frac{x^2
   \,y_1}{y_2^2})\, (1\,-\,\frac{x^2}{y_2})\, (1\,-\,\frac{x^2}{y_1 \,y_2}) (1\,-\,\frac{x^2\,
   y_1}{y_2})\, (1\,-\,\frac{y_2}{x^2})\, (1\,-\,x^2\, y_2)\, (1\,-\,\frac{x^2 \,y_2}{y_1})\,
   (1\,-\,x^2\, y_1\, y_2)$}\\
\hline
\end{array}
\end{equation*}
\caption{We exhibit the form of the W-boson 1-loop contribution of the 5D $\mathcal{N}=1^*$ $U(2)$ SYM theory for four 
arbitrarily-chosen stable fluxes on $\mathbb{F}_0$ with first Chern class $c_1 = (0,1)$ and second Chern class $c_2({\vec\sfp})$ -- see~\eqref{c2pFn} for definition -- shown in the second column. Observe that the 1-loop contribution has a double zero at $a=0$ when all entries of ${\vec\sfp}$ are non-vanishing. In contrast, it has a simple zero at $a=0$ if one of the entries of $\vec{\sfp}$ vanishes.}
\label{tab:Z1-loopZeros}
\end{table}

\medskip
\noindent
\textbf{Comments on the analytic structure of $Z_{{\rm 1-loop}}^{\,U(2)}[\mathbb{F}_n]$.} Let us start by looking more closely at the 1-loop part coming from the W-bosons and the massive adjoint hypermultiplet. More explicitly, this takes the form~\eqref{rank 2 full partition}:
\begin{equation}\label{1-loop-Fn}
    Z_{{\rm 1-loop}}^{\,U(2)}[\mathbb{F}_n]\,=\,{\rm Exp}\left[(\yadj-1)\,\chi_{\mathbb{F}_n}(x,y_1,y_2)\,\right]~,
\end{equation}
where the character $\chi_{\mathbb{F}_n}$ is defined in~\eqref{chiSN} for $N=2$. As explained in Appendix \ref{app:ZfullRes}, the zeros of the full partition function at $Z_{{\rm full},\,c_1}^{\,U(2)}[\mathbb{F}_n]$ at $a=0$ come from this 1-loop component. Moreover, following the observations made in~\cite{Bonelli:2020xps}, for a fixed flux vector $\vec{\sfp}$, the order of the zero depends on the number of vanishing entries in $\vec{\sfp}$. In Table \ref{tab:Z1-loopZeros}, we exhibit the final form of the W-boson part of~\eqref{1-loop-Fn} for the theory on $\mathbb{F}_0$ evaluated on several flux vectors. For instance, we see that:
\begin{equation}
    \operatorname*{ord}_{a=0}\,Z_{{\rm 1-loop}}^{\,U(2)}[\mathbb{F}_n]\mid_{\vec\sfp}\,=\,\begin{cases}
        2~,\qquad&\text{if }\,\zeta_{\vec\sfp}\,=\,0~,\\
        1~,\qquad&\text{if }\,\zeta_{\vec\sfp}\,=\,1~,\\
    \end{cases}
\end{equation}
where, $\zeta_{\vec\sfp}$ is the number of zero components in ${\vec\sfp}$. As we shall see below, terms with more than one vanishing entry
$\zeta_{\vec\sfp} >1$ do not contribute for Hirzebruch surfaces (although they would start to contribute for toric surfaces with $\chi\geq 5$).

Recall that, from~\eqref{ZorbitFn}, the computation of the partition function of the orbit represented by the vector $\vec{\sfp}\in\mathbb{Z}^4_{\geq 0}$ involves looking at the contribution of every single reflection ${\vec\sfp}_\eta$. Evaluating the 1-loop partition function~\eqref{1-loop-Fn} for each one of these reflections, we find that the analytic structure at $a=0$ can change dramatically; for instance, it can develop a pole rather than a zero. In Table \ref{tab:Z1-loopReflections}, we exhibit the W-boson 1-loop contribution in~\eqref{1-loop-Fn} for all possible reflections of the flux vector $\vec{\sfp} = (1,0,1,1)$. 

\begin{table}[t!]
\renewcommand{\arraystretch}{2}
\centering
\begin{equation*}
\begin{array}{|c||c|c|c|}
\hline
{~\vec{\sfp}_\eta~}&~\Delta_{c_1,\vec{\sfp}_\eta}[\mathbb{F}_0]~&~{{\rm Exp}\left[-\,\chi_{\mathbb{F}_0}(x,y_1,y_2)\,\right]\mid_{\vec{\sfp}_\eta}}~&~\text{orbit contr.}~ \\
\hline\hline
~(-1,0,-1,-1)~&~-1~&~\parbox{7cm}{\centering$(1\,-\,\frac{1}{x^2})\, (1\,-\,\frac{1}{x^2\, y_1})\, (1\,-\,\frac{y_1}{x^2})\,
   (1\,-\,\frac{1}{x^2\, y_2})\, (1\,-\,\frac{1}{x^2\, y_1\, y_2})\, (1\,-\,\frac{y_1}{x^2\, y_2})$}~&~-\,\yadj~\\
\hline
~(-1,0,1,1) ~&~0~&~\left(1\,-\,\frac{x^2}{y_1}\right) \,\left(1\,-\,\frac{x^2}{y_1\, y_2}\right)~&~\yadj~\\
\hline
~(1,0,-1,-1)~&~0~&~\left(1\,-\,\frac{1}{x^2\, y_1}\right)\,\left(1\,-\,\frac{1}{x^2\, y_1\, y_2}\right)~&~\yadj~\\
\hline
~(1,0,-1,1)~&~0~&~\left(1\,-\,x^2\, y_1\right) \,\left(1\,-\,\frac{x^2\, y_1}{y_2}\right)~&~\yadj~\\
\hline
~(-1,0,1,-1)~&~0~&~\left(1\,-\,\frac{y_1}{x^2}\right)\, \left(1\,-\,\frac{y_1}{x^2\, y_2}\right)~&~\yadj~\\
\hline
~(1,0,1,-1)~&~1~&~{\left(1\,-\,\frac{1}{x^2}\right)^{-1}\, \left(1\,-\,\frac{1}{x^2\, y_2}\right)^{-1}}~&~\yadj~\\
\hline
~(-1,0,-1,1)~&~1~&~{\left(1\,-\,x^2\right)^{-1} \,\left(1\,-\,\frac{x^2}{y_2}\right)^{-1}}~&~\yadj~\\
\hline
\end{array}
\end{equation*}
\caption{For the stable flux vector $\vec{\sfp}$ with $c_1=(0,1)$ on $\mathbb{F}_0$, we exhibit the final form of the W-boson contribution evaluated at each possible reflection $\vec{\sfp}_\eta$. We see, for instance, that the behavior at $a=0$ changes for the different $\vec{\sfp}_\eta$. In the last column, we add the final contribution to the orbit partition function~\eqref{ZorbitFn} for the $\sfq^1$ term.}
\label{tab:Z1-loopReflections}
\end{table}

\medskip
\noindent
\textbf{Comments on the analytic structure of the full partition function.} Let us now study the nature of the pole of the full partition function depending on the magnetic flux $\vec{\sfp}$. As for the $\mathbb{P}^2$ case in the previous section, let us take $\vec{\sfp}\in\mathbb{Z}^4_{\geq 0}$ to be the positive representative in the orbit under reflections. We assume that $\vec{\sfp}$ has at most one zero component, $\zeta_{\vec\sfp}\leq1$ . Let us look at those two cases separately:
\begin{itemize}
    \item \underline{For $\zeta_{\vec\sfp} = 1$}. This case is analogous to the one discussed at the beginning of Subsection~\ref{subsec:SUSYLocalP2} for $\mathbb{P}^2$: the full partition function develops a simple pole at $a=0$ and one can use the abstruse duality~\eqref{abstruce-duality-C2} to relate residues for different reflections as in~\eqref{res=-resP2}. As a result, the orbit partition function~\eqref{ZorbitFn} simplifies as follows:
    \begin{equation}
        \mathcal{Z}_{\rm orbit}[\mathbb{F}_n]\mid_{\vec\sfp}\,=\,-\,\frac{\Theta_{c_1}(\vec{\sfp})}{1\,-\,\yadj}\,\operatorname*{Res}_{a=0} \, Z_{\rm full,\,c_1}^{\,U(2)}[\mathbb{F}_n]\mid_{\vec\sfp}~,
    \end{equation}
    where $\Theta_{c_1}(\vec{\sfp})$ is defined in a similar fashion as in~\eqref{Thetac1p}, namely, it is $1$ if $\vec{\sfp}$ satisfies the strict inequalities in~\eqref{stab cond Fn}, $1/2$ if it saturates one of the inequalities, and $0$ if it violates them. The residue calculation on the r.h.s. is straightforward since the pole is simple. Upon performing it, the leading power of $\sfq$ will be the quantity $c_2(\vec\sfp)$ defined in~\eqref{c2pFn}, while higher order terms are obtained by dressing with point-like instantons. As an example for $\mathbb{F}_0$ and $c_1=(0,1)$, in the last column of Table \ref{tab:Z1-loopZeros}, we exhibit the contributions of each element in the orbit of the strictly-stable flux vector $\vec{\sfp}=(1,1,0,1)$ to the orbit partition function~\eqref{ZorbitFn}.

    Depending on which entry of the flux vector $\vec{\sfp}$ vanishes, there are four possibilities in one-to-one correspondence with the types 8--11 in~\eqref{type8}--\eqref{type11} and can be viewed as their refined version.
    
    \item  \underline{For $\zeta_{\vec\sfp}=0$}. In this case, the full partition function can have either a double pole or a simple one at $a=0$:
    \begin{table}[t!]
\centering
\centering
\renewcommand{\arraystretch}{1.5} 
\begin{tabular}{|c||c|c|}
\hline
$\vec{\sfp}_\eta$& $\Delta_{(0,1),\vec{\sfp}_{\eta}}[\mathbb{F}_0]$&$\sgn(\vec{\beta}\cdot\vec{\sfp}_\eta)\,{\rm Coeff}_{\sfq^2}\left[\operatorname*{Res}_{a=0} \, Z_{\rm full,\,(0,1)}^{\,U(2)}[\mathbb{F}_n]\mid_{\vec\sfp_\eta}\right]$ \\
\hline
\hline
$\pm(1,1,1,2)$ & $-\,3$ & $-\,y_{\mathrm{adj}}^{-2}\,
-\,2\,y_{\mathrm{adj}}^{-1}\,
-\,5\,
-\,6\,y_{\mathrm{adj}}
\,-\,3\,y_{\mathrm{adj}}^2
\,-\,y_{\mathrm{adj}}^3$\\
\hline
$\pm(1,1,1,-2)$ & $+\,1$ &$y_{\mathrm{adj}}^2\,+\,y_{\mathrm{adj}}^3$\\
\hline
$\pm(1,1,-1,2)$ & $0$ & $y_{\mathrm{adj}}^{-1}\,
+\,1\,
+\,4\,y_{\mathrm{adj}}
\,+\,2\,y_{\mathrm{adj}}^2
\,+\,y_{\mathrm{adj}}^3$\\
\hline
$\pm (1,-1,1,2)$ &$-\,1$& $y_{\mathrm{adj}}^{-2}\,
+\,2\,y_{\mathrm{adj}}^{-1}
\,+\,5
\,+\,6\,y_{\mathrm{adj}}
\,+\,2\,y_{\mathrm{adj}}^2$\\
\hline
$\pm(-1,1,1,2)$ &  $0$&$y_{\mathrm{adj}}^{-2}\,+\,y_{\mathrm{adj}}^{-1}\,
+\,4
\,+\,2\,y_{\mathrm{adj}}
\,+\,y_{\mathrm{adj}}^2$\\
\hline
$\pm(-1,1,1,-2)$ & $0$&$-\,y_{\mathrm{adj}}^{-1}
\,-\,1\,
-\,4\,y_{\mathrm{adj}}
\,-\,y_{\mathrm{adj}}^2$\\
\hline
$\pm(-1,1,-1,2)$ & $3$&$0$\\
\hline
$\pm (-1,-1,1,2)$ &$0$&$-\,y_{\mathrm{adj}}^{-2}\,
-\,y_{\mathrm{adj}}^{-1}\,
-\,4\,
-\,2\,y_{\mathrm{adj}}
\,+\,y_{\mathrm{adj}}^3$\\
\hline
\end{tabular}
\caption{For $\mathbb{F}_0$ and $c_1=(0,1)$, we consider the $\sfq^2$ term of orbit partition function~\eqref{ZorbitFn} of the strictly-stable flux vector $\vec{\sfp}=(1,1,1,2)$. For each possible reflection given in the first column, we exhibit the corresponding non-equivariant contribution in the third column. Summing up these contributions, the end result is $\yadj^2+\yadj^3$.}
\label{tab:orbit-example-F0-1112-q2}
\end{table}

    \begin{itemize}
        \item \textit{The simple-pole case.} For simplicity, let us focus on the contribution of the positive representative $\vec{\sfp}$. For each one of the local affine patches, let us expand the corresponding instanton partition function as follows:
        \begin{equation}
    Z_{\rm inst,\,adj}^{\,U(2)}[\mathbb{C}^2_\ell]\,=\,1\,+\,\sfq^{\sfp_{\ell}\sfp_{\ell+1}}\,\mathsf{Z}^{(\ell)}_{\sfp_\ell,\sfp_{\ell+1}}\,+\,\cdots~.
\end{equation}
Here, $\mathsf{Z}_{\sfp_\ell,\sfp_{\ell+1}}^{(\ell)}$ is the singular term, and the dots denote the remaining terms in the expansion which are regular at $a=0$ and carry positive powers of $\sfq$. Expanding $Z_{\rm inst,\,adj}^{\,U(2)}[\mathbb{F}_n]\mid_{\vec{\sfp}}$ in powers of $\sfq$, we obtain $4$ terms of the form:
\begin{equation}\label{4-terms-inst}
\prod_{\substack{\ell=1\\\ell\neq\ell'}}^{4}\,\sfq^{\sfp_{\ell}\sfp_{\ell+1}}\,\mathsf{Z}^{(\ell)}_{\sfp_\ell,\sfp_{\ell+1}}~,\qquad \ell'\,=\,1\,,\,\cdots\,,\,4~.
\end{equation}
Each of these terms has an order $3$ pole at $a=0$. Multiplying by the 1-loop contribution~\eqref{1-loop-Fn}, each of the four terms gives a simple pole.
Extracting the residue at the pole, we get in the end:
\begin{equation}
    \sum_{\ell'=1}^4\,\left(\sfq^{c_2(\vec\sfp)\,-\,\sfp_{\ell'}\,\sfp_{\ell'+1}}\,\operatorname*{Res}_{a=0}\,\left[Z_{{\rm 1-loop}}^{\,U(2)}[\mathbb{F}_n]\,\prod_{\substack{\ell=1\\\ell\neq\ell'}}^{4}\,\mathsf{Z}^{(\ell)}_{\sfp_\ell,\sfp_{\ell+1}}\right]\,+\,\cdots\right)~.
\end{equation}
The leading power of $\sfq$, namely $c_2(\vec\sfp)\,-\,\sfp_{\ell'}\,\sfp_{\ell'+1}$, is recognized as the second Chern class of a toric rank $2$ sheaf~\eqref{c2Etoric} with 
$v_\ell=\sfp_\ell$ and $\delta_{\ell,\ell+1}=1$ for $\ell=\ell'$, and zero otherwise. Thus these terms match the contributions of type 4-7 in the classification~\eqref{type4}--\eqref{type7}
of fixed points in the moduli space of slope-stable vector bundles. 
These arguments can be repeated for the other vectors $\vec{\sfp}_\eta$ in the orbit, where analogous behavior is expected -- i.e., that in the end we get four terms each with a simple pole. 

    As an example, in Table \ref{tab:orbit-example-F0-1112-q2}, we include the final non-equivariant contribution of all possible reflections of the strictly-stable flux vector $\vec{\sfp}=(1,1,1,2)$ to $\sfq^2$ term of the full partition function. As shown in Table \ref{tab:Z1-loopZeros}, this vector satisfies the strict inequalities 
    for $\mathbb{F}_0$ with $c_1=(0,1)$. The final result is:
    \begin{equation}\label{Zorbit1112q2}
        {\rm Coeff}_{\sfq^2}\left[Z_{\rm orbit}[\mathbb{F}_0]\mid_{(1,1,1,2)}\right]\,=\,(\yadj^2\,+\,\yadj^3)\,\sfq^2~.   
    \end{equation}
    In the unrefined limit, this corresponds to two isolated fixed points on the moduli space of sheaves, in agreement with the toric localization result as listed in Table~\ref{tab:F0_tuples}, where those two points are denoted by $(1,1,1,2)_{4+5}$.
    
    Let us remark here that unlike the relation~\eqref{res=-resP2} valid for $\IP^2$, the residues associated the various reflected fluxes $\vec{\sfp}_\eta$ 
    are in general not equal up to sign, even though the poles are simple. Yet, we observe in examples that the residue flips sign upon flipping the sign of all entries in the flux vector. Moreover, as observed below \eqref{semistabF0tuples}, although the flux $\vec{\sfp}$ solves the \textit{strict} stability conditions \eqref{stab cond Fn}, it behaves in some of those $4$ cases like a semi-stable flux. Namely, our explicit residue calculations reveal the appearance of $\sgn(0)$ weights in the corresponding orbit, as well as an overall factor of $1/2$ in the final orbit partition function \eqref{ZorbitFn}.\\

\begin{table}[t!]
\centering
\centering
\renewcommand{\arraystretch}{2} 
\begin{tabular}{|c||c|c|}
\hline
$\vec{\sfp}_\eta$& $\Delta_{(0,1),\vec{\sfp}_{\eta}}[\mathbb{F}_0]$&$\sgn(\vec{\beta}\cdot\vec{\sfp}_\eta)\,{\rm Coeff}_{\sfq^3}\left[\operatorname*{Res}_{a=0} \, Z_{\rm full,\,(0,1)}^{\,U(2)}[\mathbb{F}_n]\mid_{\vec\sfp_\eta}\right]$ \\
\hline
\hline
$\pm(1,1,1,2)$ & $-\,3$ & \parbox{7cm}{\centering$-\,y_{\mathrm{adj}}^{-2}\,
-\,2\,y_{\mathrm{adj}}^{-1}
\,-\,7\,
-\,14\,y_{\mathrm{adj}}
\,-\,24\,y_{\mathrm{adj}}^2
\,-\,25\,y_{\mathrm{adj}}^3
\,-\,19\,y_{\mathrm{adj}}^4
\,-\,10\,y_{\mathrm{adj}}^5
\,-\,2\,y_{\mathrm{adj}}^6$}\\
\hline
$\pm(1,1,1,-2)$ & $+\,1$ &$y_{\mathrm{adj}}^2
\,+\,y_{\mathrm{adj}}^3
\,+\,6\,y_{\mathrm{adj}}^4
\,+\,6\,y_{\mathrm{adj}}^5
\,+\,2\,y_{\mathrm{adj}}^6$\\
\hline
$\pm(1,1,-1,2)$ & $0$ & \parbox{7cm}{\centering$y_{\mathrm{adj}}^{-1}\,
+\,1
\,+\,6\,y_{\mathrm{adj}}
\,+\,8\,y_{\mathrm{adj}}^2
\,+\,14\,y_{\mathrm{adj}}^3
\,+\,12\,y_{\mathrm{adj}}^4
\,+\,8\,y_{\mathrm{adj}}^5
\,+\,2\,y_{\mathrm{adj}}^6$}\\
\hline
$\pm (1,-1,1,2)$ &$-\,1$& \parbox{7cm}{\centering$y_{\mathrm{adj}}^{-2}
\,+\,2\,y_{\mathrm{adj}}^{-1}
\,+\,7
\,+\,14\,y_{\mathrm{adj}}
\,+\,23\,y_{\mathrm{adj}}^2
\,+\,23\,y_{\mathrm{adj}}^3
\,+\,14\,y_{\mathrm{adj}}^4
\,+\,4\,y_{\mathrm{adj}}^5$}\\
\hline
$\pm(-1,1,1,2)$ &  $0$&\parbox{7cm}{\centering$y_{\mathrm{adj}}^{-2}
\,+\,y_{\mathrm{adj}}^{-1}
\,+\,6
\,+\,8\,y_{\mathrm{adj}}
\,+\,16\,y_{\mathrm{adj}}^2
\,+\,12\,y_{\mathrm{adj}}^3
\,+\,6\,y_{\mathrm{adj}}^4
\,+\,2\,y_{\mathrm{adj}}^5$}\\
\hline
$\pm(-1,1,1,-2)$ & $0$&\parbox{7cm}{\centering$-\,y_{\mathrm{adj}}^{-1}
\,-\,1
\,-\,6\,y_{\mathrm{adj}}
\,-\,7\,y_{\mathrm{adj}}^2
\,-\,12\,y_{\mathrm{adj}}^3
\,-\,7\,y_{\mathrm{adj}}^4
\,-\,2\,y_{\mathrm{adj}}^5$}\\
\hline
$\pm(-1,1,-1,2)$ & $3$&$-\,y_{\mathrm{adj}}^3
\,+\,y_{\mathrm{adj}}^4$\\
\hline
$\pm (-1,-1,1,2)$ &$0$&\parbox{8cm}{\centering$-\,y_{\mathrm{adj}}^{-2}
\,-\,y_{\mathrm{adj}}^{-1}
\,-\,6
\,-\,8\,y_{\mathrm{adj}}
\,-\,15\,y_{\mathrm{adj}}^2
\,-\,10\,y_{\mathrm{adj}}^3
\,-\,y_{\mathrm{adj}}^4
\,+\,4\,y_{\mathrm{adj}}^5
\,+\,2\,y_{\mathrm{adj}}^6$}\\
\hline
\end{tabular}
\caption{For $\mathbb{F}_0$ and $c_1=(0,1)$, we consider the $\sfq^3$ term of orbit partition function~\eqref{ZorbitFn} of the strictly-stable flux vector $\vec{\sfp}=(1,1,1,2)$. For each possible reflection given in the first column, we exhibit the corresponding non-equivariant contribution in the third column. Summing up these contributions, the end result is $y_{\mathrm{adj}}^2
+2y_{\mathrm{adj}}^3
+5y_{\mathrm{adj}}^4
+6y_{\mathrm{adj}}^5
+2y_{\mathrm{adj}}^6$.}
\label{tab:orbit-example-F0-1112-q3}
\end{table}

    \item \textit{The double-pole case.} Let us again focus for simplicity on the representative of the reflection orbit $\vec{\sfp}\in\mathbb{Z}^4_{>0}$. In the expansion of $Z_{\rm inst,\,adj}^{\,U(2)}[\mathbb{F}_n]$ performed above, we get the following term:
    \begin{equation}\label{one-term-inst}
\prod_{\ell=1}^{4}\,\sfq^{\sfp_{\ell}\sfp_{\ell+1}}\,\mathsf{Z}^{(\ell)}_{\sfp_\ell,\sfp_{\ell+1}}~,
\end{equation}
which has a pole of degree $4$ at $a=0$. When combining this with the corresponding 1-loop part, the double zeros will cancel two of those poles. Extracting the residue at the remaining double-pole, we get:
\begin{align}
            \sfq^{c_2(\vec\sfp)}\,\operatorname*{Res}_{a=0}\, \left[Z_{{\rm 1-loop}}^{\,U(2)}[\mathbb{F}_n]\,\prod_{\ell=1}^{4}\,\mathsf{Z}^{(\ell)}_{\sfp_\ell,\sfp_{\ell+1}}\right]\,+\,\cdots~.
        \end{align}
        Based on the results we obtained for $\mathbb{F}_0$ and $\mathbb{F}_1$ -- see Appendices \ref{app:F0Results} and \ref{app:F1Results}, respectively -- we observe that the final contribution of those cases to $\sfq^{c_2(\vec\sfp)}$ is zero. This is in agreement with the fact that the first three types in~\eqref{type1}--\eqref{type3} cancel out
        against each other. In other words, this cancellation continues to hold at the refined level. In Table \ref{tab:orbit-example-F0-1112-q3}, we list the non-equivariant contributions of all elements in the orbit of the flux $\vec{\sfp}=(1,1,1,2)$ to $\sfq^3$. As before, this is an example of a stable vector for $\mathbb{F}_0$ and $c_1=(0,1)$. The final contribution is given by:
        \begin{equation}
          {\rm Coeff}_{\sfq^3}\left[Z_{\rm orbit}[\mathbb{F}_n]\mid_{(1,1,1,2)}\right]\,=\,(y_{\mathrm{adj}}^2\,
+\,2\,y_{\mathrm{adj}}^3
\,+\,5\,y_{\mathrm{adj}}^4
\,+\,6\,y_{\mathrm{adj}}^5
\,+\,2\,y_{\mathrm{adj}}^6)\,\sfq^3~.
        \end{equation}
        In the unrefined limit, this gives us $16 = 2\times8$. The factor $2$ corresponds to the number of fixed points at order $\sfq^2$ in~\eqref{Zorbit1112q2}, while the factor $8$ comes from dressing these two points with point-like instantons encoded in the overall factor $\overline{\eta}(\sfq)^{-8}$. This indicates that the actual double-pole that is expected to contribute to $\sfq^3$ has a zero contribution. This matches with the cancellation from Table \ref{tab:F0_tuples} expressed as $(1,1,1,2)_{2-1}$.
    \end{itemize} 
\end{itemize}

%%%%%%%%%%%%%%%%%%%%%%%%%%%%%%%%%%%%%%%%%
%%%%%%%%%%%%%%%%%%%%%%%%%%%%%%%%%%%%%%%%%
\subsection{For \texorpdfstring{$\mathbb{F}_0$}{F0}}
\subsubsection{For \texorpdfstring{$c_1 = (0,1)$}{c101}}

Let us first look at the case $c_1=(0,1)$.\footnote{Recall that for $\mathbb{F}_0$, the two cases $c_1=(1,0)$ and $c_1=(0,1)$ are related by exchanging the two $\IP^1$'s, and this relation is not affected by wall-crossing.} In this case, solving the corresponding stability conditions~\eqref{stab cond Fn} with the polarization $(\boldsymbol{a},\boldsymbol{b})=(1.1,1)$, we find the following list of contributing strictly slope-stable flux vectors up to $\sfq^4$:
\begin{equation}
\begin{split}
    &\underbrace{(1,0,1,1)~,~(1,1,1,0)}_{\sfq}~,~\underbrace{(2,0,2,1)~,~(2,1,2,0)~,~(1,1,1,2)~,~(1,2,1,1)}_{\sfq^2}~,\\
    &\underbrace{(0,1,2,2)~,~(2,2,0,1)~,~(0,2,2,1)~,~(2,1,0,2)~,~(3,0,3,1)~,~(3,1,3,0)}_{\sfq^3}~,\\
    &\underbrace{(1,2,1,3)~,~(1,3,1,2)}_{\sfq^3}~,~\underbrace{(4,0,4,1)~,~(4,1,4,0)~,~(1,1,3,2)~,~(3,2,1,1)}_{\sfq^4}~,\\
    &\underbrace{(1,2,3,1)~,~(3,1,1,2)~,~(2,1,2,2)~,~(2,2,2,1)~,~(1,3,1,4)~,~(1,4,1,3)}_{\sfq^4}~.
\end{split}
\end{equation}
Meanwhile, we do not get any strictly semi-stable fluxes. In Appendix \ref{subsec:F0 c101 fixed points}, we list all the contributions of these vectors to the refined VW 
invariants in the non-equivariant limit $y_1,y_2\to 0$. Summing up these contributions, the dependence on equivariant parameters drops out, and we get:
\begin{equation}
    \begin{split}
        \mathcal{Z}_{(0,1)}[\mathbb{F}_0]\,&=\,(1\,+\,\yadj)\,\sfq\,\\
        &+\,(1\,+\,3\,\yadj\,+7\,\yadj^2\,+\,\cdots)\,\sfq^2\\
        &+\,(1\,+\,3\,\yadj\,+\,10\,\yadj^2\,+\,22\,\yadj^3\,+\,37\,\yadj^4\,+\cdots)\,\sfq^3\\
        &+\,(1\,+\,3\,\yadj\,+\,10\,\yadj^2\,+\,26\,\yadj^3\,+\,60\,\yadj^4\,+\,110\,\yadj^5\,+\,161\,\yadj^6\,+\,\cdots)\,\sfq^4\\
        &+\,\cdots~,
    \end{split}
\end{equation}
in agreement with $\mathfrak{h}_{2,(0,1)}^{\mathbb{F}_0}$ in~\eqref{F0mod2} upon setting $\yadj=y^2$ and rescaling $\sfq\mapsto \sfq\, y^4$, up to an overall factor of $y^3$.
In the unrefined limit, the above expression reduces to:
\begin{equation}
    \mathcal{Z}_{(0,1)}[\mathbb{F}_0]\mid_{\yadj\rightarrow 1}\,=\,2\,\sfq\,+\,22\,\sfq^2\,+\,146\,\sfq^3\,+\,742\,\sfq^4\,+\,\cdots~,
\end{equation}
which matches the result of toric localization in~\eqref{Kool F0 c101}.

%%%%%%%%%%%%%%%%%%%%%%%%%%%%%%%%%%%%%%%%
\subsubsection{For \texorpdfstring{$c_1= (1,1)$}{c111}}
Let us now look at the second case with $c_1=(1,1)$. Choosing the polarization to be $(\boldsymbol{a},\boldsymbol{b}) = (1.1,1)$, we find the following list of strictly slope-stable fluxes contributing up to order $\sfq^4$:\footnote{Note that these differ from those considered in~\protect\cite[Equation (6.9)]{Bonelli:2020xps} since these authors take the diagonal polarization to be $(\boldsymbol{a},\boldsymbol{b})=(1,1)$. In particular, with our choice of polarization, there are no strictly semi-stable fluxes.} 
\begin{equation}
\begin{split}
    &\underbrace{(1,0,2,1)~,~(2,1,1,0)~,~(1,1,2,0)~,~(2,0,1,1)}_{\sfq^2}~,~\underbrace{(2,0,3,1)~,~(3,1,2,0)}_{\sfq^3}~,\\
    &\underbrace{(2,1,3,0)~,~(3,0,2,1)~,~(1,1,2,2)~,~(2,2,1,1)~,~(1,2,2,1)~,~(2,1,1,2)}_{\sfq^3}\,,\\
    &\underbrace{(3,0,4,1)~,~(4,1,3,0)~,~(3,1,4,0)~,~(4,0,3,1)~,~(1,2,2,3)~,~(2,3,1,2)}_{\sfq^4}~,\\
    &\underbrace{(1,3,2,2)~,~(2,2,1,3)}_{{\sfq^4}}~.
\end{split}
\end{equation}
In Appendix \ref{app:F0 c1=(1,1)}, we list the contributions of these fluxes to the refined VW invariants in the non-equivariant limit. Summing over all these contributions, 
the dependence on equivariant parameters again drops out, and we find:
\begin{equation}\label{Z11F0}
    \begin{split}
        \mathcal{Z}_{(1,1)}[\mathbb{F}_0]\,&=\,(1\,+\,\yadj\,+\,\cdots)\,\sfq^2\\
         &+\,(1\,+\,3\,\yadj\,+\,7\,\yadj^2\,+\,9\,\yadj^3\,+\,\cdots)\,\sfq^3\\
         &+\,(1 \, + \, 3\,\yadj \, + \, 10\,\yadj^{2} \, + \, 22\,\yadj^{3} \, + \, 40\,\yadj^{4} \, + \, 50\,\yadj^{5}\,+\,\cdots)\,\sfq^4\,+\,\cdots~,
    \end{split}
\end{equation}
matching the generating series $\mathfrak{h}_{2,(1,1)}^{\mathbb{F}_0}$ in~\eqref{F0mod2}. 
In the unrefined limit, this gives:
\begin{equation}
   \mathcal{Z}_{(1,1)}[\mathbb{F}_0]\mid_{\yadj\rightarrow 1}\,=\,4\,\sfq^2\,+\,40\,\sfq^3\,+\,252\,\sfq^4\,+\,\cdots~,
\end{equation}
matching the counting in~\eqref{Kool F0 c111}.

\subsubsection{For \texorpdfstring{$c_1= (0,0)$}{c100}}
\medskip
\noindent
\textbf{Contribution from strictly-stable fluxes.}
For this choice of the first Chern class and polarization $(\boldsymbol{a},\boldsymbol{b})=(1.1,1)$, we find that the following list of 
flux vectors satisfy the strict inequalities~\eqref{stab cond Fn} and contribute up to order $\sfq^4$ to the partition function:
\begin{equation}
\begin{split}
    &\underbrace{(0,1,2,1)~,~(2,1,0,1)}_{\sfq^2}~,~\underbrace{(1,1,3,1)~,~(3,1,1,1)}_{\sfq^3}~,~\underbrace{(2,1,4,1)~,~(4,1,2,1)}_{\sfq^4}~,\\&\underbrace{(0,2,2,2)~,~(1,0,3,2)~,~(1,2,3,0)~,~(2,0,2,2)~,~(2,2,0,2)~,~(2,2,2,0)}_{\sfq^4}~,\\&\underbrace{(3,0,1,2)~,~(3,2,1,0)}_{\sfq^4}~.
\end{split}
\end{equation}
Note that they match the contributing $4$-tuples $(v_1,v_2,v_3,v_4)$ listed in Table \ref{tab:F0_tuples}. The final contributions of each  of those fluxes are listed in Appendix \ref{app:F0 c1=(0,0)} in the non-equivariant limit.
Summing up these strictly-stable contributions, we find:
\begin{equation}
    \begin{split}
       &\,(\yadj^2\,+\,\yadj^3)\,\sfq^2\,+\,(y_{\mathrm{adj}}^2\,+\,3\,y_{\mathrm{adj}}^3
\,+\,6\,y_{\mathrm{adj}}^4
\,+\,\cdots)\,\sfq^3\\&+\,(y_{\mathrm{adj}}^2\,+\,3\,y_{\mathrm{adj}}^3\,+\,10\,y_{\mathrm{adj}}^4\,+\,21\,y_{\mathrm{adj}}^5\,+\,31\,y_{\mathrm{adj}}^6\,+\,\cdots)\,\sfq^4\,+\,\cdots~.
\end{split}
\end{equation}
In the unrefined limit, 
we get $2\,\sfq^2\,+\,20\,\sfq^3\,+\,132\,\sfq^4\,+\,\cdots$,
matching the corresponding counting in~\eqref{Kool F0 c100}.

\medskip
\noindent
\textbf{Contribution of the semi-stable fluxes.} In addition, we find the following flux vectors also  satisfy the strict inequalities~\eqref{stab cond Fn}  and 
contribute up to order $\sfq^3$:
\begin{equation}\label{semistablF0c100}
    \underbrace{(1,1,1,1)}_{\sfq}~,~\underbrace{(1,2,1,2)~,~(2,1,2,1)}_{\sfq^2}~,~\underbrace{(1,3,1,3)~,~(3,1,3,1)}_{\sfq^3}~,
\end{equation}
where the power of $\sfq$ differs from the naive estimate $c_2(\vec\sfp)$ by $-\sfp_\ell \sfp_{\ell+1}$ for $\ell\in\{1,2,3,4\}$. 
Following the discussion in Subsection~\ref{subsubsec:analyzFullFuncFn}, the double pole contribution associated with these
fluxes vanish, which reflects the cancellation of the corresponding fixed loci in Table \ref{tab:F0_tuples}.
Meanwhile, the simple pole contributions behave like strictly-semi-stable fluxes as discussed in \eqref{semistabF0tuples}. 
This is reflected in our calculation by the presence of reflections $\vec{\sfp}_{\eta}$ such that $\vec{\beta}\,\cdot\,\vec{\sfp}_\eta = 0$. For instance, for the flux $(1,1,1,1)$, this happens for the reflections $\pm(-1,-1,1,1)$ and $\pm(-1,1,1,-1)$. As a second indication, the final orbit partition function for each one of those fluxes comes with an overall factor of $1/2$. This is analogous to the contribution of the semistable fluxes in the $\mathbb{P}^2$ case~\eqref{ZFullRes}.

As an example, the $\cO(\sfq)$ receives contribution only from the flux vector $(1,1,1,1)$. At the equivariant level, its contribution takes the following form:
\begin{multline}
{\rm Coeff}_{\sfq}\,\left[\mathcal{Z}_{(0,0)}[\mathbb{F}_0]\right]\,=\,\frac{\,(1\,+\,y_{\mathrm{adj}})(1\,+\,y_{\mathrm{adj}}^2)}
{2\,y_{\mathrm{adj}}}\,+\\+\,y_1\,\frac{
(y_{\mathrm{adj}}-1)^2(1+y_{\mathrm{adj}})
\left(
-y_2
+y_{\mathrm{adj}}
+2y_2y_{\mathrm{adj}}
+y_2^2y_{\mathrm{adj}}
-y_2y_{\mathrm{adj}}^2
\right)
}{
2\,y_2\,y_{\mathrm{adj}}^2}
\,+\,\cO(y_1^2)~.
\end{multline}
In the non-equivariant limit $y_1\to 0$, only the first term above survives:
\begin{equation}\label{qTermF0c100}
    {\rm Coeff}_{\sfq}\,\left[\mathcal{Z}_{(0,0)}[\mathbb{F}_0]\right]\,=\,\frac{1}{2}\left(\yadj^{-1}\,+\,1\,+\,\yadj\,+\,\yadj^2\right)~.
\end{equation}
The four terms match the earlier result in~\eqref{semistabF0tuples} that this flux contributes to $F_{4,5,6,7}$ in~\eqref{type4}--\eqref{type7}. In Appendix~\ref{app:F0 c1=(0,0)}, we list the contribution of the fluxes~\eqref{semistablF0c100} to the first four terms of the generating function~\eqref{Zc1Fn=SumOrbits} in the non-equivariant limit $y_1\rightarrow0$, $y_2\rightarrow 0$. In particular, we observe that each of the fluxes $(1,2,1,2)$ and $(2,1,2,1)$ contributes four points to the $\sfq^2$ term, leading to a total contribution of $8\,\sfq^2$. Similar remarks hold for $(1,3,1,3)$ and $(3,1,3,1)$ at $\sfq^3$. In the unrefined limit, our results indeed match~\eqref{semistabF0tuples}.

%%%%%%%%%%%%%%%%%%%%%%%%%%%%%%%%%%%%%%%%%
%%%%%%%%%%%%%%%%%%%%%%%%%%%%%%%%%%%%%%%%%
%\newpage
\subsection{For \texorpdfstring{$\mathbb{F}_1$}{F1}}

\subsubsection{For \texorpdfstring{$c_1 = (0,1)$}{c101}}
Let us start with the case with $c_1=(0,1)$. Solving the corresponding stability conditions~\eqref{stab cond Fn}, we find the following list of 
 slope-stable fluxes  contributing to the refined VW partition function up to order $\sfq^2$:
\begin{equation}
\begin{split}
&\underbrace{(1,0,1,1)}_{\sfq}~,~\underbrace{(2,0,2,1)~,~(1,1,2,2)~,~(1,2,1,1)~,~(2,1,1,2)}_{\sfq^2}~.
\end{split}
\end{equation}
The contributions of each of these fluxes in the non-equivariant limit are listed in Appendix \ref{app:F1 c1=(0,1)}. Summing those contributions up, we get the following result,
independent of equivariant parameters:
\begin{equation}
    \begin{split}
        \mathcal{Z}_{(0,1)}[\mathbb{F}_1]\,&=\,\sfq\,+\,(1\,+\,3\,\yadj\,+\,5\,\yadj^2\,+\,\cdots)\,\sfq^2\,+\,\cdots~.\\
    \end{split}
\end{equation}
This matches $\mathfrak{h}^{\mathbb{F}_1}_{2,(1,1)}$ in~\eqref{F1mod2}. In the unrefined limit, we get:
\begin{equation}
   \mathcal{Z}_{(1,0)}[\mathbb{F}_1]\mid_{\yadj\rightarrow 1}\,=\,\sfq\,+\,13\,\sfq^2\,+\,\cdots~.
\end{equation}
which matches~\eqref{Kool F1 c101}.

%%%%%%%%%%%%%%%%%%%%%%%%%%%%%%%%%%%%%%%%%%%%%
\subsubsection{For \texorpdfstring{$c_1 = (1,0)$}{c101}}
In this case, we find the following  fluxes contributing to the refined VW partition function up to order $\sfq^2$:
\begin{equation}
    \underbrace{(1,1,1,1)}_{\sfq}~,~\underbrace{(1,0,2,2)~,~(2,0,1,2)~,~(2,1,2,1)~,~(1,2,2,2)~,~(2,2,1,2)}_{\sfq^2}~.
\end{equation}
The contributions of those fluxes in the non-equivariant limit are listed in Appendix~\ref{app:F1 c1=(1,0)}. Summing up these contributions, we find the following 
refined VW invariants, independent of equivariant parameters:
\begin{equation}
\begin{split}
        \mathcal{Z}_{(1,0)}[\mathbb{F}_1]\,&=\,(1\,+\,\yadj)\,\sfq\,+\,(1 \, + \, 3\,\yadj \, + \, 7\,\yadj^{2} \, + \,\cdots)\,\sfq^2\,+\,\cdots~.\\
\end{split}
\end{equation}
matching \eqref{F1mod2} for the case $\mathfrak{h}^{\mathbb{F}_1}_{2,(1,0)}$. 
In the unrefined limit, we get:
\begin{equation}
    \mathcal{Z}_{(1,0)}[\mathbb{F}_1]\mid_{\yadj\rightarrow 1}\,=\,2\,\sfq\,+\,22\,\sfq^2\,+\,\cdots~,
\end{equation}
matching with the toric localization result in~\eqref{Kool F1 c110}.

\subsubsection{For \texorpdfstring{$c_1= (1,1)$}{c111}}
In this case, we find the following fluxes contributing to the refined VW partition function up to order $\sfq^3$:
\begin{equation}
\begin{split}
      &\underbrace{(1,0,2,1)~,~(1,1,1,0)~,~(2,0,1,1)}_{\sfq^2},\\
    &\underbrace{(0,2,1,1)~,~(1,2,0,1)~,~(2,0,3,1)~,~(2,1,2,0)~,~(3,0,2,1)~,~(1,1,3,2)~,~(3,1,1,2)}_{\sfq^3}~.
\end{split}
\end{equation}
The contributions of those fluxes to the refined VW invariants are listed in Appendix~\ref{app:F1 c1=(1,1)}. Summing these up, we find:
\begin{equation}
\begin{split}
        \mathcal{Z}_{(1,1)}[\mathbb{F}_1]\,&=\,(1\,+\,\yadj\,+\,\yadj^2)\,\sfq^2\\
        &+\,(1 \, + \, 3\,\yadj \, + \, 7\,\yadj^{2} \, + \, 9\,\yadj^{3} \,+\,\cdots)\,\sfq^3\,+\,\cdots~.
\end{split}
\end{equation}
Note that this matches~\eqref{F1mod2} for $\mathfrak{h}_{2,(1,1)}^{\mathbb{F}_1}$. In the unrefined limit, we get:
\begin{equation}
    \mathcal{Z}_{(1,1)}[\mathbb{F}_1]\mid_{\yadj\rightarrow 1}\,=\,3\,\sfq^2\,+\,31\,\sfq^3\,+\,\cdots~,
\end{equation}
which matches the toric localization result in~\eqref{Kool F1 c111}.

\subsubsection{For \texorpdfstring{$c_1= (0,0)$}{c100}}
For this case, we find the following list of contributing strictly-stable flux vectors to the $\sfq^2$ term
term of the generating function:
\begin{equation}\label{strictstF1c100}
        \underbrace{(1,1,2,1)~,~(2,1,1,1)}_{\sfq^2}~.
\end{equation}
Similar to the earlier discussion for the $\mathbb{F}_0$ case with $c_1 = (0,0)$~\eqref{semistablF0c100}, the same flux vector can contribute as a strictly-stable and as semistable to different powers of $\sfq$. With this in mind, we find the following contributing semistable flux vectors up to order $\sfq^2$:
\begin{equation}\label{semistF1c100}
\begin{split}
    &\underbrace{(0,1,1,1)~,~(1,1,0,1)~,~(1,1,2,1)~,~(2,1,1,1)}_{\sfq}~,\\
    &\underbrace{(1,2,3,2)~,~(1,2,1,2)~,~(2,1,3,1)~,~(3,1,2,1)~,~(3,2,1,2)}_{\sfq^2}~.
\end{split}
\end{equation}

\begin{table}[t!]
\centering
\centering
\renewcommand{\arraystretch}{2} 
\begin{tabular}{|c||c|c|}
\hline
$\vec{\sfp}$& ${\rm Coeff}_{\sfq}\,\left[Z_{\rm orbit}[\mathbb{F}_1]\mid_{\vec\sfp}\right]$ \\
\hline
\hline
$(0,1,1,1)$ & $\frac{
(y_2\,-\,1)\,(y_1\,-\,y_{\mathrm{adj}})
\,\left(y_1\,y_2\,y_{\mathrm{adj}}\,-\,1\right)
}{2\,
(y_1\,-\,1)\,(y_1\,y_2\,-\,1)\,(y_2\,-\,y_{\mathrm{adj}})
}$\\
\hline
$(1,1,0,1)$ & $\frac{
\left(y_1\,y_2\,-\,1\right)
\left(y_1\,y_{\mathrm{adj}}\,-\,1\right)
\left(y_2\,y_{\mathrm{adj}}\,-\,1\right)
}{2\,
(y_1\,-\,1)\,(y_2\,-\,1)\,\left(y_1\,y_2\,-\,y_{\mathrm{adj}}\right)
}$\\
\hline
$(1,1,2,1)$ & $\frac{
(y_1\,-\,y_2)\,(y_2\,-\,y_{\mathrm{adj}})\,\left(y_1\,y_{\mathrm{adj}}\,-\,1\right)
}{
2\,(y_1\,-\,1)\,(y_2\,-\,1)\,\left(y_1\,-\,y_2\,y_{\mathrm{adj}}\right)
}$ \\
\hline
$(2,1,1,1)$ & $\frac{
\left(y_1^2\,y_2\,-\,1\right)\,
\left(y_1\,-\,y_{\mathrm{adj}}\right)
\left(y_1\,y_2\,-\,y_{\mathrm{adj}}\right)
}{2\,
(y_1\,-\,1)\,
\left(y_1\,y_2\,-\,1\right)\,
\left(y_1^2\,y_2\,y_{\mathrm{adj}}\,-\,1\right)
}$ \\
\hline
\end{tabular}
\caption{The equivariant contribution of the four fluxes listed in the first line~\eqref{semistF1c100} to the $\sfq$ term of the generating function $\mathcal{Z}_{(0,0)}[\mathbb{F}_1]$.}
\label{tab:equivF1c100q}
\end{table}

The contributions of each of those fluxes to the generating function~\eqref{ZfullFn} are shown in Appendix~\ref{app:F1 c1=(0,0)} in the non-equivariant limit. 
For the leading $\cO(\sfq)$ term, we show the full equivariant contribution of the fluxes in the first line in~\eqref{semistF1c100} in Table~\ref{tab:equivF1c100q}. 
Unlike cases where $c_1\neq (0,0) \mod 2$, the dependence on equivariant parameters does not disappear after summing up these contributions, even though the power of $\sfq$ is odd.
Moreover, the non-equivariant limit depends on the order in which we take the limits $y_1\to 0$ and $y_2\to 0$. 
Taking the limit $y_2\to 0$ first, we get:
\begin{multline}\label{CoeffqZ00F}
   {\rm Coeff}_{\sfq}\,\left[\mathcal{Z}_{(0,0)}[\mathbb{F}_1]\right]\,=\, \frac{1\,+\,y_{\mathrm{adj}}\,+\,y_{\mathrm{adj}}^2\,+\,y_{\mathrm{adj}}^3}
{2\,y_{\mathrm{adj}}}\,+
\\+\,y_2\, 
\frac{
(1+y_1)\,(y_{\mathrm{adj}}-1)
\left(
-2y_1
+y_1y_{\mathrm{adj}}
+y_{\mathrm{adj}}^2
+y_1^2y_{\mathrm{adj}}^2
-y_1y_{\mathrm{adj}}^3
\right)
}{
2y_1y_{\mathrm{adj}}
}
\,+\,\cO(y_2^2)~.
\end{multline}
In the non-equivariant limit $y_2\to 0$, this reduces to the same $\cO(\sfq)$ term in $\mathbb{F}_0$~\eqref{qTermF0c100}. Note also that the dependence on 
equivariant parameters disappear in the unrefined limit $\yadj\to 1$. Taking the limit $y_1\to 0$ first, the leading order result still
depends on $y_2$: 
\be
\begin{split}
 {\rm Coeff}_{\sfq}\,\left[\mathcal{Z}_{(0,0)}[\mathbb{F}_1]\right]\,=&
 \frac{\left(1\,+\,2y_{\mathrm{adj}}\,+\,y_{\mathrm{adj}}^3\right)}{2\yadj}
 + \frac{y_2(1-\yadj)}{2(\yadj-y_2}
\\ &-\frac{
y_1\,(y_{\mathrm{adj}}-1)}{
2y_2\,(y_2-y_{\mathrm{adj}})\,y_{\mathrm{adj}}^2
}
\left(
-y_2^2
+2y_2y_{\mathrm{adj}}
+2y_2^2y_{\mathrm{adj}}
+y_2^3y_{\mathrm{adj}}
-y_{\mathrm{adj}}^2 \right.
\\ &\left. \qquad -y_2y_{\mathrm{adj}}^2
-y_2^2y_{\mathrm{adj}}^2
-y_2y_{\mathrm{adj}}^3
-2y_2^2y_{\mathrm{adj}}^3
+y_2y_{\mathrm{adj}}^4
+y_2^2y_{\mathrm{adj}}^4
\right)
+\,\cO(y_1^2)~.
\end{split}
\ee
Further taking the limit $y_2\rightarrow 0$, we obtain $\frac{1}{2}(\yadj^{-1}+2+\yadj^2)$ which differs from the $y_1\to 0$ limit of \eqref{CoeffqZ00F}, even though the
two results coincide in the unrefined limit. Indeed, when $\yadj=1$, the dependence on the equivariant parameters drops out entirely.

\medskip
\noindent
\textbf{In the unrefined limit.}  
Let us now look at the unrefined limit of the final results given in Appendix \ref{app:F1 c1=(0,0)}. The goal is to connect our results with the ones reported in Subsection~\ref{subsec:F1Literature}. For starters, 
let us look more closely at the flux $\vec{\sfp} = (1,1,2,1)$. From the referenced appendix, we find that in the unrefined limit, it has the following contribution:
\begin{equation}
   \lim_{\yadj\rightarrow 1}\, Z_{\rm orbit}[\mathbb{F}_1]\mid_{(1,1,2,1)}\,=\,\frac{1}{2}\,\sfq\,+\,\frac{11}{2}\,\sfq^2\,+\,\cdots~.
\end{equation}
For the first term, we see that this flux contributes a semi-stable vector bundle to $\sfq$, and this matches the first line of~\eqref{semistabF1c100tuples}. For the $\sfq^2$ term above, we can expand its coefficient as follows:
\begin{equation}
    \frac{1}{2}\,\times\, 11\,=\,1\,+\,\frac{1}{2}\,+\,\frac{1}{2}\,\times\,8~.
\end{equation}
The first term on the r.h.s. indicates that the flux vector $(1,1,2,1)$ also contributes a strictly-stable vector bundle in agreement with Table~\ref{tab:F1_tuples}. The second term indicates a new semi-stable vector bundle which agrees with the first line in~\eqref{semistabF1c100tuples}. As for the last term, it comes from dressing the $\sfq$ fixed point with a point-like instanton. The two new fixed vector bundles are indeed the ones anticipated in \eqref{semistabF1c100tuples}.

Similar reasoning can be followed for the contributions of the flux $(2,1,1,1)$. In the end, we find that the contribution of the strictly-stable fluxes in the unrefined limit is of the form $2\,\sfq^2\,+\,\cdots$ in agreement with~\eqref{Kool F1 c100}. Meanwhile, the contribution of the semi-stable sheaves is given by $\frac{1}{2}(4\,\sfq\,+\,40\,\sfq^2\,+\,\cdots)$ in agreement with \eqref{F0sssF1} upon dressing the latter with point-like instantons and including the $1/2$ factor for the semistable sheaves.
%%%%%%%%%%%%%%%%%%%%%%%%%%%%%%%%%%%%%%%%%
%%%%%%%%%%%%%%%%%%%%%%%%%%%%%%%%%%%%%%%%%
%%%%%%%%%%%%%%%%%%%%%%%%%%%%%%%%%%%%%%%%%

\subsubsection*{Acknowledgments}

We thank Mikhail Bershtein, Giulio Bonelli, Aleksei Bykov, Cyril Closset, Elias Furrer, Jan Manschot, Martijn Kool, Massimiliano Ronzani, Ekaterina Sysoeva, and Dirk van Bree for valuable discussions, and Ekaterina Sysoeva for sharing Mathematica code from the project \cite{Bonelli:2020xps}. 
AT is grateful to the F\'ed\'eration de Recherche Interactions Fondamentales (FRIF) for support during a visit at LPTHE in Fall 2025. 
AT is partly supported by the INFN Iniziativa Specifica GAST, Indam GNFM, the EU project Caligola HORIZON-MSCA-2021-SE-01; Project ID: 101086123, and CA21109 - COST Action CaLISTA.  
 \textit{For the purpose of Open Access, a CC-BY public copyright license has been applied by the authors to the present document and will be applied to all subsequent versions up to the Author Accepted Manuscript arising from this submission.}

%%%%%%%%%%%%%%%%%%%%%%%%%%%%%%%%%%%%%%%%%
%%%%%%%%%%%%%%%%%%%%%%%%%%%%%%%%%%%%%%%%%
%%%%%%%%%%%%%%%%%%%%%%%%%%%%%%%%%%%%%%%%%
\appendix
\section{Some explicit derivations}
In this Appendix, we derive some results that were quoted in the main text.

\subsection{Deriving the perturbative part of \texorpdfstring{\eqref{ZSU(N) full}}{ZSUN}}\label{app: perturbative part}
The perturbative part of the 5D Nekrasov partition function is given by~\cite[Equation (9.66)]{Kim:2025fpz},~\cite[Section 4.2 ]{Nakajima:2005fg},~\cite[Equation (1.28)]{Gottsche:2006bm}:
\begin{equation}\label{W-boson SU(N)}
    \log\,Z_{\rm pert}^{\,U(N)}[\mathbb{C}^2]\,=\, -\,\sum_{\substack{\alpha,\beta=1\\\alpha\neq\beta}}^{N}\,\widetilde{\gamma}_{\eps_1,\eps_2}(a_{\alpha,\beta}|\boldsymbol{\beta},\Lambda)~,
\end{equation}
with $a_{\alpha,\beta}\equiv a_\alpha-a_\beta$. The function $\widetilde{\gamma}$ is defined as:
\begin{equation}\label{tilde gamma}
\begin{split}
        \widetilde{\gamma}_{\eps_1,\eps_2}(a|\boldsymbol{\beta},\Lambda)\,:=\,&\gamma_{\eps_1,\eps_2}(a|\boldsymbol{\beta},\Lambda)\,+\,\frac{1}{\eps_1\,\eps_2} \left(\frac{\pi^2\,a}{6\,\boldsymbol{\beta}}\,-\,\frac{\zeta(3)}{\boldsymbol{\beta}^2}\right)\,\\
        &+\,\frac{\eps_1\,+\,\eps_2}{2\,\eps_1\,\eps_2}\left(a\,\log(\mathcal{R}) \,+\,\frac{\pi^2}{6\,\boldsymbol{\beta}}\right)+\frac{\eps_1^2\,+\,\eps_2^2\,+\,3\,\eps_1\,\eps_2}{12\,\eps_1\,\eps_2}\,\log(\mathcal{R})~,
\end{split}
\end{equation}
and, 
\begin{equation}\label{gamma}
\begin{split}
        \gamma_{\eps_1,\eps_2}(a|\boldsymbol{\beta}, \Lambda) \,:=\,&\frac{1}{2\,\eps_1\,\eps_2}\,\left(-\,\frac{\boldsymbol{\beta}}{6}\left(a\,+\,\frac{\eps_1\,+\,\eps_2}{2}\right)^3\,+\,a^2\,\log(\mathcal{R})\right)\\
        &+\sum_{n\geq 1}\frac{1}{n}\frac{x^n}{(y_1^{-n}\,-\,1)(y_2^{-n}\,-\,1)}~.
\end{split}
\end{equation}
Here, $\mathcal{R}$ is related to $\sfq$ by~\cite[Equation (9.71)]{Kim:2025fpz}:
\begin{equation}\label{q R relation}
    \mathsf{q} \,\equiv\, \mathcal{R}^4\,y_1\,y_2~.
\end{equation}
Recall from~\eqref{K to coh} that $y_{1,2}\equiv e^{-\boldsymbol{\beta}\,\eps_{1,2}}$. Our aim is to show
how these terms simplify when plugging back into~\eqref{W-boson SU(N)} and summing over the $\chi(S)$ affine patches of a compact toric surface $S$. We do this in two stages, first summing over the roots of the gauge group, and then taking the product over  affine patches.

\subsubsection*{Summing over the roots of the $U(N)$ gauge algebra} 
Examining first the sum over $\alpha,\beta$ in~\eqref{W-boson SU(N)} , we make the following observations:

\begin{itemize}
    \item Starting with the second and third terms appearing in~\eqref{tilde gamma}, we note that, upon taking the sum over $\alpha\neq\beta$, the terms involving the Coulomb branch parameters $a_{\alpha,\beta}$ cancel.\footnote{This is because $\sum_{\alpha\neq \beta} a_{\alpha,\beta}^n = 0$ for any odd integer $n$.\label{foosuman}} Therefore, we are left with:
    \begin{equation}
        {N\,(N\,-\,1)}\,\left(\frac{\pi^2}{6\,\boldsymbol{\beta}}\,\frac{\eps_1\,+\,\eps_2}{2\,\eps_1\,\eps_2}\,-\,\frac{\zeta(3)}{\boldsymbol{\beta}^2}\,\frac{1}{\eps_1\,\eps_2}\right)~.
    \end{equation}
    Furthermore, from~\eqref{toric-weights-ident}, these two terms cancel out when taking the product over affine patches.
    
    \item Using~\eqref{q R relation}, we can expand the last term in~\eqref{tilde gamma} as follows:
        \begin{align}
        \frac{\eps_1^2\,+\,\eps_2^2\,+\,3\,\eps_1\,\eps_2}{48\,\eps_1\,\eps_2}\,\log\sfq\,\,+\,\frac{\boldsymbol{\beta}}{4}\,\frac{\eps_1^2\,+\,\eps_2^2\,+\,3\,\eps_1\,\eps_2}{12\,\eps_1\,\eps_2}\,(\eps_1\,+\,\eps_2)~.
    \end{align}
    Pluggeing back into the sum over roots~\eqref{W-boson SU(N)}, this term gets multiplied by a factor of $-\,N(N-1)$.
     \item The second term appearing in~\eqref{gamma} can be simplified via~\eqref{q R relation} as follows:
    \begin{equation}
        \begin{split}
            \frac{a^2}{2\,\eps_1\,\eps_2}\log \mathcal{R}\,=\,\frac{a^2}{8\,\eps_1\,\eps_2}\,\log\sfq \,+\,\frac{\boldsymbol{\beta}\,a^2}{8\,\eps_1\,\eps_2}\,(\eps_1\,+\,\eps_2)~.
        \end{split}
    \end{equation}
    Note that, when plugging this back into the r.h.s. of~\eqref{W-boson SU(N)}, we get:
     \begin{equation}\label{second term contribution}
        \log\left( \sfq^{-\frac{\sum_{\alpha\neq\beta}\,a_{\alpha,\beta}^2}{8\,\eps_1\,\eps_2}} \right)\,-\,\frac{\boldsymbol{\beta}\,(\eps_1\,+\,\eps_2)}{8\,\eps_1\,\eps_2}\,\sum_{\alpha\neq\beta}\,a_{\alpha,\beta}^2~.
    \end{equation}
    Note that the first term is precisely the classical contribution appearing in~\cite[Equation (3.2)]{Bonelli:2020xps}.
    
     \item Let us now look at the first term in~\eqref{gamma}. As pointed out earlier (see footnote \ref{foosuman}), when taking the sum over the roots $\alpha\neq\beta$ in~\eqref{W-boson SU(N)}, only even powers of $a_{\alpha,\beta}$ contribute. Thus, this term gives us the following contribution:
    \begin{equation}
        {N\,(N\,-\,1)}\,\frac{\boldsymbol{\beta}}{6\,\eps_1\,\eps_2}\left(\frac{\eps_1\,+\,\eps_2}{2}\right)^3\,+\,\frac{\boldsymbol{\beta}\,(\eps_1\,+\,\eps_2)}{8\,\eps_1\,\eps_2}\,\sum_{\alpha\neq\beta}\,a_{\alpha,\beta}^2~,
    \end{equation}
    Note that the second term cancels against the one appearing in~\eqref{second term contribution}. 
\end{itemize}

To sum up, upon taking the sum over roots $\alpha\neq\beta$ of the gauge algebra,  the perturbative contribution~\eqref{W-boson SU(N)} becomes:
\begin{equation}\label{Zwbos C2 simplified}
\begin{split}
      \log\,Z_{\rm pert}^{\,U(N)}[\mathbb{C}^2]\,&=\, \left(-\,\frac{\sum_{\alpha\neq\beta}\,a_{\alpha,\beta}^2}{8\,\eps_1\,\eps_2}\,-\,{N\,(N\,-\,1)}\,\frac{\eps_1^2\,+\,\eps_2^2\,+\,3\,\eps_1\,\eps_2}{48\,\eps_1\,\eps_2}\right)\,\log\sfq\,\\&-\,\mathsf{F}_N[\mathbb{C}^2]\,+\,\log\,{\rm Exp}[ -\,\chi_N^{\mathbb{C}^2}(
      \{a_{\alpha,\beta}\},\eps_1,\eps_2)]~.
\end{split}
\end{equation}
Here, we defined
\begin{multline}\label{FNC2}
    \mathsf{F}_N[\mathbb{C}^2]\,:=\,{N\,(N\,-\,1)}\,\Bigg[\frac{\pi^2}{6\,\boldsymbol{\beta}}\,\frac{\eps_1\,+\,\eps_2}{2\,\eps_1\,\eps_2}\,-\,\frac{\zeta(3)}{\boldsymbol{\beta}^2}\,\frac{1}{\eps_1\,\eps_2}\,+\,\frac{\boldsymbol{\beta}}{6\,\eps_1\,\eps_2}\left(\frac{\eps_1\,+\,\eps_2}{2}\right)^3\\
    \,+\,\frac{\boldsymbol{\beta}}{4}\,\frac{\eps_1^2\,+\,\eps_2^2\,+\,3\,\eps_1\,\eps_2}{12\,\eps_1\,\eps_2}\,(\eps_1\,+\,\eps_2)\Bigg]~.
\end{multline}
while the last term in~\eqref{Zwbos C2 simplified} is 
the plethystic exponential of
\begin{equation}\label{chiC2N}
    \chi_N^{\mathbb{C}^2}(
\{a_{\alpha,\beta}\},,\eps_1,\eps_2)\,:=\,\sum_{\substack{\alpha,\beta=1\\\alpha\neq \beta}}^N\,\frac{x_{\alpha,\beta}}{(1\,-\,y_1^{-1})\,(1\,-\,y_2^{-1})}~,
\end{equation}
where $x_{\alpha,\beta} = e^{-\boldsymbol{\beta}\,a_{\alpha,\beta}}$, see~\eqref{K to coh}. Using the definition of the plethystic exponential in~\eqref{pexp defn}, we have, more explicitly
\be
{\rm Exp}[ -\,\chi_N^{\mathbb{C}^2}
(\{a_{\alpha,\beta}\},\eps_1,\eps_2)]
= \exp\left(- \sum_{n\geq 1} \frac{1}{n}
\sum_{\substack{\alpha,\beta=1\\\alpha\neq \beta}}^N\,\frac{x^n_{\alpha,\beta}}{(1\,-\,y_1^{-n})\,(1\,-\,y_2^{-n})}
\right)~.
\ee

\medskip
\noindent
\textbf{Comparing with the 4D $1$-loop contribution.} Recall that, in the 4D $\mathcal{N}=2$ theory, the 1-loop contribution is given in terms of the Barnes function $\Gamma_2(x)$ -- see e.g.~\cite[App. B]{Bonelli:2020xps} for its definition and properties. More explicitly, the 4D 1-loop partition function is given by~\cite[Equation (3.3)]{Bonelli:2020xps}:
\begin{equation}
    Z_{{\rm 1-loop,\,W}}^{{\rm 4D}, \,SU(N)}[\mathbb{C}^2]\,=\,\prod_{\substack{\alpha, \beta=1\\\alpha\neq\beta}}^N\,\Gamma_2(a_{\alpha,\beta})^{-1}~.
\end{equation}
Assuming that the equivariant parameters ${\rm Re}(\eps_1), {\rm Re}(\eps_2)$ are positive, one can show that~\cite[Equation (B.3)]{Bonelli:2020xps}
\begin{equation}
   \Gamma_2(a_{\alpha,\beta})^{-1}\,=\,\prod_{i,j\geq 0}(a_{\alpha,\beta}\,+\,i\,\eps_1\,+\,j\,\eps_2)~.
\end{equation}
From this expression, one can deduce the corresponding K-theoretic uplift -- i.e., the corresponding expression in the 5D $\mathcal{N}=1$ theory. In our convention, this amounts to considering the following expression:
\begin{equation}
\begin{split}
    Z_{\rm 1-loop}^{\rm 5D}[\mathbb{C}^2]\,&=\,\prod_{\substack{\alpha, \beta=1\\\alpha\neq\beta}}^N\,\prod_{i,j\geq 0}\frac{2\sinh\left(\frac{\boldsymbol{\beta}}{2}\,(a_{\alpha,\beta}\,+\,i\,\eps_1\,+\,j\,\eps_2)\right)}{\exp\left(-\,\frac{\boldsymbol{\beta}}{2}\,(a_{\alpha,\beta}\,+\,i\,\eps_1\,+\,j\,\eps_2)\right)}\\
    &=\,\prod_{\substack{\alpha, \beta=1\\\alpha\neq\beta}}^N\,\prod_{i,j\geq 0} \left(\,1\,-\,x_{\alpha,\beta}^{2}\,y_1^{-i}\,y_2^{-j}\,\right)~.
\end{split}
\end{equation}
Note that the last expression reproduces the plethystic exponential of~\eqref{chiC2N} in the regime where ${\rm Re}(\eps_1), {\rm Re}(\eps_2)\geq 0$. A similar analysis works for
other choices of $\sgn({\rm Re}(\eps_i))$.  
This elucidates the definition given in~\eqref{Z Wbos C2 SUN}.

\subsubsection*{Gluing along the affine patches}
Let us now turn to the second step, namely, taking the
product over the affine patches of the toric 
surface $S$. Note that:
\begin{itemize}
    \item From the second term in~\eqref{Zwbos C2 simplified} we get:
    \begin{equation}\label{extraPower}
        -\,{N\,(N\,-\,1)}\,\sum_{\ell=1}^{\chi}\left(\frac{\eps_{1,\ell}^2\,+\,\eps_{2,\ell}^2\,+\,3\,\eps_{1,\ell}\,\eps_{2,\ell}}{48\,\eps_{1,\ell}\,\eps_{2,\ell}}\right)\,\log\sfq\,=\,-\frac{N\,(N\,-\,1)}{4}\,\log\sfq~,
    \end{equation}
    where we used the identities~\eqref{sum inter numbs} and 
  the second line in~\eqref{toric-weights-ident}.
    \item The first and second terms in~\eqref{FNC2} cancel out due to the first identity in~\eqref{toric-weights-ident}.
    \item The last two terms in~\eqref{FNC2} give the following contribution:
    \begin{equation}
        {N\,(N\,-\,1)}\,\frac{\boldsymbol{\beta}}{24}\,\sum_{\ell=1}^\chi\,\left(\frac{\eps_{1,\ell}^2}{\eps_{2,\ell}}\,+\,\frac{\eps_{2,\ell}^2}{\eps_{1,\ell}}\right)\,=\,-\,\frac{N\,(N\,-\,1)\,\boldsymbol{\beta}}{24}\,\sum_{\ell=1}^\chi\,h_{\ell}\,(\eps_{1,\ell}\,+\,\eps_{2,\ell-1})~,
    \end{equation}
   where we used the recursive property~\eqref{defn toric weights} of the equivariant weights.
\end{itemize}

\subsubsection*{Putting everything together}
Based on these observations, we can write down the full perturbative part of the partition function on a toric surface $S$ as follows:
\begin{equation}\label{pert Z S SUN}
    Z_{\rm pert}^{\,U(N)}[S]\,=\,Z_{\rm class}^{\,U(N)}\,\times\,e^{\mathsf{F}_N[S]}\,\times\, Z_{\rm 1-loop,\,W}^{\,U(N)}[S]~,
\end{equation}
where
\begin{equation}\label{log Zclass S SUN}
    \log Z_{\rm class}^{\,U(N)}[S]\,:=\,\left(\,-\,\frac{N\,(N\,-\,1)}{4}\,-\,\sum_{\ell=1}^\chi\frac{\sum_{\alpha\neq\beta}\,(a^{(\ell)}_{\alpha,\beta})^2}{8\,\eps_{1,\ell}\,\eps_{2,\ell}}\right)\,\log\sfq~.
\end{equation}
The Coulomb branch parameters $a_{\alpha,\beta}^{(\ell)}$ are given in~\eqref{a ell alpha beta}. The second term in~\eqref{pert Z S SUN} is defined by 
\begin{equation}\label{FNS}
    {\sf F}_N[S]\,:=\,\sum_{\ell=1}^\chi\,{\sf F}_N[\mathbb{C}^2_\ell]\,=\,\frac{N\,(N\,-\,1)\,\boldsymbol{\beta}}{24}\,\sum_{\ell=1}^\chi\,h_{\ell}\,(\eps_{1,\ell}\,+\,\eps_{2,\ell-1})~.
\end{equation}
Note that this is an overall factor depending only on the equivariant parameters $y_{1,\ell}$ and $y_{2,\ell}$, which we omit in our calculations in the main text. As for the last term in~\eqref{pert Z S SUN}, it arises from the adjoint W-bosons in the vector multiplet, and is of the explicit form:  
\begin{equation}\label{W-bos SUN S}
    Z_{\rm 1-loop,\,W}^{\,U(N)}[S]\,:=\,{\rm Exp}[-\,\chi^S_N(x,y_1,y_2)]~,
\end{equation}
where $\chi_N^S$ is defined in terms of $\chi_N^{\mathbb{C}^2}$~\eqref{chiC2N} as:
\begin{eqnarray}\label{chiSN}
    \chi_N^S(x,y_1, y_2)\,:=\,\sum_{\ell=1}^\chi \chi_N^{\mathbb{C}^2}(x_\ell,y_{1,\ell}, y_{2,\ell})~.
\end{eqnarray}
Here, $x_\ell$ stands for $\{x_{\alpha,\beta}^{(\ell)}\}$.

\subsection{Deriving the overall \texorpdfstring{$\mathsf{Z}$}{Z} factor in \texorpdfstring{\eqref{ZU(2) decomp}}{ZU2decomp}}\label{app:Z factor}
To obtain the overall factor $\mathsf{Z}$ in~\eqref{ZU(2) decomp}, we follow the same line of reasoning as in~\cite[Section 4.2]{Sysoeva:2022syp} and 
study the asymptotic behavior of the 
instanton partition function ~\eqref{K inst adj U(N)} 
in the limit $x_1\ll x_2$, restricting to rank $N=2$ 
for simplicity. From~\eqref{adj K instanton U(N)}, we have:
\begin{equation}\label{start point}
    Z_{\rm inst,\,adj}^{ SU(2)}[\mathbb{C}^2]\,=\,\sum_{\lambda_1,\lambda_2}\,\mathsf{q}^{|\lambda_1|\,+\,|\lambda_2|}\,Z_{\vec \lambda,\, {\rm adj}}^{\,U(2)}~.
\end{equation}
Each contribution on the r.h.s. decomposes as a product of 4 factors:
\begin{equation}\label{prod Z alpha beta}
    Z_{\vec \lambda,\, {\rm adj}}^{\,U(2)}\,=\, Z_{\lambda_1, \lambda_2}^{1,1}\,Z_{\lambda_1, \lambda_2}^{1,2}\,Z_{\lambda_1, \lambda_2}^{2,1}\,Z_{\lambda_1, \lambda_2}^{2,2}~,
\end{equation}
with
\begin{equation}\label{Z alpha beta}
    Z_{\lambda_1,\lambda_2}^{\alpha,\beta}\,:=\,\prod_{\square\in \lambda_\alpha}\, \frac{\left(\,1\,-\,\frac{x_\beta}{x_\alpha}\,\yadj\,y_1^{-{\rm L}_\beta(\square)}\,y_2^{{\rm A}_\alpha(\square)+1}\,\right)}{\left(\,1\,-\,\frac{x_\beta}{x_\alpha}\,y_1^{-{\rm L}_\beta(\square)}\,y_2^{{\rm A}_\alpha(\square)+1}\,\right)}\,\prod_{\square\in \lambda_\beta}\,\frac{\left(\,1\,-\,\frac{x_\beta}{x_\alpha}\,\yadj\,y_1^{{\rm L}_\alpha(\square)+1}\,y_2^{-{\rm A}_\beta(\square)}\,\right)}{\left(\,1\,-\,\frac{x_\beta}{x_\alpha}\,y_1^{{\rm L}_\alpha(\square)+1}\,y_2^{-{\rm A}_\beta(\square)}\,\right)}~.
\end{equation}
We now observe that the first and last factors in~\eqref{prod Z alpha beta}, having $\alpha=\beta$, are independent of the Coulomb branch parameters, and thus are preserved upon taking the limit   $x_1/x_2\,\to\,0$. Meanwhile the third factor trivializes,  $Z_{\lambda_1,\lambda_2}^{2,1}\,\to\,1$, while the second factor simplifies to:
\begin{equation}
    Z_{\lambda_1, \lambda_2}^{1,2}\,\longrightarrow\,\prod_{\square\in\lambda_1} \yadj\,\prod_{\square\in\lambda_2}\,\yadj \,=\,\yadj^{|\lambda_1|\,+\,|\lambda_2|}~.
\end{equation}
Combining these observations, we find that~\eqref{start point}
reduce sto 
\begin{equation}
    Z_{\rm inst,\,adj}^{\,U(2)}\,\xrightarrow{\frac{x_1}{x_2}\to 0} \,\sum_{\lambda_1,\lambda_2}\,(\yadj\,\mathsf{q})^{|\lambda_1|\,+\,|\lambda_2|}\,Z_{\lambda_1,\lambda_2}^{1,1}\,Z_{\lambda_1,\lambda_2}^{2,2}~.
\end{equation}
Inserting the expression of $Z_{\lambda_1,\lambda_2}^{\alpha, \alpha}$ in terms of  $\square\in\lambda_\alpha$, we can write the above limit as:
\begin{equation}
\begin{split}
     \left[\sum_{\lambda}\,(\yadj\,\mathsf{q})^{|\lambda|} \,\prod_{\square\in \lambda} \frac{ \left(\,1\,-\,\yadj\,y_1^{-{\rm L}_\lambda(\square)}\,y_2^{{\rm A}_\lambda(\square)+1}\,\right)}{ \left(\,1\,-\,y_1^{-{\rm L}_\lambda(\square)}\,y_2^{{\rm A}_\lambda(\square)+1}\,\right)} \,\frac{\left(\,1\,-\,\yadj\,y_1^{{\rm L}_\lambda(\square)+1}\,y_2^{-{\rm A}_\lambda(\square)}\,\right)}{\left(\,1\,-\,y_1^{{\rm L}_\lambda(\square)+1}\,y_2^{-{\rm A}_\lambda(\square)}\,\right)}\right]^2~.\\
    \end{split}
\end{equation}
Upon using the identity relating~\eqref{K inst U(1)} to~\eqref{Zinst K U(1) C2}, we recognize the above limit as the expression written earlier in~\eqref{prefactor for ZU(2)}. 

\subsection{Deriving the final expression in \texorpdfstring{\eqref{ZFullRes}}{(ZFull)}}\label{app:ZfullRes}
 In order to derive \eqref{ZFullRes}, 
 we start by analyzing the structure of the perturbative and instanton contributions appearing in~\eqref{rank 2 full partition} for $S = \mathbb{P}^2$. Next, we extract the residue of the full partition function at $a=0$ to get the final formula quoted in~\eqref{ZFullRes}.

\subsubsection*{The analytic structure of the perturbative part}
Let us start with the 1-loop contribution coming from the W-bosons and the massive adjoint hypermultiplet, given by~\eqref{rank 2 full partition}:
\begin{equation}\label{Z pert P2}
    Z_{\rm 1-loop,\,adj}^{\,U(2)}[\mathbb{P}^2]\mid_{\vec{\sfp}}\,=\, {\rm Exp}\left[(\yadj-1)\,\chi_{\mathbb{P}^2}(x,y_1,y_2)\,\right]~,
\end{equation}
with $\chi_{\mathbb{P}^2}(x,y_1,y_2)$ given by~\eqref{chiC2N}:
\begin{equation}
    \chi_{\mathbb{P}^2}(x,y_1,y_2)\,=\,\sum_{\ell=1}^{3}\,\frac{x_\ell^2\,+\,x_\ell^{-2}}{(1\,-\,y_{1,\ell}^{-1})\,(1\,-\,y_{2,\ell}^{-1})}~,
\end{equation}
where the equivariant parameters in the various patches are listed in Table \ref{tab:toricDataP2}. On the l.h.s. of~\eqref{Z pert P2}, we indicate the dependence, via \eqref{2 a ell}, on the flux vector $\vec{\sfp}\in\mathbb{Z}^3$. 
% satisfying the stability conditions~\eqref{stab cond for P2}.

Following the analysis performed in~\cite[Section 3.2]{Bershtein:2015xfa}, the plethystic exponential in \eqref{Z pert P2} can be written as a finite product depending on the total sum $|\vec{\sfp}|:=\sfp_1+\sfp_2+\sfp_3$ as follows~\cite[Equation (3.24)--(3.26)]{Bershtein:2015xfa},~\cite[Equation (K.16)--(K.17)]{Kim:2025fpz}:
\begin{multline}\label{Exp-chiP2 explicit}
    {\rm Exp}[-\,\chi_{\mathbb{P}^2}(x,y_1,y_2)]\,=\\\begin{cases}
        \mathsf{G}^{(+)}_{\sfp_1,\sfp_2}\,\times \,\mathsf{G}_{-\sfp_1,-\sfp_2}^{(-)}~, \qquad &|\vec{\sfp}|\,=\,0~,\\
        \prod_{\substack{(i,j)\in\mathbb{N}^2\\i+j\leq|\vec{\mathsf{p}}|}}\,\mathsf{G}_{\sfp_1-j,\sfp_2-i}^{(+)}\,\times\,\prod_{\substack{(i,j)\in\mathbb{N}^2\\i+j\leq |\vec{\mathsf{p}}|-3}}\,\mathsf{G}^{(-)}_{1+j-\sfp_1,1+i-\sfp_2}~, \qquad &|\vec{\sfp}|\,>\,0~,\\
        \prod_{\substack{(i,j)\in\mathbb{N}^2\\i+j\leq||\vec{\mathsf{p}}||}}\mathsf{G}^{(-)}_{-\sfp_1-j,-\sfp_2-i}\,\times\,\prod_{\substack{(i,j)\in\mathbb{N}^2\\i+j\leq ||\vec{\mathsf{p}}||-3}}\,\mathsf{G}^{(+)}_{1+j+\sfp_1,1+i+\sfp_2}~, \quad&|\vec{\sfp}|\,<\,0~,
    \end{cases}
\end{multline}
with $\mathsf{G}^{(\pm)}_{i,j}$ defined in~\eqref{G+-}, and we introduced the absolute value of the toal sum $||\vec{\sfp}||:=|\sfp_1+\sfp_2+\sfp_3|$. Indeed, in the 4D limit~\eqref{K to coh}, we recover the corresponding expressions in~\cite{Bershtein:2015xfa} at leading order in $\boldsymbol{\beta}$. It is straightforward to apply this analysis for the contribution coming from the massive adjoint hypermultiplet in~\eqref{Z pert P2}.

Let us now analyze the order of the zero or pole of~\eqref{Z pert P2} at $a=0$. From the above explicit expressions, we see that the perturbative part can only develop a zero at $a=0$ with order depending on the flux vector $\vec{\sfp}$. Following the discussion at the beginning of subsection \ref{subsec:SUSYLocalP2}, we will focus on the case where $|\vec{\sfp}|\geq 0$ and $\sfp_1,\sfp_2,\sfp_3\in\mathbb{Z}_{\geq 0}$. In this case, we see that the order of the zero at $a=0$ is:
\begin{equation}\label{deg of zero P2}
    \operatorname*{ord}_{a=0}\,Z_{\rm 1-loop,\,adj}^{\,U(2)}[\mathbb{P}^2]\mid_{\vec{\sfp}}\,=\,\begin{cases}
        1~, \qquad &\text{if }~\zeta_{\vec{\mathsf{p}}} \,=\,1\,,\,2\,~,\\
        2~, \qquad &\text{if }~\zeta_{\vec{\sfp}}\,=\,0\,,\,3~,
    \end{cases}
\end{equation}
where $\zeta_{\vec{\sfp}}$ here denotes the number of zero components in the flux vector $\vec{\sfp}$. Indeed, for 
zero flux one can see from~\eqref{Exp-chiP2 explicit} that the W-boson perturbative contribution for  is of the form:
\begin{eqnarray}
     {\rm Exp}[-\,\chi_{\mathbb{P}^2}(x,y_1,y_2)]\mid_{(0,0,0)}\,=\,-\,\frac{(x^2\,-\,1)^2}{x^2}~,
\end{eqnarray}
giving a zero of order 2. 
For the case with $\zeta_{\vec\sfp} = 0$, the same factor appears
in the plethystic exponential, coming from the term $(i,j) = (\sfp_2, \sfp_1)$ of the first product and $(i,j) = (\sfp_2-1,\sfp_1-1)$ of the second product in the corresponding expression in~\eqref{Exp-chiP2 explicit}.

\subsubsection*{The analytic structure of the instanton part}
Let us now do a similar analysis for the zeros and poles of the instanton part of the partition function~\eqref{rank 2 full partition}:
\begin{equation}\label{Zinst U(2) P2}
    Z_{\rm inst,\,adj}^{\,U(2)}[\mathbb{P}^2]\mid_{\vec{\sfp}}\,=\,\prod_{\ell=1}^3 Z_{\rm inst,\,adj}^{\,U(2)}[\mathbb{C}^2_\ell]\mid_{\vec{\sfp}}~,
\end{equation}
where $\mathbb{C}^2_\ell$ is to indicate that one needs to use the corresponding equivariant parameters appearing in Table \ref{tab:toricDataP2}. We now expand each term in the product using the recurrence relation~\eqref{HKadj recurrence}:
\begin{equation}\label{Zinst P2 explicit}
    \begin{split}
        Z_{\rm inst}^{{\rm K}, \,U(2)}[\mathbb{P}^2]\,&=\,\mathsf{Z}_1^2\,\left(1\,-\,\sum_{m,n=1}^\infty \frac{\mathsf{q}^{mn}\,T^{\rm adj}_{m,n}(y_{1,1}, y_{2,1})\,H^{\rm K}(y_{1,1}^{\frac{m}{2}}y_{2,1}^{-\frac{n}{2}})}{(y_{1,1}\,y_{2,1})^{mn}\,(x_1^{-2}\,-\,y_{1,1}^m\,y_{2,1}^n)\,(1\,-\,x_1^2\,y_{1,1}^{-m}\,y_{2,1}^{-n})}\right)\\
        &\times\,\mathsf{Z}_2^2\,\left(1\,-\,\sum_{m,n=1}^\infty \frac{\mathsf{q}^{mn}\,T^{\rm adj}_{m,n}(y_{1,2}, y_{2,2})\,H^{\rm K}(y_{1,2}^{\frac{m}{2}}y_{2,2}^{-\frac{n}{2}})}{(y_{1,2}\,y_{2,2})^{mn}\,(x_2^{-2}\,-\,y_{1,2}^m\,y_{2,2}^n)\,(1\,-\,x_2^2\,y_{1,2}^{-m}\,y_{2,2}^{-n})}\right)\\
        &\times\,\mathsf{Z}_3^2\,\left(1\,-\,\sum_{m,n=1}^\infty \frac{\mathsf{q}^{mn}\,T^{\rm adj}_{m,n}(y_{1,3}, y_{2,3})\,H^{\rm K}(y_{1,3}^{\frac{m}{2}}y_{2,3}^{-\frac{n}{2}})}{(y_{1,3}\,y_{2,3})^{mn}\,(x_3^{-2}\,-\,y_{1,3}^m\,y_{2,3}^n)\,(1\,-\,x_3^2\,y_{1,3}^{-m}\,y_{2,3}^{-n})}\right)~.
    \end{split}
\end{equation}
Here the overall factors $\mathsf{Z}_\ell$ are defined as in~\eqref{prefactor for ZU(2)}:
\begin{equation}
    \mathsf{Z}_\ell \,:= \,{\rm Exp}\left[\frac{\yadj\,\mathsf{q}\,(1\,-\,\yadj\,y_{1,\ell})\,(1\,-\,\yadj\,y_{2,\ell})}{(1\,-\,y_{1,\ell})\,(1\,-\,y_{2,\ell})\,(1\,-\,\yadj^2\,\mathsf{q})}\right]~.
\end{equation}

For a fixed flux vector $\vec{\sfp}\in\mathbb{Z}_{\geq 0}^3$, we see from~\eqref{Zinst P2 explicit} that the instanton partition function can develop a pole at $a=0$ with maximal degree depending on the number of zero entries of the flux vector:
\begin{equation}\label{deg of pole P2}
  \operatorname*{ord}_{a=0}\, Z_{\rm inst}^{\,U(2)}[\mathbb{P}^2]\,=\, \begin{cases}
        0~, \qquad &\text{if }~\zeta_{\vec{\sfp}}\,=\,{2\,,\,3}~,\\
        -\,1~, \qquad &\text{if }~\zeta_{\vec{\sfp}}\,=\,1~,\\
        -\,3~, \qquad &\text{if }~\zeta_{\vec{\sfp}}\,=\,0~.
    \end{cases}
\end{equation}
In particular, for the last case, this pole can be obtained by looking at the terms with $(m,n)= (\sfp_1,\sfp_2),(\sfp_2,\sfp_3),(\sfp_3,\sfp_1)$ of the three sums appearing in~\eqref{Zinst P2 explicit} respectively. If one of the components, say $\sfp_2$, vanishes, then the first two choices for $(m,n)$ are not possible anymore since the sums run over $m,n\geq 1$.

In the generic case $\zeta_{\vec{\sfp}}=0$, the instanton partition function~\eqref{Zinst P2 explicit} can be expanded into the following form:
\begin{equation}\label{zinst P2 at x2=1}
    Z_{\rm inst}^{\,U(2)}[\mathbb{P}^2]\,=\,\frac{\mathsf{q}^{\sum_{\ell}\mathsf{p}_\ell\mathsf{p}_{\ell+1}}}{(x^2\,-\,1)^3}\,\prod_{\ell=1}^3\frac{\mathsf{Z}_\ell^2\,T_{\mathsf{p}_\ell,\mathsf{p}_{\ell+1}}^{\rm adj}(y_{1,\ell},y_{2,\ell})\,H^{\rm K}(y_{1,\ell}^{\frac{\mathsf{p}_\ell}{2}}y_{2,\ell}^{-\frac{\mathsf{p}_{\ell+1}}{2}})}{(y_{1,\ell}\,y_{2,\ell})^{\mathsf{p}_\ell\mathsf{p}_{\ell+1}}\left(y_{1,\ell}^{-\mathsf{p}_\ell}\,y_{2,\ell}^{-\mathsf{p}_{\ell+1}}\,-\,y_{1,\ell}^{\mathsf{p}_\ell}\,y_{2,\ell}^{\mathsf{p}_{\ell+1}}\right)} \, \,+\,\cdots~,
\end{equation}
with the lapses denoting terms of higher order in $(1-x^2)$.

\subsubsection*{Extracting the residue at \texorpdfstring{$a=0$}{x21}}
Let us now combine our observations together. According to~\eqref{deg of zero P2} and~\eqref{deg of pole P2}, 
the residue of the full partition function of the $U(2)$ theory on $\mathbb{P}^2$ is non-trivial for the cases where $\zeta_{\vec\sfp} = 0$. In such a case, the full partition function has a simple pole at $a=0$ and one can straightforwardly evaluate the residue in~\eqref{ZP2 residue}. Doing so, we indeed end up with the expression appearing on the r.h.s. of~\eqref{ZFullRes}.

%%%%%%%%%%%%%%%%%%%%%%%%%%%%%%%%%%%%%%%%%
%%%%%%%%%%%%%%%%%%%%%%%%%%%%%%%%%%%%%%%%%
%%%%%%%%%%%%%%%%%%%%%%%%%%%%%%%%%%%%%%%%%

\section{Explicit results for refined VW invariants of \texorpdfstring{$\mathbb{F}_0$}{F0}}\label{app:F0Results}

Here, we record the explicit expressions for the  orbit contributions of the different magnetic fluxes $\vec{\sfp}$  to the first four terms in the refined VW invariants of $\mathbb{F}_0$ for the $3$ different choices of the first Chern class $c_1=(0,1),(1,1)$ and $(0,0)$, restricting to the non-equivariant limit for simplicity. In all cases, we choose the polarization to be 
$(\boldsymbol{a},\boldsymbol{b}) = (1.1,1)$.

\subsection{For \texorpdfstring{$c_1 = (0,1)$}{c101}}\label{subsec:F0 c101 fixed points}

\begin{center}
    \renewcommand{\arraystretch}{2.5}
    \begin{longtable}[!h]{|c||c|c|c|c|c|}
    \hline
    $\vec{\sfp}$& $\sfq$ &$\sfq^2$ &$\sfq^3$& $\sfq^4$  \\
    \hline
    \hline
      $(1,0,1,1)$ &$\yadj$ & \parbox{4cm}{\centering$\yadj\,+\,2\,\yadj^2\,+\,3\,\yadj^3\,+\,\yadj^4\,+\,\yadj^5$} & \parbox{4cm}{\centering$\yadj\,+\,2\,\yadj^2\,+\,7\,\yadj^3\,+\,9\,\yadj^4\,+\,11\,\yadj^5\,+\,7\,\yadj^6\,+\,5\,\yadj^7\,+\,\yadj^8\,+\,\yadj^9$}&\parbox{5cm}{\centering$y_{\mathrm{adj}} 
\,+\, 2\,y_{\mathrm{adj}}^{2} 
\,+\, 7\,y_{\mathrm{adj}}^{3} 
\,+\, 15\,y_{\mathrm{adj}}^{4} 
\,+\, 28\,y_{\mathrm{adj}}^{5} 
\,+\, 35\,y_{\mathrm{adj}}^{6} 
\,+\, 39\,y_{\mathrm{adj}}^{7} 
\,+\, 29\,y_{\mathrm{adj}}^{8} 
\,+\, 20\,y_{\mathrm{adj}}^{9} 
\,+\, 9\,y_{\mathrm{adj}}^{10} 
\,+\, 5\,y_{\mathrm{adj}}^{11} 
\,+\, y_{\mathrm{adj}}^{12} 
\,+\, y_{\mathrm{adj}}^{13}$}\\
%%%%%%%%%%%
\hline
$(1,1,1,0)$ &1&\parbox{4cm}{\centering$1 \,+\, \yadj\, +\, 3\,\yadj^{2}\, +\, 2\,\yadj^{3}\, +\, \yadj^{4}$}&\parbox{4cm}{\centering$1 \,+\, \yadj \,+\, 5\,\yadj^{2}\, +\, 7\,\yadj^{3} \,+\, 11\,\yadj^{4} \,+\, 9\,\yadj^{5}\, +\, 7\,\yadj^{6}\, +\, 2\,\yadj^{7}\, +\, \yadj^{8}$}&\parbox{5cm}{\centering$1\, +\, \yadj \,+\, 5\,\yadj^{2} \,+\, 9\,\yadj^{3} \,+\, 20\,\yadj^{4} \,+\, 29\,\yadj^{5} \,+\, 39\,\yadj^{6}\, +\, 35\,\yadj^{7} \,+\, 28\,\yadj^{8}\, +\, 15\,\yadj^{9}\, +\, 7\,\yadj^{10}\, +\, 2\,\yadj^{11} \,+\, \yadj^{12}$}\\
%%%%%%%%%%%
\hline
$(2,0,2,1)$ & $0$ & $\yadj^4$ &\parbox{4cm}{\centering$\yadj^{4}\, +\, 2\,\yadj^{5} \,+\, 3\,\yadj^{6}\, +\, \yadj^{7} \,+\, \yadj^{8}$} &\parbox{5cm}{\centering$\yadj^{4}\, +\, 2\,\yadj^{5}\, +\, 7\,\yadj^{6} \,+\, 9\,\yadj^{7} \,+\, 11\,\yadj^{8} \,+\, 7\,\yadj^{9} \,+\, 5\,\yadj^{10} \,+\, \yadj^{11} \,+\, \yadj^{12}$}\\
%%%%%%%%%
\hline
$(2,1,2,0)$&$0$&$\yadj$&\parbox{4cm}{\centering $\yadj \, + \, \yadj^{2} \, + \, 3\,\yadj^{3} \, + \, 2\,\yadj^{4} \, + \, \yadj^{5}$}&\parbox{5cm}{\centering$\yadj \, + \, \yadj^{2} \, + \, 5\,\yadj^{3} \, + \, 7\,\yadj^{4} \, + \, 11\,\yadj^{5} \, + \, 9\,\yadj^{6} \, + \, 7\,\yadj^{7} \, + \, 2\,\yadj^{8} \, + \, \yadj^{9}$}\\
%%%%%%%%%
\hline
$(1,1,1,2)$ &$0$&$\yadj^2\,+\,\yadj^3$&\parbox{4cm}{\centering$\yadj^{2} \, + \, 2\,\yadj^{3} \, + \, 5\,\yadj^{4} \, + \, 6\,\yadj^{5} \, + \, 2\,\yadj^{6}$}&\parbox{5cm}{\centering$\yadj^{2} \, + \, 2\,\yadj^{3} \, + \, 7\,\yadj^{4} \, + \, 14\,\yadj^{5} \, + \, 23\,\yadj^{6} \, + \, 24\,\yadj^{7} \, + \, 13\,\yadj^{8} \, + \, 4\,\yadj^{9}$}\\
%%%%%%%%%
\hline
$(1,2,1,1)$&$0$&$\yadj^2\,+\,\yadj^3$&\parbox{4cm}{\centering$2\,\yadj^{3} \, + \, 6\,\yadj^{4} \, + \, 5\,\yadj^{5} \, + \, 2\,\yadj^{6} \, + \, \yadj^{7}$}&\parbox{5cm}{\centering$4\,\yadj^{4} \, + \, 13\,\yadj^{5} \, + \, 24\,\yadj^{6} \, + \, 23\,\yadj^{7} \, + \, 14\,\yadj^{8} \, + \, 7\,\yadj^{9} \, + \, 2\,\yadj^{10} \, + \, \yadj^{11}$}\\
%%%%%%%%%
\hline
$(0,1,2,2)$ &$0$&$0$&$\yadj^4$&\parbox{5cm}{\centering$\yadj^{4} \, + \, \yadj^{5} \, + \, 4\,\yadj^{6} \, + \, 2\,\yadj^{7}$}\\
%%%%%%%%%
\hline
$(2,2,0,1)$&$0$&$0$&$\yadj^5$&\parbox{5cm}{\centering$2\,\yadj^{6} \, + \, 4\,\yadj^{7} \, + \, \yadj^{8} \, + \, \yadj^{9}$}\\
%%%%%%%%%
\hline
$(0,2,2,1)$ &$0$&$0$&$\yadj^6$& \parbox{5cm}{\centering$2\,\yadj^{7} \, + \, 4\,\yadj^{8} \, + \, \yadj^{9} \, + \, \yadj^{10}$}\\
%%%%%%%%%
\hline
$(2,1,0,2)$&$0$&$0$&$\yadj^3$&\parbox{5cm}{\centering$\yadj^{3} \, + \, \yadj^{4} \, + \, 4\,\yadj^{5} \, + \, 2\,\yadj^{6}$}\\
%%%%%%%%%
\hline
$(3,0,3,1)$&$0$&$0$&$\yadj^7$&\parbox{5cm}{\centering$\yadj^{7} \, + \, 2\,\yadj^{8} \, + \, 3\,\yadj^{9} \, + \, \yadj^{10} \, + \, \yadj^{11}$}\\
%%%%%%%%%
\hline
$(3,1,3,0)$ &$0$&$0$&$\yadj^2$&\parbox{5cm}{\centering$\yadj^{2} \, + \, \yadj^{3} \, + \, 3\,\yadj^{4} \, + \, 2\,\yadj^{5} \, + \, \yadj^{6}$}\\
%%%%%%%%%
\hline
$(1,2,1,3)$ &$0$ &$0$ &$\yadj^4\,+\,\yadj^5$ &\parbox{5cm}{\centering$2\,\yadj^{5} \, + \, 6\,\yadj^{6} \, + \, 6\,\yadj^{7} \, + \, 2\,\yadj^{8}$}\\
%%%%%%%%%
\hline
$(1,3,1,2)$ &$0$ &$0$ &$\yadj^4\,+\,\yadj^5$ &\parbox{5cm}{\centering$2\,\yadj^{5} \, + \, 6\,\yadj^{6} \, + \, 6\,\yadj^{7} \, + \, 2\,\yadj^{8}$}\\
%%%%%%%%%
\hline
$(4,0,4,1)$ &$0$&$0$&$0$&$\yadj^{10}$\\
%%%%%%%%%
\hline
$(4,1,4,0)$ &$0$&$0$&$0$ &$\yadj^3$\\
%%%%%%%%%
\hline
$(1,1,3,2)$&$0$&$0$&$0$&$\yadj^5$\\
%%%%%%%%%
\hline
$(3,2,1,1)$&$0$&$0$&$0$&$\yadj^8$\\
%%%%%%%%%
\hline
$(1,2,3,1)$&$0$&$0$&$0$&$\yadj^9$\\
%%%%%%%%%
\hline
$(3,1,1,2)$&$0$&$0$&$0$&$\yadj^4$\\
%%%%%%%%%
\hline
$(2,1,2,2)$&$0$&$0$&$0$&$\yadj^7\,+\,\yadj^8$\\
\hline
$(2,2,2,1)$&$0$&$0$&$0$&$\yadj^5\,+\,\yadj^6$\\
%%%%%%%%%
% \hline
% $(3,1,1,2)$&$0$&$0$&$0$&$\yadj^4$\\
%%%%%%%%%
% %%%%%%%%%
% \hline
% $(3,1,1,2)$&$0$&$0$&$0$&$\yadj^8$\\
%%%%%%%%%
\hline
$(1,3,1,4)$&$0$&$0$&$0$&$\yadj^6\,+\,\yadj^7$\\
%%%%%%%%%
\hline
$(1,4,1,3)$&$0$&$0$&$0$&$\yadj^6\,+\,\yadj^7$\\
\hline
\end{longtable}
\end{center}
%%%%%%%%%%%%%%%%%%%%%%%%%%%%%%%%%%%%%%%%%%%%%%%%%%%%%%
%%%%%%%%%%%%%%%%%%%%%%%%%%%%%%%%%%%%%%%%%%%%%%%%%%%%%%
\subsection{For \texorpdfstring{$c_1 = (1,1)$}{c100}}\label{app:F0 c1=(1,1)}
\begin{center}
    \renewcommand{\arraystretch}{2.5}
    \begin{longtable}[!h]{|c||c|c|c|c|c|}
    \hline
    $\vec{\sfp}$& $\sfq^2$ &$\sfq^3$& $\sfq^4$ & \text{unrefined limit} \\
    \hline
    \hline
    $(1,0,2,1)$ & $\yadj^2$& \parbox{5cm}{\centering$\yadj^{2} \, + \, 2\,\yadj^{3} \, + \, 3\,\yadj^{4} \, + \, \yadj^{5} \, + \, \yadj^{6}$}& \parbox{5cm}{\centering$\yadj^{2} \, + \, 2\,\yadj^{3} \, + \, 7\,\yadj^{4} \, + \, 9\,\yadj^{5} \, + \, 11\,\yadj^{6} \, + \, 7\,\yadj^{7} \, + \, 5\,\yadj^{8} \, + \, \yadj^{9} \, + \, \yadj^{10}$}&$\sfq^2\,+\,8\,\sfq^3\,+\,44\,\sfq^4$\\
    \hline
    $(2,1,1,0)$ & $\yadj$ & \parbox{5cm}{\centering$\yadj \, + \, \yadj^{2} \, + \, 3\,\yadj^{3} \, + \, 2\,\yadj^{4} \, + \, \yadj^{5}$} & \parbox{5cm}{\centering$\yadj \, + \, \yadj^{2} \, + \, 5\,\yadj^{3} \, + \, 7\,\yadj^{4} \, + \, 11\,\yadj^{5} \, + \, 9\,\yadj^{6} \, + \, 7\,\yadj^{7} \, + \, 2\,\yadj^{8} \, + \, \yadj^{9}$} & $\sfq^2\,+\,8\,\sfq^3\,+\,44\,\sfq^4$\\
    \hline
    $(1,1,2,0)$ & $1$ & \parbox{5cm}{\centering$1 \, + \, \yadj \, + \, 3\,\yadj^{2} \, + \, 2\,\yadj^{3} \, + \, \yadj^{4}$}&\parbox{5cm}{\centering$1 \, + \, \yadj \, + \, 5\,\yadj^{2} \, + \, 7\,\yadj^{3} \, + \, 11\,\yadj^{4} \, + \, 9\,\yadj^{5} \, + \, 7\,\yadj^{6} \, + \, 2\,\yadj^{7} \, + \, \yadj^{8}$}&$\sfq^2\,+\,8\,\sfq^3\,+\,44\,\sfq^4$\\
   \hline
   $(2,0,1,1)$ & $\yadj^3$ & \parbox{5cm}{\centering$\yadj^{3} \, + \, 2\,\yadj^{4} \, + \, 3\,\yadj^{5} \, + \, \yadj^{6} \, + \, \yadj^{7}$}&\parbox{5cm}{\centering$\yadj^{3} \, + \, 2\,\yadj^{4} \, + \, 7\,\yadj^{5} \, + \, 9\,\yadj^{6} \, + \, 11\,\yadj^{7} \, + \, 7\,\yadj^{8} \, + \, 5\,\yadj^{9} \, + \, \yadj^{10} \, + \, \yadj^{11}$}&$\sfq^2\,+\,8\,\sfq^3\,+\,44\,\sfq^4$\\
   \hline
   $(2,0,3,1)$ & $0$ & {$\yadj^5$} & \parbox{5cm}{\centering$\yadj^{5} \, + \, 2\,\yadj^{6} \, + \, 3\,\yadj^{7} \, + \, \yadj^{8} \, + \, \yadj^{9}$} & $\sfq^3\,+\,8\,\sfq^4$\\
   \hline
   $(3,1,2,0)$ & $0$ & {$\yadj^2$} & \parbox{5cm}{\centering$\yadj^{2} \, + \, \yadj^{3} \, + \, 3\,\yadj^{4} \, + \, 2\,\yadj^{5} \, + \, \yadj^{6}$} & $\sfq^3\,+\,8\,\sfq^4$\\
   \hline
   $(2,1,3,0)$ & $0$ & {$\yadj$} & \parbox{5cm}{\centering$\yadj \, + \, \yadj^{2} \, + \, 3\,\yadj^{3} \, + \, 2\,\yadj^{4} \, + \, \yadj^{5}$} & $\sfq^3\,+\,8\,\sfq^4$\\
   \hline
   $(3,0,2,1)$ & $0$ & {$\yadj^6$} & \parbox{5cm}{\centering$\yadj^{6} \, + \, 2\,\yadj^{7} \, + \, 3\,\yadj^{8} \, + \, \yadj^{9} \, + \, \yadj^{10}$} & $\sfq^3\,+\,8\,\sfq^4$\\
       \hline
    $(1,1,2,2)$ & $0$ & {$\yadj^4$} & \parbox{5cm}{\centering$\yadj^{4} \, + \, 2\,\yadj^{5} \, + \, 4\,\yadj^{6} \, + \, 2\,\yadj^{7}$} & $\sfq^3\,+\,9\,\sfq^4$\\
     \hline
    $(2,2,1,1)$&$0$&$\yadj^3$&\parbox{5cm}{\centering$2\,\yadj^{4} \, + \, 4\,\yadj^{5} \, + \, 2\,\yadj^{6} \, + \, \yadj^{7}$}&$\sfq^3\,+\,9\,\sfq^4$\\
    \hline
    $(1,2,2,1)$ & $0$ & {$\yadj^2$} & \parbox{5cm}{\centering$2\,\yadj^{3} \, + \, 4\,\yadj^{4} \, + \, 2\,\yadj^{5} \, + \, \yadj^{6}$}& $\sfq^3\,+\,9\,\sfq^4$\\
    \hline
      $(2,1,1,2)$ & $0$ & {$\yadj^5$} & \parbox{5cm}{\centering$\yadj^{5} \, + \, 2\,\yadj^{6} \, + \, 4\,\yadj^{7} \, + \, 2\,\yadj^{8}$}& $\sfq^3\,+\,9\,\sfq^4$\\
   \hline
   $(3,0,4,1)$ & $0$ & {$0$} & {$\yadj^8$} & $\sfq^4$\\
    \hline
     $(4,1,3,0)$ & $0$ & {$0$} & {$\yadj^3$} & $\sfq^4$\\
   \hline
    $(3,1,4,0)$ & $0$ & {$0$} & {$\yadj^2$} & $\sfq^4$\\
    \hline
    $(4,0,3,1)$ & $0$ & {$0$} & {$\yadj^9$} & $\sfq^4$\\
    \hline
    $(1,2,2,3)$&$0$&$0$&$\yadj^6$&$\sfq$\\
       \hline
    $(2,3,1,2)$&$0$&$0$&$\yadj^5$&$\sfq$\\
      \hline
    $(1,3,2,2)$&$0$&$0$&$\yadj^4$&$\sfq$\\
      \hline
    $(2,2,1,3)$&$0$&$0$&$\yadj^7$&$\sfq$\\
    \hline
\end{longtable}
\end{center}

\subsection{For \texorpdfstring{$c_1 = (0,0)$}{c100}}\label{app:F0 c1=(0,0)}

\begin{center}
    \renewcommand{\arraystretch}{2.5}
    \begin{longtable}[!h]{|c||c|c|c|c|c|}
    \hline
    $\vec{\sfp}$& $\sfq^2$ &$\sfq^3$& $\sfq^4$ \\
    \hline
    \hline
    $(0,1,2,1)$ & $\yadj^3$&\parbox{5cm}{\centering$\yadj^{3} \, + \, \yadj^{4} \, + \, 4\,\yadj^{5} \, + \, \yadj^{6} \, + \, \yadj^{7}$} &\parbox{5cm}{\centering$\yadj^{3} \, + \, \yadj^{4} \, + \, 6\,\yadj^{5} \, + \, 7\,\yadj^{6} \, + \, 14\,\yadj^{7} \, + \, 7\,\yadj^{8} \, + \, 6\,\yadj^{9} \, + \, \yadj^{10} \, + \, \yadj^{11}$}\\     
\hline
$(2,1,0,1)$ & $\yadj^2$ &\parbox{5cm}{\centering $\yadj^{2} \, + \, \yadj^{3} \, + \, 4\,\yadj^{4} \, + \, \yadj^{5} \, + \, \yadj^{6}$}&\parbox{5cm}{\centering $\yadj^{2} \, + \, \yadj^{3} \, + \, 6\,\yadj^{4} \, + \, 7\,\yadj^{5} \, + \, 14\,\yadj^{6} \, + \, 7\,\yadj^{7} \, + \, 6\,\yadj^{8} \, + \, \yadj^{9} \, + \, \yadj^{10}$}\\
\hline
{$(1,1,1,1)$} &\parbox{4cm}{\centering$\frac{1}{2}({\yadj^{-1}}
\,+\,2
\,+\,6\,\yadj
\,+\,7\,\yadj^{2}
\,+\,7\,\yadj^{3}
\,+\,6\,\yadj^{4}
\,+\,2\,\yadj^{5}
\,+\,\yadj^{6})$}&\parbox{5cm}{\centering$\frac{1}{2}({\yadj^{-1}}
\,+\,2
\,+\,8\,\yadj
\,+\,15\,\yadj^{2}
\,+\,28\,\yadj^{3}
\,+\,34\,\yadj^{4}
\,+\,34\,\yadj^{5}
\,+\,28\yadj^{6}
\,+\,15\yadj^{7}
\,+\,8\yadj^{8}
\,+\,2\yadj^{9}
\,+\,\yadj^{10})$}&\parbox{5cm}{\centering$\frac{1}{2}({\yadj^{-1}}
\,+\,2
\,+\,8\,\yadj
\,+\,17\,\yadj^{2}
\,+\,41\,\yadj^{3}
\,+\,71\,\yadj^{4}
\,+\,111\,\yadj^{5}
\,+\,133\,\yadj^{6}
\,+\,133\,\yadj^{7}
\,+\,111\,\yadj^{8}
\,+\,71\,\yadj^{9}
\,+\,41\,\yadj^{10}
\,+\,17\,\yadj^{11}
\,+\,8\,\yadj^{12}
\,+\,2\,\yadj^{13}
\,+\,\yadj^{14})$}\\
\hline
{$(1,2,1,2)$} &\parbox{4cm}{\centering$\frac{1}{2}(\yadj
\,+\,\yadj^{2}
\,+\,\yadj^{3}
\,+\,\yadj^{4})$}&\parbox{5cm}{\centering$\frac{1}{2}(2\,\yadj^{2}
\,+\,6\,\yadj^{3}
\,+\,8\,\yadj^{4}
\,+\,8\,\yadj^{5}
\,+\,6\,\yadj^{6}
\,+\,6\,\yadj^{7}
\,+\,2\,\yadj^{8})$}&\parbox{5cm}{\centering$\frac{1}{2}(4\,\yadj^{3}
\,+\,13\,\yadj^{4}
\,+\,31\,\yadj^{5}
\,+\,40\,\yadj^{6}
\,+\,40\,\yadj^{7}
\,+\,31\,\yadj^{8}
\,+\,13\,\yadj^{9}
\,+\,4\,\yadj^{10})$}\\
\hline
{$(2,1,2,1)$} &\parbox{4cm}{\centering$\frac{1}{2}(1
\,+\,\yadj
\,+\,\yadj^{4}
\,+\,\yadj^{5})$}&\parbox{5cm}{\centering$\frac{1}{2}(1
\,+\,2\,\yadj
\,+\,5\,\yadj^{2}
\,+\,5\,\yadj^{3}
\,+\,3\,\yadj^{4}
\,+\,3\,\yadj^{5}
\,+\,5\,\yadj^{6}
\,+\,5\,\yadj^{7}
\,+\,2\yadj^{8}
\,+\,\yadj^{9})$}&\parbox{5cm}{\centering$\frac{1}{2}(1
\,+\,2\,\yadj
\,+\,7\,\yadj^{2}
\,+\,13\,\yadj^{3}
\,+\,22\,\yadj^{4}
\,+\,23\,\yadj^{5}
\,+\,20\,\yadj^{6}
\,+\,20\,\yadj^{7}
\,+\,23\,\yadj^{8}
\,+\,22\,\yadj^{9}
\,+\,13\,\yadj^{10}
\,+\,7\,\yadj^{11}
\,+\,2\,\yadj^{12}
\,+\,\yadj^{13})$}\\
\hline
$(1,1,3,1)$ &$0$&$\yadj^4\,+\,\yadj^6$&\parbox{5cm}{\centering$\yadj^{4} \, + \, \yadj^{5} \, + \, 4\,\yadj^{6} \, + \, 3\,\yadj^{7} \, + \, 5\,\yadj^{8} \, + \, \yadj^{9} \, + \, \yadj^{10}$}\\
\hline
$(3,1,1,1)$ &$0$&$\yadj^3\,+\,\yadj^5$&\parbox{5cm}{\centering$\yadj^{3} \, + \, \yadj^{4} \, + \, 5\,\yadj^{5} \, + \, 3\,\yadj^{6} \, + \, 4\,\yadj^{7} \, + \, \yadj^{8} \, + \, \yadj^{9}$}\\
\hline
{$(1,3,1,3)$}&$0$&\parbox{5cm}{\centering$\frac{1}{2}({\yadj^{3}} \, + \, {\yadj^{4}} \, + \, {\yadj^{5}} \, + \, {\yadj^{6}})$}&\parbox{5cm}{\centering$\frac{1}{2}(2\,\yadj^{4}\, +\, 6\,\yadj^{5} \,+\, 8\,\yadj^{6} \,+\, 8\,\yadj^{7} \,+\, 6\,\yadj^{8}\, + \,6\,\yadj^{9}\, +\, 2\,\yadj^{10})$}\\
\hline
{$(3,1,3,1)$} &$0$&\parbox{5cm}{\centering$\frac{1}{2}({\yadj} \,+\, {\yadj^{2}} \,+\, {\yadj^{7}}\, +\,{\yadj^{8}})$}&\parbox{5cm}{\centering$\frac{1}{2}(\yadj \, + \,2\, \yadj^{2} \, + \, {5}\,\yadj^{3} \, + \, 
{5}\,\yadj^{4} \, + \, 2\,\yadj^{5} \, + \, \yadj^{6} \, + \,\yadj^{7} \, + \,2\, \yadj^{8} \, + \, {5}\,\yadj^{9} \, + \, {5}\,\yadj^{10} \, + \,2\, \yadj^{11} \, + \, \yadj^{12})$}\\
\hline
$(2,1,4,1)$ &$0$&$0$&$\yadj^5\,+\,\yadj^9$\\
\hline
$(4,1,2,1)$ &$0$&$0$&$\yadj^4\,+\,\yadj^8$\\
\hline
$(0,2,2,2)$ &$0$&$0$&$\yadj^7$\\
\hline
$(2,2,2,0)$ &$0$&$0$&$\yadj^6$\\
\hline
$(1,0,3,2)$ &$0$&$0$&$\yadj^6$\\
\hline
$(3,2,1,0)$ &$0$&$0$&$\yadj^7$\\
\hline
$(1,2,3,0)$ &$0$&$0$&$\yadj^5$\\
\hline
$(3,0,1,2)$&$0$&$0$&$\yadj^8$\\
\hline
$(2,0,2,2)$ &$0$&$0$&$\yadj^7$\\
\hline
$(2,2,0,2)$ &$0$&$0$&$\yadj^6$\\
\hline
\end{longtable}
\end{center}

\section{Explicit results for refined VW invariants of \texorpdfstring{$\mathbb{F}_1$}{F1}}\label{app:F1Results}
Here, we record the explicit expressions for the  orbit contributions of the different magnetic fluxes $\vec{\sfp}$ to the first two non-trivial terms in the refined VW invariants of $\mathbb{F}_1$. 
This is done for the four possible choices of the first Chern class $c_1=(1,0),(0,1),(1,1)$, and $(0,0)$. For simplicity, we restrict to the non-equivariant limit $y_2\to 0$. In all cases, choose the polarization
to be $(\boldsymbol{a},\boldsymbol{b}) = (1.1,1)$.

\subsection{For \texorpdfstring{$c_1 = (0,1)$}{c100}}\label{app:F1 c1=(0,1)}
\begin{center}
    \renewcommand{\arraystretch}{2.5}
    \begin{longtable}[!h]{|c||c|c|c|c|c|}
    \hline
    $\vec{\sfp}$& $\sfq$& $\sfq^2$ & unrefined limit \\
    \hline
    \hline
    $(1,0,1,1)$& $1$ & $1\,+\,2\,\yadj\,+\,3\,\yadj^2\,+\,\yadj^3\,+\,\yadj^4$ & $\sfq\,+\,8\,\sfq^2$\\
    \hline
    $(2,0,2,1)$ & $0$ & $\yadj^3$ & $0\,\sfq\,+\,\sfq^2$\\
    \hline 
    $(1,1,2,2)$ & $0$ & $\yadj$ & $0\,\sfq\,+\,\sfq^2$\\
    \hline
    $(1,2,1,1)$ & $0$ & $\yadj^2\,+\,\yadj^3$ & $0\,\sfq\,+\,2\,\sfq^2$\\
    \hline
    $(2,1,1,2)$ & $0$ & $\yadj^2$ & $0\,\sfq\,+\,\sfq^2$\\
\hline
\end{longtable}
\end{center}

\subsection{For \texorpdfstring{$c_1 = (1,0)$}{c100}}
\label{app:F1 c1=(1,0)}
\begin{center}
    \renewcommand{\arraystretch}{2.5}
    \begin{longtable}[!h]{|c||c|c|c|c|c|}
    \hline
    $\vec{\sfp}$& $\sfq$& $\sfq^2$ & unrefined limit \\
    \hline
    \hline
    $(1,1,1,1)$ & $1\,+\,\yadj$ & $1\,+\,2\,\yadj\,+\,5\,\yadj^2\,+\,5\,\yadj^3\,+\,2\,\yadj^4\,+\,\yadj^5$&$2\,\sfq\,+\,16\,\sfq^2$\\
    \hline
    $(1,0,2,2)$ & $0$ & $\yadj$ & $0\,\sfq\,+\,\sfq^2$\\
    \hline
    $(2,0,1,2)$ & $ 0$ & $\yadj^2$ & $0\,\sfq\,+\,\sfq^2$\\
    \hline
    $(2,1,2,1)$ & $0$ & $\yadj^3\,+\,\yadj^4$ & $0\,\sfq\,+\,2\,\sfq^2$ \\
    \hline
    $(1,2,2,2)$ & $0$ & $\yadj^2$ & $0\,\sfq\,+\,\sfq^2$\\
    \hline
    $(2,2,1,2)$ & $0$ & $\yadj^3$ & $0\,\sfq\,+\,\sfq^2$\\
\hline
\end{longtable}
\end{center}

\subsection{For \texorpdfstring{$c_1 = (1,1)$}{c111}}
\label{app:F1 c1=(1,1)}
\begin{center}
    \renewcommand{\arraystretch}{2.5}
    \begin{longtable}[!h]{|c||c|c|c|c|c|}
    \hline
    $\vec{\sfp}$& $\sfq^2$& $\sfq^3$ & unrefined limit \\
    \hline
      \hline
    $(1,1,1,0)$ & $1$ & $1\,+\,\yadj\,+\,3\,\yadj^2\,+\,2\,\yadj^3\,+\,\yadj^4$ & $\sfq^2\,+\,8\,\sfq^2$\\
    \hline
    $(1,0,2,1)$ & $\yadj$ & $\yadj\,+\,2\,\yadj^2\,+\,3\,\yadj^3\,+\,\yadj^4\,+\,\yadj^5$&$\sfq^2\,+\,8\,\sfq^3$\\
    \hline
    $(2,0,1,1)$ & $\yadj^2$ & $\yadj^2\,+\,2\,\yadj^3\,+\,3\,\yadj^4\,+\,\yadj^5\,+\,\yadj^6$ & $\sfq^2\,+\,8\,\sfq^3$\\
      \hline
    $(0,2,1,1)$ & $0$&$\yadj^3$& $0\,\sfq^2\,+\,\sfq^3$\\
    \hline
    $(1,2,0,1)$ & $0$ & $\yadj^2$ & $0\,\sfq^2\,+\,\sfq^3$\\
    \hline
    $(2,0,3,1)$ & $0$ & $\yadj^4$ & $0\,\sfq^2\,+\,\sfq^3$\\
    \hline
    $(2,1,2,0)$ & $0$ & $\yadj$ & $0\,\sfq^2\,+\,\sfq^3$\\
    \hline
    $(3,0,2,1)$ & $0$ & $\yadj^5$ & $0\,\sfq^2\,+\,\sfq^3$\\
    \hline
    $(1,1,3,2)$ & $0$ & $\yadj^3$ & $0\,\sfq^2\,+\,\sfq^3$\\
    \hline
    $(3,1,1,2)$ & $0$ & $\yadj^4$ & $0\,\sfq^2\,+\,\sfq^3$ \\
\hline
\end{longtable}
\end{center}

\subsection{For \texorpdfstring{$c_1 = (0,0)$}{c100}}\label{app:F1 c1=(0,0)}
\begin{center}
    \renewcommand{\arraystretch}{2.5}
    \begin{longtable}[!h]{|c||c|c|c|c|c|}
    \hline
    $\vec{\sfp}$& $\sfq$& $\sfq^2$ &\text{unrefined limit} \\
    \hline
      \hline
    $(0,1,1,1)$&$\frac{1}{2}$&$\frac{1}{2}(1
\,+\,y_{\mathrm{adj}}
\,+\,4\,y_{\mathrm{adj}}^2
\,+\,y_{\mathrm{adj}}^3
\,+\,y_{\mathrm{adj}}^4)$&$\frac{1}{2}\,\sfq\,+\,\frac{8}{2}\,\sfq^2$\\
\hline
$(1,1,0,1)$ & $\frac{1}{2\,\yadj}$ &  $\frac{1}{2}(y_{\mathrm{adj}}^{-1}
\,+\,1
\,+\,4\,y_{\mathrm{adj}}
\,+\,y_{\mathrm{adj}}^2
\,+\,y_{\mathrm{adj}}^3)$& $\frac{1}{2}\,\sfq\,+\,\frac{8}{2}\,\sfq^2$\\
\hline
$(1,1,2,1)$ & $\frac{\yadj}{2}$ & $ \frac{1}{2}(2\,y_{\mathrm{adj}}
\,+\,2\,y_{\mathrm{adj}}^2
\,+\,5\,y_{\mathrm{adj}}^3
\,+\,y_{\mathrm{adj}}^4
\,+\,y_{\mathrm{adj}}^5)$&$\frac{1}{2}\,\sfq\,+\,\frac{11}{2}\,\sfq^2$\\
\hline
$(2,1,1,1)$ &$\frac{\yadj^2}{2}$& $\frac{1}{2}(1
\,+\,2\,y_{\mathrm{adj}}^2
\,+\,2\,y_{\mathrm{adj}}^3
\,+\,4\,y_{\mathrm{adj}}^4
\,+\,y_{\mathrm{adj}}^5
\,+\,y_{\mathrm{adj}}^6)$&$\frac{1}{2}\,\sfq\,+\,\frac{11}{2}\,\sfq^2$\\
\hline
$(1,2,1,2)$ & $0$&$\frac{1}{2}(\yadj\,+\,\yadj^2)$ & $\frac{2}{2}\,\sfq^2$\\
\hline
$(1,2,3,2)$ &$0$&$\frac{\yadj}{2}$&$\frac{1}{2}\,\sfq^2$\\
\hline
$(2,1,3,1)$ &$0$&$\frac{\yadj^4}{2}$&$\frac{1}{2}\,\sfq^2$\\
\hline
$(3,1,2,1)$ &$0$&$\frac{\yadj^5}{2}$& $\frac{1}{2}\,\sfq^2$\\
\hline
$(3,2,1,2)$ &$0$&$\frac{\yadj^4}{2}$&$\frac{1}{2}\,\sfq^2$\\
\hline
\end{longtable}
\end{center}
%%%%%%%%%%%%%%%%%%%%%%%%%%%%%%%%%%%%%%%%%
%%%%%%%%%%%%%%%%%%%%%%%%%%%%%%%%%%%%%%%%%
%%%%%%%%%%%%%%%%%%%%%%%%%%%%%%%%%%%%%%%%%
% \newpage
\addcontentsline{toc}{section}{References}
\bibliographystyle{utphys}
\bibliography{Ktheory}
\end{document}